\documentclass[vecphys]{svmult}

\usepackage{amsmath}

\usepackage{makeidx}         
\usepackage{graphicx}        
\usepackage{multicol}        
\usepackage[bottom]{footmisc}

\usepackage{cite}

\usepackage{float}

\makeindex             

\newcommand\beq{\begin{equation}}

\newcommand\eeq{\end{equation}}
\newcommand\beqa{\begin{eqnarray}}
\newcommand\eeqa{\end{eqnarray}}
\newcommand{\nn}{\nonumber\\}

\newcommand{\GG}{\mathcal{G}}

\newcommand{\openone}{\mathsf{I}}

\newcommand{\Sn}{\Omega}

\newcommand{\lambdak}{\vartheta_1}

\newcommand{\lambdakk}{\vartheta_2}

\newcommand{\ee}{\E}

\newcommand{\hs}{\text{HS}}

\newcommand{\hd}{\text{HD}}

\newcommand{\hr}{\text{HR}}

\newcommand{\LJ}{\text{LJ}}

\newcommand{\NN}{\rho}

\newcommand{\yy}{u}

\newcommand{\KK}{J}

\newcommand{\aQ}{Q_0}

\newcommand{\aK}{K}

\newcommand{\aKQ}{Q}

\newcommand{\SE}{E}

\newcommand{\aC}{W}

\newcommand{\diam}{\sigma_0}

\newcommand{\muM}{M}
\newcommand{\mo}{\muM_1}
\newcommand{\mt}{\muM_2}
\newcommand{\mth}{\muM_3}
\newcommand{\mn}{\muM_n}

\newcommand{\pure}{{\text{s}}}
\newcommand\gh{\widetilde{\gamma}}
\newcommand\ph{\widetilde{p}}
\newcommand\Deltah{\widetilde{\Delta}}
\newcommand\etah{\widetilde{\eta}}
\def\zero{{(0)}}
\def\one{{(1)}}
\def\two{{(2)}}
\def\three{{(3)}}

\def\n{{(n)}}

\begin{document}

\title*{Alternative Approaches to the Equilibrium Properties of Hard-Sphere Liquids}
\titlerunning{Alternative Approaches to  Hard-Sphere Liquids}
\author{M. L\'opez de Haro\inst{1}, S. B. Yuste\inst{2} \and
A. Santos\inst{3}}
\institute{Centro de Investigaci\'{o}n en Energ\'{\i}a, Universidad
Nacional Aut\'onoma de M\'exico (U.N.A.M.), Temixco, Morelos 62580,
M{e}xico \\
\texttt{malopez@servidor.unam.mx} \and Departamento de
F\'{\i}sica, Universidad de Extremadura, E-06071 Badajoz, Spain
\texttt{santos@unex.es} \and Departamento de F\'{\i}sica,
Universidad de Extremadura, E-06071 Badajoz, Spain
\texttt{andres@unex.es}}
%
%
\maketitle

An overview of some analytical approaches to the computation of the
structural and thermodynamic properties of single component and
multicomponent  hard-sphere fluids is provided. For the structural
properties, they yield a thermodynamically consistent formulation,
thus improving and extending the known analytical results of the
Percus--Yevick theory. Approximate expressions for the contact
values of the radial distribution functions and the corresponding
analytical equations of state  are also discussed. Extensions of
this methodology to related systems, such as sticky hard spheres and
square-well fluids, as well as its use in connection with the
perturbation theory of fluids are briefly addressed.

\section{Introduction}
\label{sec1}
In the statistical thermodynamic approach to the theory of simple
liquids, there is a close connection between the thermodynamic and
structural properties \cite{BH76,M76,F85,HM86}. These properties
depend on the intermolecular potential of the system, which is
generally assumed to be well represented by pair interactions. The
simplest model pair potential   is that of a hard-core fluid (rods,
disks, spheres, hyperspheres) in which attractive forces are
completely neglected. In fact, it is a model that has been most
studied and has rendered some analytical results, although up to
this day no general (exact) explicit expression for the equation of
state is available, except for the one-dimensional case. Something
similar applies to the structural properties. An interesting feature
concerning the thermodynamic properties is that in hard-core systems
the equation of state depends only on the contact values of the
radial distribution functions. In the absence of a completely
analytical approach, the most popular methods to deal with both
kinds of properties of these systems are integral equation theories
and computer simulations.

It is well known that in real gases and liquids at high temperatures
the state and thermodynamic properties are determined almost
entirely by the repulsive forces among molecules. At lower
temperatures, attractive forces become significant, but even in this
case they affect very little the configuration of the system at
moderate and high densities. These facts are taken into account in
the application of the perturbation theory of fluids, where
hard-core fluids are used as the reference systems in the
computation of the thermodynamic and structural properties of real
fluids. However, successful results using perturbation theory are
rather limited due to the fact that, as mentioned above, there are
in general no exact (analytical) expressions for the thermodynamic
and structural properties of the reference systems which are in
principle required in the calculations. On the other hand, in the
realm of soft condensed matter the use of the hard-sphere model in
connection, for instance, with sterically stabilized colloidal
systems is quite common. This is due to the fact that nowadays it is
possible to prepare (almost) monodisperse spherical colloidal
particles with short-ranged harshly repulsive interparticle forces
that may be well described theoretically with the hard-sphere
potential.

This chapter presents an overview of the efforts we have made over
the last few years to compute the thermodynamic and structural
properties of hard-core systems using relatively simple
(approximate) analytical methods. It is structured as follows. In
Section \ref{sec2} we describe our proposals to derive the contact
values of the radial distribution functions of a multicomponent
mixture (with an arbitrary size distribution, either discrete or
continuous) of $d$-dimensional hard spheres from the use of some
consistency conditions and the knowledge of the contact value of the
radial distribution function of the corresponding  single component
system. In turn, these contact values lead to equations of state
both for additive and non-additive hard spheres. Some consequences
of such equations of state, in particular the demixing transition,
are briefly analyzed. This is followed in Section \ref{sec3} by the
description of the Rational Function Approximation method to obtain
analytical expressions for the structural quantities of
three-dimensional single component  and multicomponent fluids. The
only required inputs in this approach are the contact values of the
radial distribution functions and so the connection with the work of
the previous section follows naturally. Structural properties of
related systems, like sticky hard spheres or square-well fluids,
that may also be tackled with the same philosophy  are also
discussed in Section \ref{sec3bis}. Section \ref{sec4} provides an
account of the reformulation of the perturbation theory of liquids
using the results of the Rational Function Approximation method for
a single component hard-sphere fluid and its illustration in the
case of the Lennard--Jones fluid. In the final section, we provide
some perspectives of the achievements obtained so far and of the
challenges that remain ahead.

\section{Contact Values and Equations of State for Mixtures}
\label{sec2} As stated in the Introduction,  a nice feature of
hard-core fluids is that the expressions of all their thermodynamic
properties in terms of the radial distribution functions (RDF) are
particularly simple. In fact, for these systems the internal energy
reduces to that of the ideal gas and in the pressure equation it is
only the contact values rather than the full RDF which appear
explicitly. In this section we present our approach to the
derivation of the contact values of hard-core fluid mixtures in $d$
dimensions.

\subsection{Additive Systems in $d$ Dimensions}
\label{ss2.1}

If $\sigma_{ij}$ denotes the distance of separation at contact
between the centers of two interacting fluid particles, one of
species $i$ and the other of species $j$, the mixture is said to be
\emph{additive} if $\sigma_{ij}$ is just the arithmetic mean of the
hard-core diameters of each species. Otherwise, the system is
\emph{non-additive}. We deal in this subsection and in Subsection
\ref{sss2.1.2} with additive systems, while non-additive hard-core
mixtures will be treated in Subsection \ref{ss2.2}.

\subsubsection{Definitions}

Let us consider an additive mixture of hard spheres (HS) in $d$
dimensions with an arbitrary number $N$ of components. In fact, our
discussion will remain valid for $N\to\infty$, {i.e.}, for
polydisperse mixtures with a continuous distribution of sizes.

The {additive}  hard core of the interaction between a sphere of
species $i$ and a sphere of species $j$ is $\sigma_{ij}=\frac{1}{2
}(\sigma _{i}+\sigma _{j})$, where the diameter of a sphere of
species $i$ is $\sigma _{ii}=\sigma _{i}$. Let the number density of
the mixture be $\rho $ and the mole fraction of species $i$ be
$x_{i}=\rho _{i}/\rho $, where $\rho_i$ is the number density of
species $i$. {}From these quantities one can define the packing
fraction $\eta =v_{d}\rho \muM_d $, where $v_{d}=(\pi
/4)^{d/2}/\Gamma (1+d/2)$ is the volume of a $d$-dimensional sphere
of unit diameter and
\beq
\muM_n \equiv \langle \sigma^n\rangle=\sum_{i=1}^N x_{i}\sigma
_{i}^{n}
\label{moments}
\eeq
 denotes the $n$th moment of the diameter distribution.

In a HS mixture, the knowledge of the contact values
$g_{ij}(\sigma_{ij})$ of the RDF $g_{ij}(r)$, where $r$ is the
distance, is important for a number of reasons. For example, the
availability of $g_{ij}(\sigma _{ij})$ is sufficient to get the
equation of state (EOS) of the mixture via the virial expression
\begin{equation}
Z(\eta )=1+\frac{2^{d-1}}{\muM_d} \eta\sum_{i,j=1}^N
x_{i}x_{j}{\sigma _{ij}^{d} } g_{ij}(\sigma_{ij}),
\label{1}
\end{equation}
where $Z=p/\rho k_{B}T$ is the compressibility factor of the
mixture, $p$ being the pressure, $k_{B}$ the Boltzmann constant, and
$T$ the absolute temperature.

The exact form of $g_{ij}(\sigma _{ij})$ as functions of the packing
fraction $\eta$, the set of diameters $\{\sigma _{k}\}$, and the set
of mole fractions $\{x_{k}\}$ is only known in the one-dimensional
case, where one simply has \cite{LZ71}
\beq
g_{ij}(\sigma _{ij})=\frac{1}{1-\eta }, \quad (d=1).
\label{exact1D}
\eeq
Consequently, {for $d\geq 2$} one has to resort to approximate
theories or empirical expressions. For hard-disk mixtures, an
accurate expression is provided by Jenkins and Mancini's (JM)
approximation \cite{JM87,BS01},
\beq
g_{ij}^{\text{JM}}(\sigma_{ij})=\frac{1}{1-\eta}+\frac{9}{16}\frac{\eta}{(1-\eta)^2}\frac{\sigma_i\sigma_j
\muM_1}{\sigma_{ij}\muM_2}, \quad (d=2).
\label{JM}
\eeq
The associated compressibility factor is
\beq
Z_\text{JM}(\eta)=\frac{1}{1-\eta}+\frac{\muM_1^2}{\muM_2}\eta\frac{1+\eta/8}{(1-\eta)^2},\quad
(d=2).
\label{ZJM}
\eeq
 In the case of three-dimensional systems, some important analytical expressions for the
contact values and the corresponding compressibility factor also
exist. For instance, the expressions which follow from the solution
of the Percus--Yevick (PY) equation of additive HS mixtures by
Lebowitz \cite{L64} are
\beq
g_{ij}^{\text{PY}}(\sigma_{ij})=\frac{1}{1-\eta
}+\frac{3}{2}\frac{\eta }{(1-\eta )^{2}}\frac{\sigma_i\sigma_j
\muM_2}{\sigma_{ij}\muM_3},\quad (d=3),
\label{15PY}
\eeq
\beq
Z_\text{PY}(\eta)=\frac{1}{1-\eta}+\frac{\muM_1\muM_2}{\muM_3}\frac{3\eta}{(1-\eta)^2}+\frac{\muM_2^3}{\muM_3^2}
\frac{3\eta^2}{(1-\eta)^2},\quad (d=3).
\label{ZPY}
\eeq
Also analytical are the results obtained from the Scaled Particle
Theory (SPT) \cite{RFL59,MR75,R88,HC04},
\beq
g_{ij}^{\text{SPT}}(\sigma_{ij})=\frac{1}{1-\eta
}+\frac{3}{2}\frac{\eta }{(1-\eta )^{2}}\frac{\sigma_i\sigma_j
\muM_2}{\sigma_{ij}\muM_3}+\frac{3}{4}\frac{\eta^{2}}{(1-\eta)^{3}}\left(\frac{\sigma_i\sigma_j
\muM_2}{\sigma_{ij}\muM_3}\right)^{2},\quad (d=3),
\label{15SPT}
\eeq
\beq
Z_\text{SPT}(\eta)=\frac{1}{1-\eta}+\frac{\muM_1\muM_2}{\muM_3}\frac{3\eta}{(1-\eta)^2}+\frac{\muM_2^3}{\muM_3^2}
\frac{3\eta^2}{(1-\eta)^3},\quad (d=3).
\label{ZSPT}
\eeq
 Neither the PY nor the SPT lead to particularly accurate
values and so Boubl\'{\i}k \cite{B70} and, independently, Grundke
and Henderson \cite{GH72} and Lee and Levesque \cite{LL73} proposed
an interpolation between the PY and the SPT contact values, that we
will refer to as the BGHLL values:
\beq
g_{ij}^{\text{BGHLL}}(\sigma_{ij})=\frac{1}{1-\eta
}+\frac{3}{2}\frac{\eta }{(1-\eta )^{2}}\frac{\sigma_i\sigma_j
\muM_2}{\sigma_{ij}\muM_3}+\frac{1}{2}\frac{\eta^{2}}{(1-\eta)^{3}}\left(\frac{\sigma_i\sigma_j
\muM_2}{\sigma_{ij}\muM_3}\right)^{2},\quad (d=3).
\label{15BGHLL}
\eeq
This  leads through Eq.\ \eqref{1} to the widely used and rather
accurate Boubl\'{\i}k--Mansoori--Carnahan--Starling--Leland (BMCSL)
EOS \cite{B70,MCSL71} for  HS mixtures:
\begin{equation}
Z_{\text{BMCSL}}(\eta )=\frac{1}{1-\eta
}+\frac{\muM_1\muM_2}{\muM_3}\frac{3\eta }{(1-\eta
)^{2}}+\frac{\muM_2^3}{\muM_3^2}\frac{\eta^{2}(3-\eta )}{(1-\eta
)^{3}},\quad (d=3).
 \label{BMCSL}
\end{equation}
 Refinements of the BGHLL
values have been subsequently introduced, among others, by Henderson
{et al.} \cite{HMLC96,YCH96,HC98,HBCW98,MHC99,CCHW00}, Matyushov and
Ladanyi \cite{ML97}, and Barrio and Solana \cite{BS00} to eliminate
some drawbacks of the BMCSL EOS in the so-called colloidal limit of
{binary} HS mixtures. On a different path, but also having to do
with the colloidal limit, Viduna and Smith \cite{VS02} have proposed
a method to obtain contact values of the RDF of HS mixtures from a
given EOS. However, none of these proposals may be easily
generalized so as to be valid for any dimensionality and any number
of components. Therefore, if one wants to have a more general
framework able to deal with arbitrary $d$ and $N$ an alternative
strategy is called for.

\subsubsection{Universality Ansatz}

In order to follow our alternative strategy, it is useful to make
use of exact limit results that can help one in the construction of
approximate expressions for $g_{ij}(\sigma _{ij})$. Let us consider
first the limit in which one of the species, say $i$, is made of
point particles, {i.e.}, $\sigma _{i}\rightarrow 0$. In that case,
$g_{ii}(\sigma _{i})$ takes the ideal gas value, except that one has
to take into account that the available volume fraction is $1-\eta$.
Thus,
\begin{equation}
\lim_{\sigma_{i}\rightarrow
0}g_{ii}(\sigma_{i})=\frac{1}{1-\eta}.
\label{2}
\end{equation}
An even simpler situation occurs when all the species have the same
size, $\{\sigma _{k}\}\rightarrow \sigma $, so that the system
becomes equivalent to a single component system. Therefore,
\begin{equation}
\lim_{\{ \sigma _{k}\}\rightarrow \sigma
}g_{ij}(\sigma_{ij})=g_{\pure}, \label{3}
\end{equation}
where $g_{\pure}$ is the contact value of the RDF of the single
component fluid at the same packing fraction $\eta$ as that of the
mixture. Table \ref{Tableg_s} lists some of the most widely used
proposals for the contact value $g_\pure$ and the associated
compressibility factor
\beq
Z_\pure=1+2^{d-1}\eta g_\pure
\label{Z_s}
\eeq
in the case of the single component HS fluid.
\begin{table}
\centering \caption{Some expressions of $g_\pure$ and $Z_\pure$ for
the single component HS fluid. In the SHY proposal,
$\eta_{\text{cp}}=(\sqrt{3}/6)\pi$ is the crystalline close-packing
fraction for hard disks. In the LM proposal, $b_3$ and $b_4$ are the
(reduced) third and fourth virial coefficients, $\zeta(\eta
)=1.2973(59)-0.062(13)\eta/\eta_{\text{cp}}$ for $d=4$, and
$\zeta(\eta )=1.074(16)+0.163(45)\eta/\eta_{\text{cp}}$ for $d=5$,
where the values of the close-packing fractions are
$\eta_{\text{cp}}=\pi^2/16\simeq 0.617$ and
$\eta_{\text{cp}}=\pi^2\sqrt{2}/30\simeq 0.465$ for $d=4$ and $d=5$,
respectively. }
\label{Tableg_s}       %
\begin{tabular}{ccccc}
\hline
$d$& $g_\pure$&$Z_\pure$&Label&Ref.\\
\hline
\noalign{\smallskip}
 $2$&$\dfrac{1-7\eta/16}{(1-\eta)^2}$&$\dfrac{1+\eta^2/8}{(1-\eta)^2}$&H&\protect\cite{H75}\\
\noalign{\smallskip}
 $2$&$\dfrac{1-\eta(2\eta_{\text{cp}}-1)/2\eta_{\text{cp}}^2}{1-2\eta+\eta^2(\eta_{\text{cp}}-1)/2\eta_{\text{cp}}^2}$&
$\dfrac{1}{1-2\eta+\eta^2(\eta_{\text{cp}}-1)/2\eta_{\text{cp}}^2}$&SHY&\protect\cite{SHY95}\\
 \noalign{\smallskip}
 $2$&$g_{\pure}^{\text{H}}-\dfrac{\eta^3}{2^7(1-\eta)^4}$&$Z_{\pure}^{\text{H}}-\dfrac{\eta^4}{2^6(1-\eta)^4}$&L&\protect\cite{L01}\\
\noalign{\smallskip}
 $3$&$\dfrac{1+\eta/2}{(1-\eta)^2}$&$\dfrac{1+2\eta+3\eta^2}{(1-\eta)^2}$&PY&\protect\cite{W63}\\
\noalign{\smallskip}
 $3$&$\dfrac{1-\eta /2+\eta ^{2}/4}{(1-\eta
 )^{3}}$&$\dfrac{1+\eta+\eta^2}{(1-\eta)^3}$&SPT&\protect\cite{RFL59}\\
\noalign{\smallskip}
 $3$&$\dfrac{1-\eta /2}{(1-\eta )^{3}}$&$\dfrac{1+\eta+\eta^2-\eta^3}{\left(1-\eta\right)^3}$&CS&\protect\cite{CS69}\\
\noalign{\smallskip}
 $4,5$&$\dfrac{1+[
2^{1-d}b_{3}-\zeta (\eta )b_{4}/b_{3}] \eta }{1-\zeta (\eta
)(b_{4}/b_{3})\eta +\left[ \zeta (\eta )-1\right]
2^{1-d}b_{4}\eta ^{2}}$&$1+2^{d-1}\eta g_\pure^{\text{LM}}$ &LM&\protect\cite{LM90}\\
\noalign{\smallskip}
\hline
\end{tabular}
\end{table}

Equations (\ref{2}) and (\ref{3}) represent the simplest and most
basic conditions that $g_{ij}(\sigma _{ij})$ must satisfy. There is
a number of other {less trivial} consistency conditions
\cite{R88,HMLC96,HC98,HBCW98,ML97,BS00,H94,V98,THM99}, some of which
will be used later on.

In order to proceed, in line with a property shared by earlier
proposals [see, in particular, Eqs.\ \eqref{JM}, \eqref{15PY},
\eqref{15SPT}, and \eqref{15BGHLL}], we assume that, at a given
packing fraction $\eta$, the dependence of $g_{ij}(\sigma_{ij})$ on
the parameters $\{\sigma _{k}\}$ and $\{x_{k}\}$ takes place
\textit{only} through the scaled quantity
\begin{equation}
z_{ij}\equiv \frac{\sigma _{i}\sigma_{j}}{\sigma
_{ij}}\frac{\muM_{d-1}}{\muM_d}.
\label{zij}
\end{equation}
More specifically, we assume
\begin{equation} g_{ij}(\sigma
_{ij})=\GG(\eta,z_{ij}),
\label{5}
\end{equation}
where the function $\GG(\eta,z)$ is \textit{universal} in the sense
that it is a common function for all the pairs $(i,j)$, regardless
of the {composition and} number of components of the mixture. Of
course, the function $\GG(\eta ,z)$ is in principle different for
each dimensionality $d$. To clarify the implications of this
universality ansatz, let us imagine two  mixtures $\mathcal{M}$ and
$\mathcal{M}'$ having the same packing fraction $\eta$ but strongly
differing in the set of mole fractions, the sizes of the particles,
and even the number of components. Suppose now that there exists  a
pair $(i,j)$ in mixture $\mathcal{M}$ and another pair $(i',j')$ in
mixture $\mathcal{M}'$ such that $z_{ij} = z_{i'j'}$. Then,
according to Eq.\ (\ref{5}), the contact value of the RDF for the
pair $(i,j)$ in mixture $\mathcal{M}$ is the same as that for the
pair $(i',j')$ in mixture $\mathcal{M}'$, {i.e.},
$g_{ij}(\sigma_{ij})=g_{i'j'}(\sigma_{i'j'})$. In order to ascribe a
physical meaning to the parameter $z_{ij}$, note that the ratio
$\muM_{d-1}/\muM_d$ can be understood as a ``typical'' inverse
diameter (or curvature) of the particles of the mixture. Thus,
$z_{ij}^{-1}=\frac{1}{2}(\sigma_i^{-1}+\sigma_j^{-1})/(\muM_{d-1}/\muM_d)$
represents  the arithmetic mean curvature, in units of
$\muM_{d-1}/\muM_d$, of a particle of species $i$ and a particle of
species $j$.

Once the ansatz (\ref{5}) is adopted, one may use the limits in
(\ref{2}) and (\ref{3}) to get $\GG(\eta,z)$ at $z=0$ and $z=1$,
respectively. Since $ z_{ii}\rightarrow 0$ in the limit $\sigma
_{i}\rightarrow 0$, insertion of Eq.~(\ref{2}) into (\ref{5}) yields
\begin{equation}
\GG(\eta ,0)=\frac{1}{1-\eta }\equiv \GG_0(\eta). \label{6}
\end{equation}
Next, if all the diameters are equal, $z_{ij}\rightarrow 1$, so that
Eq.~(\ref{3}) implies that
\begin{equation}
\GG(\eta ,1)=g_{\pure}.
\label{7}
\end{equation}

\subsubsection{Linear Approximation}

As the simplest approximation \cite{SYH99}, one may assume  a linear
dependence of $\GG$ on $z$ that satisfies the basic requirements
(\ref{6}) and (\ref{7}), namely
\begin{equation}
\GG(\eta,z)=\frac{1}{1-\eta }+\left( g_{\pure}-\frac{1}{1-\eta }
\right) z.
\label{9}
\end{equation}
Inserting this into Eq.\ \eqref{5}, one has
\beq
g_{ij}^{\text{e1}}(\sigma_{ij})=\frac{1}{1-\eta }+\left(
g_{\pure}-\frac{1}{1-\eta
}\right)\frac{\muM_{d-1}}{\muM_d}\frac{\sigma_i\sigma_j}{\sigma_{ij}}.
\label{gije1}
\eeq
Here, the label ``e1'' is meant to indicate that (i) the contact
values used are an \emph{extension} of the single component contact
value $g_{\pure}$ and that (ii) $\GG(\eta,z)$ is a \emph{linear}
polynomial in $z$. This notation will become handy below. Although
the proposal (\ref{gije1}) is rather crude and does not produce
especially accurate results for $g_{ij}(\sigma _{ij})$ when $d\geq
3$, it nevertheless leads to an EOS that exhibits an excellent
agreement with simulations in 2, 3, 4, and 5 dimensions, provided
that an accurate $g_{\pure}$ is used as input
\cite{SYH99,MV99,SYH01,GAH01,HYS02}. This EOS may be written as
\begin{equation}
Z_\text{e1}(\eta
)=1+\frac{\eta}{1-\eta}2^{d-1}(\Sn_0-\Sn_1)+\left[Z_{\pure}(\eta
)-1\right]\Sn_1,
 \label{Ze1}
\end{equation}
where   the coefficients $\Sn_{m}$ depend only on the composition of
the mixture and are defined by
\begin{equation}
\Sn_{m}=2^{-(d-m)}\frac{\muM_{d-1}^{m}}{\muM_{d}^{m+1}}\sum_{n=0}^{d-m}\binom{d-m}{n}
{\muM_{n+m} }{\muM_{d-n}}.
\label{Omega}
\end{equation}
In particular, for $d=2$ and $d=3$,
\begin{equation}
Z_\text{e1}(\eta )=\frac{1}{1-\eta }+\frac{\mo ^{2}}{ \mt}\left[
Z_{\pure}(\eta)-\frac{1}{1-\eta } \right] ,\quad (d=2),
\label{xnew2}
\end{equation}
\beqa
 Z_\text{e1}(\eta )&=&\frac{1}{1-\eta }+\frac{\mo \mt}{2\mth}\left\{ \left[ Z_{\pure}(\eta )-\frac{1}{1-\eta
 }\right]\left(1+\frac{\mt^2}{\mo\mth}\right)\right.\nn
 &&\left.+\frac{3\eta}{1-\eta}\left(1-\frac{\mt^2}{\mo\mth}\right)\right\} ,\quad (d=3).
 \label{Ze1bis}
\eeqa

As an extra asset, from Eq.\ (\ref{Ze1}) one may write the virial
coefficients of the mixture $B_n$, defined by
\beq
Z=1+\sum_{n=1}^\infty B_{n+1} \rho^{n},
\label{virial}
\eeq
 in terms of the
(reduced) virial coefficients of the single component fluid $b_n$
defined by
\beq
Z_\pure=1+\sum_{n=1}^\infty b_{n+1} \eta^{n}.
\label{virial_s}
\eeq
The result is
\begin{equation}
B_n=v_d^{n-1} \muM_{d}^{n-1} \left[\Sn_1
b_n+2^{d-1}(\Sn_0-\Sn_1)\right].
\label{Virial}
\end{equation}
In the case of binary mixtures, these coefficients are in very good
agreement with the available exact and simulation results
\cite{SYH99,SYH01}, except when the mixture involves components of
very disparate sizes, especially for high dimensionalities. One may
perform a slight modification such that this deficiency is avoided
and thus get a modified EOS \cite{SYH01,S99}. For $d=2$ and $d=3$ it
reads
\begin{eqnarray}
Z(\eta ) &=&Z_{\text{s}}(\eta )+x_{1}\left[ \frac{1}{1-\eta _{2}}
Z_{\text{s}}\left( \frac{\eta _{1}}{1-\eta _{2}}\right)
-Z_{\text{s}}(\eta ) \right] \left( \frac{\sigma _{2}-\sigma
_{1}}{\sigma _{2}}\right) ^{d-1}
\nonumber \\
&&+x_{2}\left[ \frac{1}{1-\eta _{1}}Z_{\text{s}}\left( \frac{\eta
_{2}}{ 1-\eta _{1}}\right) -Z_{\text{s}}(\eta )\right] \left(
\frac{\sigma _{1}-\sigma _{2}}{\sigma _{1}}\right) ^{d-1},\quad
(d=2,3),\nn && \label{x50}
\end{eqnarray}
where $\eta _{i}=v_{d}\rho _{i}\sigma _{i}^{d}$ is the
\emph{partial} volume packing fraction due to species $i $. In
contrast to most of the approaches (PY, SPT, BMCSL, e1, \ldots), the
proposal (\ref{x50}) expresses $Z(\eta )$ in terms not only of
$Z_{\text{s}}(\eta )$ but also involves $Z_{\text{s}}\left(
\frac{\eta _{1}}{ 1-\eta _{2}}\right) $ and $Z_{\text{s}}\left(
\frac{\eta _{2}}{1-\eta _{1}} \right) $. Equation  \eqref{x50}
should in principle be useful in particular for binary mixtures
involving components of very disparate sizes. However, it is
slightly less accurate than the one given in Eq.\ (\ref{Ze1}) for
ordinary mixtures \cite{SYH01}.

\subsubsection{Quadratic Approximation}

In order to improve the proposal contained in Eq.\ (\ref{gije1}), in
addition to the consistency requirements (\ref{2}) and (\ref{3}),
one may consider the condition stemming from a binary mixture in
which one of the species (say $i=1$) is much larger than the other
one ({i.e.}, $\sigma _{1}/\sigma _{2}\rightarrow \infty $), but
occupies a negligible volume ({i.e.}, $x_{1}(\sigma _{1}/\sigma
_{2})^{d}\rightarrow 0$). In that case, a sphere of species 1 is
felt as a wall by particles of species 2, so that
\cite{HMLC96,HBCW98,RDA01}
\begin{equation}
\lim_{\stackrel{\sigma _{1}/\sigma _{2}\rightarrow \infty}{
x_{1}(\sigma _{1}/\sigma _{2})^{d}\rightarrow 0}}\left[g_{12}(\sigma
_{12})-2^{d-1}\eta g_{22}(\sigma_{2})\right] =1.
\label{4}
\end{equation}
Hence, in the limit considered in Eq.~(\ref{4}), we have
$z_{22}\rightarrow 1$, $z_{12}\rightarrow 2$. Consequently, under
the universality ansatz \eqref{5}, one may rewrite Eq.\ \eqref{4} as
\begin{equation} \GG(\eta ,2)=1+2^{d-1}
\eta \GG(\eta ,1).
\label{8}
\end{equation}
Thus, {Eqs.\ (\ref{6}), (\ref{7}), and (\ref{8}) provide  complete
information on the function $\GG$ at $z=0$, $z=1$, and $z=2$,
respectively, in terms of the contact value $g_{\pure}$ of the
single component RDF.}

The simplest functional form of $\GG$ that complies with the above
consistency conditions is a quadratic function of $z$ \cite{SYH02}:
\begin{equation}
\GG(\eta ,z)=\GG_{0}(\eta )+\GG_{1}(\eta )z+\GG_{2}(\eta )z^{2},
\label{10}
\end{equation}
where the coefficients $\GG_{1}(\eta )$ and $\GG_{2}(\eta )$ are
explicitly given by
\begin{equation}
\GG_{1}(\eta )= (2-2^{d-2}\eta )g_{\pure}-\frac{2-\eta/2}{1-\eta } ,
\label{11b}
\end{equation}
\begin{equation}
\GG_{2}(\eta )= \frac{1-\eta/2}{1-\eta }-(1-2^{d-2}\eta )g_{\pure}.
\label{11c}
\end{equation}
Therefore, the explicit expression for the contact values is
\beqa
g_{ij}^{\text{e2}}(\sigma_{ij})&=&\frac{1}{1-\eta }+\left[
(2-2^{d-2}\eta )g_{\pure}-\frac{2-\eta/2}{1-\eta
}\right]\frac{\muM_{d-1}}{\muM_d}\frac{\sigma_i\sigma_j}{\sigma_{ij}}\nn
&&+\left[\frac{1-\eta/2}{1-\eta }-(1-2^{d-2}\eta
)g_{\pure}\right]\left(\frac{\muM_{d-1}}{\muM_d}\frac{\sigma_i\sigma_j}{\sigma_{ij}}\right)^2.
\label{gije2}
\eeqa
Following the same criterion as the one used in connection with Eq.\
\eqref{gije1}, the label ``e2'' is meant to indicate that (i) the
resulting contact values represent an \emph{extension} of the single
component contact value $g_{\pure}$ and that (ii) ${\GG}(\eta, z)$
is a \emph{quadratic} polynomial in $z$. Of course, the quadratic
form (\ref{10}) is not the only choice compatible with conditions
(\ref{6}), (\ref{7}), and (\ref{8}). For instance, a rational
function was also considered in Ref.\ \cite{SYH02}. However,
although it is rather accurate, it does not lead to a closed form
for the EOS. In contrast, when Eq.~(\ref{gije2}) is inserted into
Eq.~(\ref{1}), one gets a closed expression for the compressibility
factor in terms of the packing fraction $\eta $ and the first few
moments $\mn$, $n\leq d$.  The result is
\begin{eqnarray}
Z_{\text{e2}}(\eta) &=&1+2^{d-2}\frac{\eta }{1-\eta }\left[
2(\Sn_{0}-2\Sn_{1}+\Sn_{2})+(\Sn_{1}-\Sn_{2})\eta \right]  \nonumber \\
&&+\left[Z_{\pure}(\eta )-1\right] \left[
2\Sn_{1}-\Sn_{2}+2^{d-2}(\Sn_{2}-\Sn_{1})\eta \right] ,
\label{14}
\end{eqnarray}
where the quantities $\Omega_m$ are defined in Eq.\ \eqref{Omega}.
Quite interestingly, in the two-dimensional case Eq.\ \eqref{14}
reduces to Eq.\ \eqref{xnew2}, {i.e.},
\beq
Z_{\text{e1}}(\eta)=Z_{\text{e2}}(\eta),\quad (d=2).
\label{e1e2}
\eeq
This illustrates the fact that two different proposals for the
contact values $g_{ij}(\sigma _{ij})$ can yield the same EOS when
inserted into Eq.~(\ref{1}). On the other hand, for
three-dimensional mixtures  Eq.~(\ref{14}) becomes
\beq
 Z_\text{e2}(\eta )=\frac{1}{1-\eta }+\frac{\mo \mt}{\mth}\left( 1-\eta +\frac{
 \mt^{2}}{\mo \mth }\eta \right) \left[ Z_{\pure}(\eta )-\frac{1}{1-\eta
 }\right] ,\quad (d=3),
 \label{e23D}
\eeq
which differs from Eq.\ \eqref{Ze1bis}. In fact,
\beq
Z_\text{e1}(\eta)-
Z_\text{e2}(\eta)=\frac{\muM_1\muM_2}{2\muM_3}\left(1-\frac{\muM_2^2}{\muM_1\muM_3}\right)\left[\frac{1+\eta}{1-\eta}-(1-2\eta)
Z_\pure(\eta)\right],\quad (d=3).
\label{e1-e2}
\eeq

\subsubsection{Specific Examples}
In this subsection, rather than carrying out an exhaustive
comparison with the wealth of results available in the literature,
we will consider only a few representative examples. In particular,
for $d=3$, we will  restrict ourselves to a comparison with
classical proposals (say BGHLL, PY, and SPT for the contact values).
The comparison with more recent ones may be found in Refs.\
\cite{SYH99,SYH02,SYH05}.

 Thus far the development has been rather
general since  $g_\pure$ remains free in Eqs.\ \eqref{gije1} and
\eqref{gije2}. In order to get specific results, it is necessary to
fix $g_\text{s}$ [cf.\ Table \ref{Tableg_s}]. In the one-dimensional
case, one has $g_\pure=1/(1-\eta)$ and so one gets the exact result
\eqref{exact1D} after substitution into Eq.\ (\ref{gije1}).
Similarly Eqs.~(\ref{11b}) and (\ref{11c}) lead to
$\GG_{1}=\GG_{2}=0$ and so  we recover again the exact result.

If in the two-dimensional case we take Henderson's value \cite{H75}
$g_\pure=g_\pure^{\text{H}}$, then the linear approximation
(\ref{gije1}) reduces to the JM  approximation, Eq.\ \eqref{JM}.
This equivalence can be symbolically represented as
$g_{ij}^{\text{eH1}}=g_{ij}^{\text{JM}}$, where the label ``eH1''
refers to the extension of Henderson's \emph{single component} value
in the linear approximation. While $g_{ij}^{\text{JM}}$ is  very
accurate, even better results are provided by the quadratic form
\eqref{gije2}, especially if Luding's value \cite{L01}
$g_\pure=g_\pure^{\text{L}}$ is used \cite{LS04}.

In the three-dimensional case, Eq.~(\ref{gije1}) is  of the form of
the solution of the PY equation \cite{L64}. In fact, insertion of
$g_\pure=g_\pure^{\text{PY}}$ leads to Eq.\ \eqref{15PY}, {i.e.},
$g_{ij}^{\text{ePY1}}=g_{ij}^{\text{PY}}$. Similarly, if the SPT
expression \cite{RFL59} $g_\pure=g_\pure^{\text{SPT}}$ is used for
the single component contact value in the quadratic approximation
(\ref{gije2}), we reobtain the SPT expression for the mixture, Eq.\
\eqref{15SPT}. In other words,
$g_{ij}^{\text{eSPT2}}=g_{ij}^{\text{SPT}}$.  On the other hand, if
the much more accurate CS \cite{CS69} expression
$g_\pure=g_\pure^{\text{CS}}$ is used as input, we arrive at the
following expression:
\begin{equation}
g_{ij}^{\text{eCS2}}=\frac{1}{1-\eta }+\frac{3}{2}\frac{ \eta
(1-\eta /3)}{(1-\eta )^{2} }\frac{\sigma_i\sigma_j
\muM_2}{\sigma_{ij}\muM_3}+\frac{\eta ^{2}(1-\eta /2)}{(1-\eta
)^{3}}\left(\frac{\sigma_i\sigma_j
\muM_2}{\sigma_{ij}\muM_3}\right)^{2},\quad (d=3),
\label{16}
\end{equation}
which is different from the BGHLL one, Eq.\ \eqref{15BGHLL},
improves the latter {for $z_{ij}>1$}, and leads to similar results
for $z_{ij}<1$, as comparison with computer simulations shows
\cite{SYH02}. The four approximations \eqref{15PY}, \eqref{15SPT},
\eqref{15BGHLL}, and \eqref{16} are consistent with conditions
(\ref{2}) and (\ref{3}), but only the SPT and eCS2 are also
consistent with condition (\ref{4}). It should also be noted that if
one considers a binary mixture in the infinite solute dilution
limit, namely  $x_1 \rightarrow 0$, so that $z_{12} \rightarrow
2/(1+\sigma_2/\sigma_1)$, Eq.~(\ref{16}) yields the same result for
$g_{12}(\sigma_{12})$ as the one proposed by Matyushov and Ladanyi
\cite{ML97} for this  quantity on the basis of exact geometrical
relations. However, the extension that the same authors propose when
there is a non-vanishing solute concentration, {{i.e.,}} for
$x_1\neq 0$, is different from Eq.\ (\ref{16}).

Equation \eqref{gije2} can also be used in the case of hyperspheres
($d\geq 4$) \cite{SYH02}. In particular, a very good agreement with
available computer simulations \cite{GAH01} is obtained for $d=4$
and $d=5$ by using Luban and Michels \cite{LM90} value
$g_\pure=g_\pure^{\text{LM}}$.

%
\begin{figure}[h]
\centering
    \includegraphics[height=8cm]{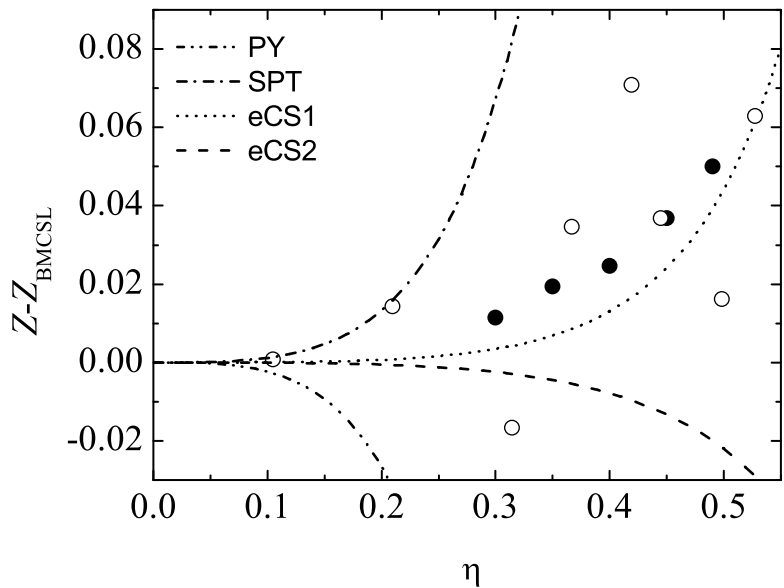}
%
%
\caption{Deviation of the compressibility factor from the BMCSL
value,
  as a
 function of the packing fraction $\eta$ for an equimolar
three-dimensional binary mixture with $\sigma_2/\sigma_1=0.6$.
 The open (Ref.\ \protect\cite{YCH96}) and closed (Ref.\ \protect\cite{BMLS96}) circles are simulation data.
 The lines are the PY EOS (-- $\cdot$ $\cdot$ --), the SPT EOS (-- $\cdot$ -- $\cdot$), the eCS1
 EOS ($\cdots$), and the eCS2 EOS (-- -- --).}
\label{z-zBMCSL}       
\end{figure}
%
%
\begin{figure}[h]
\centering
    \includegraphics[height=8cm]{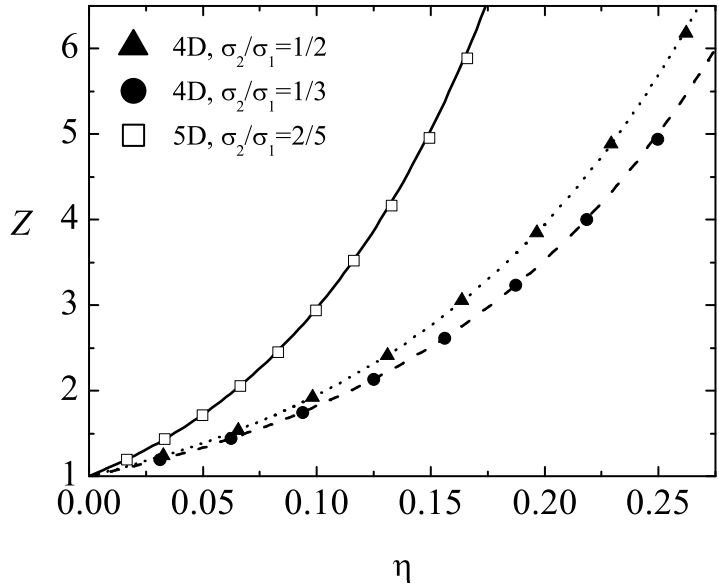}
%
%
\caption{Compressibility factor for three equimolar mixtures in 4D
and 5D systems. Lines are the eLM1 predictions, while symbols are
simulation data \protect\cite{GAH01}.}
\label{Z4D5D}       
\end{figure}

Now we turn to the compressibility factors \eqref{Ze1} and
\eqref{14}, which are obtained from the contact values (\ref{gije1})
and (\ref{gije2}), respectively. Since they depend on the details of
the composition through the $d$ first moments, they are meaningful
even for continuous polydisperse mixtures.

As said above, in the two-dimensional case both Eqs.\ \eqref{Ze1}
and \eqref{14} reduce to Eq.\ \eqref{xnew2}, which yield very
accurate results when a good $Z_\pure$ is used as input
\cite{HYS02,SYH02,LS04}. For three-dimensional mixtures, insertion
of $Z_\pure=Z_\pure^\text{CS}$ in Eqs.\ \eqref{Ze1bis} and
\eqref{e23D} yields
\begin{equation}
Z_{\text{eCS1}}(\eta )=Z_{\text{BMCSL}}(\eta )+\frac{\eta
^{3}\mt}{(1-\eta)^{3}\mth ^{2}} \left( \mo\mth-\mt ^{2}\right) ,
\quad (d=3),
\label{eCS1}
\end{equation}
\begin{equation}
Z_{\text{eCS2}}(\eta )=Z_{\text{BMCSL}}(\eta)-\frac{\eta
^{3}\mt}{(1-\eta)^{2}\mth ^{2}}\left( \mo \mth-\mt ^{2}\right),
\quad (d=3),
\label{eCS2}
\end{equation}
where $Z_{\text{BMCSL}}(\eta )$ is given by Eq.\ \eqref{BMCSL}. Note
that  $Z_{\text{eCS1}}(\eta )>Z_{\text{BMCSL}}(\eta
)>Z_{\text{eCS2}}(\eta )$. Since simulation data indicate that the
BMCSL EOS tends to underestimate the compressibility factor, it
turns out that, as illustrated in Fig.\ \ref{z-zBMCSL} for an
equimolar binary mixture with $\sigma_2/\sigma_1=0.6$, the
performance of $Z_{\text{eCS1}}$ is, {paradoxically}, better than
that of $Z_{\text{eCS2}}$ \cite{SYH02}, despite the fact that the
underlying linear approximation for the contact values is much less
accurate than the quadratic approximation. This shows that a rather
crude approximation such as Eq.~(\ref{gije1}) may lead to an
extremely good EOS \cite{SYH99,SYH01,GAH01,HYS02}, which, as clearly
seen in Fig.\ \ref{z-zBMCSL}, represents a substantial improvement
over the classical proposals. Interestingly, the EOS corresponding
to $Z_{\text{eCS1}}$ has recently been independently derived as the
second order approximation of the Fundamental Measure Theory for the
HS fluid by Hansen-Goos and Roth \cite{H-GR06}.

In the case of $d=4$ and $d=5$, use of
$Z_\pure(\eta)=Z_\pure^\text{LM}(\eta)$ in Eq.~(\ref{Ze1}) produces
a simple extended EOS of a mixture of hard additive hyperspheres in
these dimensionalities. The accuracy of these two EOS for hard
hypersphere mixtures in the fluid region has been confirmed by
simulation data \cite{GAH01} for a wide range of compositions and
size ratios. In Fig. \ref{Z4D5D}, this accuracy is explicitly
exhibited in the case of three equimolar mixtures, two in $4$D and
one in $5$D.

\subsection{A More Consistent Approximation for Three-Dimensional Additive Mixtures}
\label{sss2.1.2}
Up to this point, we have considered an  arbitrary dimensionality
$d$ and have constructed, under the universality assumption
\eqref{5}, the acurate quadratic approximation \eqref{gije2}, which
fulfills the consistency conditions \eqref{2}, \eqref{3}, and
\eqref{4}. However, there exist extra consistency conditions that
are not necessarily satisfied by \eqref{gije2}. In particular, when
the mixture is in contact with a hard wall, the state of equilibrium
imposes that the pressure evaluated near the wall by considering the
impacts with the wall must be the same as the pressure in the bulk
evaluated from the particle-particle collisions. This consistency
condition is especially important if one is interested in deriving
accurate expressions for the contact values of the particle-wall
correlation functions.

 Since a hard wall can be seen as a sphere of infinite diameter, the
contact value $g_{wj}$ of the correlation function  of a sphere of
diameter $\sigma_j$ with the wall can be obtained from
$g_{ij}(\sigma_{ij})$ as
\beq
g_{wj}=\lim_{\stackrel{\sigma_i\to\infty}{
x_{i}\sigma_{i}^d\rightarrow 0}} g_{ij}(\sigma_{ij}).
\label{3p}
\eeq
Note that $g_{wj}$ provides the ratio between the density of
particles of species $j$ adjacent to the wall and the density of
those particles far away from the wall. The sum rule connecting the
pressure of the fluid and the above contact values is \cite{E90}
\beq
Z_w(\eta)=\sum_{j=1}^N  x_j g_{wj},
\label{4p}
\eeq
where the subscript $w$ in $Z_w$ has been used to emphasize that
Eq.\ (\ref{4p}) represents a route alternative to the virial one,
Eq.\ (\ref{1}), to get the EOS of the HS mixture.  The condition
$Z=Z_w$ is equivalent to \eqref{4} in the special case where one has
a \emph{single} fluid in the presence of the wall. However, in the
general case of a mixture plus a wall, the condition $Z=Z_w$ is
stronger than Eq.\ \eqref{4}. In the two-dimensional case, it turns
out that the quadratic approximation \eqref{gije2} already satisfies
the requirement $Z=Z_w$, regardless of the density and composition
of the mixture \cite{LS04}. However, this is not the case for $d\geq
3$.

Our problem now consists of computing $g_{ij}(\sigma_{ij})$ and the
associated $g_{wj}$ for the  HS mixture in the presence of a hard
wall, so that the condition $Z=Z_w$ is satisfied for an arbitrary
mixture \cite{SYH05}. Due to the mathematical complexity of the
problem, here we will restrict ourselves to three-dimensional
systems ($d=3$). Similarly to what we did in the preceding
subsection, we consider a class of approximations of the universal
type \eqref{5}, so that conditions \eqref{2} and \eqref{3} lead
again to Eqs.\ \eqref{6} and \eqref{7}, respectively. Notice that
Eq.\ \eqref{5} implies in particular that
 \beq
g_{wj}={\GG}(\eta, z_{wj}), \quad z_{wj}=2
\sigma_j\frac{\muM_2}{\muM_3}.
\eeq

 Assuming that $z=0$ is a regular point and taking into account condition
(\ref{6}),  ${\GG}(\eta, z)$ can be expanded in a power series in
$z$:
\beq
{\GG}(\eta, z)={\GG}_0(\eta)+\sum_{n=1}^\infty {\GG}_n(\eta) z^n.
\label{n4}
\eeq
After simple algebra, using the ansatz (\ref{5}) and Eq.\ (\ref{n4})
in Eqs.\ (\ref{1}) (with $d=3$) and \eqref{4p} one gets
\beq
Z={\GG}_0+3\eta\frac{\mo\mt}{\mth}{\GG}_0+4\eta\sum_{n=1}^\infty
{\GG}_n\frac{\mt^n}{\mth^{n+1}}\sum_{i,j=1}^N
x_ix_j\sigma_i^n\sigma_j^n\sigma_{ij}^{3-n},
\label{n5}
\eeq
\beq
Z_w={\GG}_0+\sum_{n=1}^\infty
2^{n}{\GG}_n\frac{\mt^n}{\mth^{n}}\muM_n.
\label{n5bis}
\eeq
Notice that if the series \eqref{n4} is truncated after a given
order $n\geq 3$, $Z_w$ is given by the first $n$ moments of the size
distribution only. On the other hand, $Z$ still involves an infinite
number of moments if the truncation is made after $n\geq 4$ due to
the presence of terms like
$\sum_{i,j}x_ix_j\sigma_i^4{\sigma_j}^4/\sigma_{ij}$,
$\sum_{i,j}x_ix_j\sigma_i^5{\sigma_j}^5/\sigma_{ij}^2$, \ldots.
Therefore, if we want the consistency condition $Z=Z_w$ to be
satisfied for \emph{any} discrete or continuous polydisperse
mixture, either the whole infinite series \eqref{n4} needs to be
considered or it must be truncated after $n=3$. The latter is of
course the simplest possibility and thus we make the approximation
\beq
{\GG}(\eta, z)={\GG}_0(\eta)+{\GG}_1(\eta) z+{\GG}_2(\eta)
z^2+{\GG}_3(\eta) z^3.
\label{5p}
\eeq
As a consequence, $Z$ and $Z_w$ depend functionally on the size
distribution of the mixture  only through the first three moments
(which is in the spirit of Rosenfeld's Fundamental Measure Theory
\cite{R89}).

Using the approximation (\ref{5p}) in Eqs.\ (\ref{n5}) and
(\ref{n5bis}) we are led to
\beq
Z={\GG}_0+\eta\left[\frac{\mo \mt}{\mth}
\left(3{\GG}_0+2{\GG}_1\right)+2\frac{\mt^3}{\mth^2}\left({\GG}_1+2{\GG}_2+2{\GG}_3\right)\right],
\label{11p}
\eeq
\beq
Z_w={\GG}_0+2\frac{\mo \mt}{\mth}{\GG}_1
+4\frac{\mt^3}{\mth^2}\left({\GG}_2+2{\GG}_3\right).
\label{14bis}
\eeq
Thus far, the dependence of both $Z$ and $Z_w$ on the moments
$\muM_1$, $\muM_2$, and $\muM_3$ is explicit and we only lack the
packing-fraction dependence of ${\GG}_1$, ${\GG}_2$, and ${\GG}_3$.
From Eqs.\ (\ref{11p}) and (\ref{14bis}) it follows that the
difference between $Z$ and $Z_w$ is given by
\beq
Z-Z_w=\frac{\mo \mt}{\mth}\left[3\eta
{\GG}_0-2(1-\eta){\GG}_1\right] +2\frac{\mt^3}{\mth^2}\left[\eta
{\GG}_1-2(1-\eta){\GG}_2-2(2-\eta){\GG}_3\right].
\label{15x}
\eeq
Therefore, $Z=Z_w$ for \emph{any} dispersity provided that
\beq
{\GG}_1(\eta)=\frac{3 \eta}{2 \left(1-\eta\right)^2},
\label{n7a}
\eeq
\beq
{\GG}_2(\eta)=\frac{3 \eta^2}{4
\left(1-\eta\right)^3}-\frac{2-\eta}{1-\eta}{\GG}_3(\eta),
\label{n7b}
\eeq
where use has been made of the definition of ${\GG}_0$, Eq.\
(\ref{6}). To close the problem, we use the equal size limit given
in Eq.\ (\ref{7}), which yields $\GG_0+\GG_1+\GG_2+\GG_3=g_\pure$.
After a little algebra we are led to
\beq
{\GG}_2(\eta)=(2-\eta)g_{\pure}-\frac{2+\eta^2/4}{\left(1-\eta\right)^2},
\label{n6a}
\eeq
\beq
{\GG}_3(\eta)=(1-\eta)\left(g_{\pure}^{\text{SPT}}-g_{\pure}\right).
\label{n6b}
\eeq
 This completes the
derivation of our improved approximation, which we will call ``e3'',
following the same criterion as the one used to call ``e1'' and
``e2'' to the approximations \eqref{gije1} and \eqref{gije2},
respectively. In Eq.\ \eqref{n6b}, $g_{\pure}^{\text{SPT}}$ is the
SPT contact value for a single fluid, whose expression appears in
Table \ref{Tableg_s}. {}From Eq.\ \eqref{n6b} it is obvious that the
choice $g_\pure=g_{\pure}^{\text{SPT}}$ makes our e3 approximation
to become the e2 approximation, both reducing to the SPT for
mixtures, Eq.\ \eqref{15SPT}. This means that the SPT is fully
internally consistent with the requirement $Z=Z_w$, although it has
the shortcoming of not being too accurate in the single component
case. The e3 proposal, on the other hand, satisfies the condition
$Z=Z_w$ and has the flexibility of accommodating any desired
$g_\pure$.

For the sake of concreteness, let us write explicitly the contact
values in the e3 aproximation:
\beqa
g_{ij}^\text{e3}(\sigma_{ij})&=&\frac{1}{1-\eta}+\frac{3 \eta}{2
\left(1-\eta\right)^2}\frac{\mt}{\mth}\frac{\sigma_i
\sigma_j}{\sigma_{ij}}+\left[(2-\eta)g_{\pure}-\frac{2+\eta^2/4}{\left(1-\eta\right)^2}\right]\nn
&&\times\left(\frac{\mt}{\mth}\frac{\sigma_i
\sigma_j}{\sigma_{ij}}\right)^2+(1-\eta)\left(g_{\pure}^{\text{SPT}}-g_{\pure}\right)\left(\frac{\mt}{\mth}\frac{\sigma_i
\sigma_j}{\sigma_{ij}}\right)^3,
\label{e3}
\eeqa
\beqa
g_{wj}^\text{e3}&=&\frac{1}{1-\eta}+\frac{3 \eta}{
\left(1-\eta\right)^2}\frac{\mt}{\mth}\sigma_j+4\left[(2-\eta)g_{\pure}-\frac{2+\eta^2/4}{\left(1-\eta\right)^2}\right]\left(\frac{\mt}{\mth}\sigma_j\right)^2\nn
&&+8(1-\eta)\left(g_{\pure}^{\text{SPT}}-g_{\pure}\right)\left(\frac{\mt}{\mth}\sigma_j\right)^3.
\label{e3w}
\eeqa
 With the above results the compressibility factor may be finally written
in terms of $Z_{\pure}$ as
\beq
Z_\text{e3}(\eta)=\frac{1}{\left(1-\eta\right)}+\left(\frac{\mo\mt}{\mth}-\frac{\mt^3}{\mth^2}\right)\frac{3
\eta}{
\left(1-\eta\right)^2}+\frac{\mt^3}{\mth^2}\left[Z_\pure(\eta)-\frac{1}{1-\eta}\right].
\label{ZZ}
\eeq

A few comments are in order at this stage. {}First, from Eq.\
(\ref{11p}) we can observe that, for the class of approximations
(\ref{5p}), the compressibility factor $Z$ does not depend on the
individual values of the coefficients $\GG_2$ and $\GG_3$, but only
on their sum. As a consequence, two different approximations of the
form (\ref{5p}) sharing the same density dependence of $\GG_1$ and
$\GG_2+\GG_3$ also share the same virial EOS. For instance, if one
makes the choice $g_\pure=g_\pure^{\text{PY}}$, then
 $Z_{\text{ePY3}}=Z_{\text{PY}}$, even though
$g_{ij}^{\text{ePY3}}(\sigma_{ij})\neq
g_{ij}^{\text{PY}}(\sigma_{ij})$. Furthermore, if one makes the more
accurate choice $g_\pure=g_\pure^{\text{CS}}$, then
 $Z_{\text{eCS3}}=Z_{\text{BMCSL}}$, but again
$g_{ij}^{\text{eCS3}}(\sigma_{ij})\neq
g_{ij}^{\text{BGHLL}}(\sigma_{ij})$. The eCS3 contact values are
\beqa
g_{ij}^{\text{eCS3}}(\sigma_{ij})&=&\frac{1}{1-\eta}+\frac{3 \eta}{2
\left(1-\eta\right)^2}\frac{\mt}{\mth}\frac{\sigma_i
\sigma_j}{\sigma_{ij}}+\frac{\eta^2(1+\eta)}{4(1-\eta)^3}\left(\frac{\mt}{\mth}\frac{\sigma_i
\sigma_j}{\sigma_{ij}}\right)^2\nn
&&+\frac{\eta^2}{4(1-\eta)^2}\left(\frac{\mt}{\mth}\frac{\sigma_i
\sigma_j}{\sigma_{ij}}\right)^3,
\label{eCS3}
\eeqa
\beqa
g_{wj}^{\text{eCS3}}&=&\frac{1}{1-\eta}+\frac{3 \eta}{
\left(1-\eta\right)^2}\frac{\mt}{\mth}\sigma_j
+\frac{\eta^2(1+\eta)}{(1-\eta)^3}\left(\frac{\mt}{\mth}\sigma_j\right)^2\nn
&&+\frac{2\eta^2}{(1-\eta)^2}\left(\frac{\mt}{\mth}\sigma_j\right)^3.
\label{eCS3w}
\eeqa

%
\begin{figure}[t]
\centering
    \includegraphics[height=8cm]{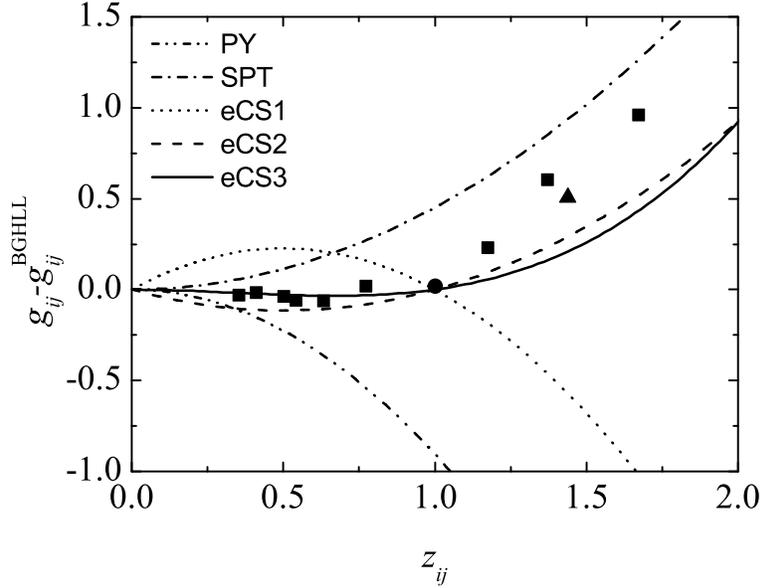}
%
%
\caption{ Plot of the difference  $g_{ij}(\sigma
_{ij})-g_{ij}^\text{BGHLL}(\sigma _{ij})$ as a function of the
parameter $z_{ij}=(\sigma _{i}\sigma_{j}/\sigma_{ij})\mt/\mth$ for
hard spheres ($d=3$) at a packing fraction $\eta =0.49$. The symbols
are simulation data for the single fluid (circle, Ref.\
\protect\cite{MV99}), three binary mixtures (squares, Ref.\
\protect\cite{MBS97}) with $\sigma_2/\sigma_1=0.3$ and $x_1=0.0625$,
0.125, and 0.25, and a ternary mixture (triangles, Ref.\
\protect\cite{MMYSH02}) with $\sigma_2/\sigma_1=\frac{2}{3}$,
$\sigma_3/\sigma_1=\frac{1}{3}$, and $x_1=0.1$, $x_2=0.2$. The lines
are the PY approximation (-- $\cdot$ $\cdot$ --), the SPT
approximation (\mbox{--~$\cdot$~--~$\cdot$}), the eCS1
 approximation ($\cdots$), the eCS2 approximation (-- -- --), and the eCS3 approximation (---).}
\label{gvsz049d3_1}       
\end{figure}
%
%
%
\begin{figure}[t]
\centering
    \includegraphics[height=8cm]{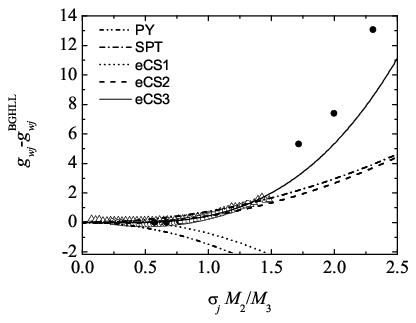}
%
%
\caption{ Plot of the difference  $g_{wj}-g_{wj}^\text{BGHLL}$ as a
function of the parameter $z_{wj}/2=\sigma_{j}\mt/\mth$ for hard
spheres ($d=3$) at a packing fraction $\eta =0.4$. The symbols are
simulation data for a polydisperse mixture with a narrow top-hat
distribution (open squares, Ref.\ \protect\cite{BPW04}), a
polydisperse mixture with a wide top-hat distribution (open circles,
Ref.\ \protect\cite{BPW04}), a polydisperse mixture with a Schulz
distribution (open triangles, Ref.\ \protect\cite{BPW04}), and a
binary mixture (closed circles, Ref.\ \protect\cite{Alexander}). The
lines are the PY approximation (-- $\cdot$ $\cdot$ --), the SPT
approximation (\mbox{--~$\cdot$~--~$\cdot$}), the eCS1
 approximation ($\cdots$), the eCS2 approximation (-- -- --), and the eCS3 approximation (---).}
\label{dgWPolidisHS.Escalado}       
\end{figure}

In Figs.\ \ref{gvsz049d3_1} and \ref{dgWPolidisHS.Escalado} we
display the performance of the contact values as given by Eqs.\
\eqref{eCS3} and \eqref{eCS3w}, respectively, by comparison with
results of computer simulations for both discrete and polydisperse
mixtures. In both figures we have also included the results that
follow from the classical proposals as well as those of the eCS1 and
eCS2 approximations. It is clear that for the wall-particle contact
values the eCS3 approximation yields the best performance, while for
the particle-particle contact values both the eCS2 and eCS3 are of
comparable accuracy. A further feature to be pointed out is that the
practical  collapse on a common curve of the simulation data in
Figs.\ \ref{gvsz049d3_1} and \ref{dgWPolidisHS.Escalado} provide
\emph{a posteriori} support for the universality ansatz made in Eq.\
\eqref{5}.

As mentioned earlier, there exist extra consistency conditions
(see for instance Ref.\ \cite{HC04}) that one might use as well
within our approach. Assuming that the ansatz (\ref{5}) still
holds, some of these conditions are related to the derivatives of
${\GG}$ with respect to $z$, namely \beq \left.\frac{\partial
{\GG}(\eta, z)}{\partial
z}\right|_{z=0}=\frac{3\eta}{2(1-\eta)^2}, \label{exC1} \eeq \beq
\left.\frac{\partial^2 {\GG}(\eta, z)}{\partial
z^2}\right|_{z=0}=\frac{3\eta}{1-\eta}\left(g_{\pure}^{\text{PY}}-\frac{1}{2}g_\pure\right),
\label{exC2} \eeq \beq \left.\frac{\partial^3 {\GG}(\eta,
z)}{\partial z^3}\right|_{z=2}=0. \label{exC3} \eeq Interestingly
enough, as shown by Eq.\ \eqref{n7a}, condition (\ref{exC1}) is
already satisfied by our e3 approximation without having to be
imposed. On the other hand, condition (\ref{exC3}) implies $\GG_3=0$
in the e3 scheme and thus it is only satisfied if
$g_{\pure}=g_{\pure}^\text{SPT}$, in which case we recover the SPT.
Condition \eqref{exC2} is not fulfilled either by the SPT or by the
e3 approximation (except for a particular expression of $g_\pure$
which is otherwise not very accurate). Thus, fulfilling the extra
conditions \eqref{exC2} and \eqref{exC3} with a free $g_{\pure}$
requires either considering a higher order polynomial in $z$ (in
which case the consistency condition $Z=Z_w$ cannot be satisfied for
arbitrary mixtures, as discussed before) or not using the
universality ansatz at all. In the first case, we have checked
 that a quartic or even
a quintic polynomial does not improve matters, whereas giving up the
universality assumption increases significantly the number of
parameters to be determined and seems not to be adequate in view of
the behavior observed in the simulation data.

An additional comment has to do with the restriction to $d=3$ in
this subsection. As noted before,  the approximation e1 reduces to
the exact result \eqref{exact1D} for $d=1$. For $d=2$, the
approximation e2 already fulfills the condition $Z=Z_w$ and so there
is no real need to go further in that case. Since we have needed the
approximation e3 to satisfy $Z=Z_w$ for $d=3$, it is tempting to
speculate that a polynomial form for $\GG(z)$ of degree $d$ could be
found to be consistent with the condition $Z=Z_w$ for $d\geq 4$.
However, a detailed analysis shows that this is not the case for an
\emph{arbitrary} mixture, since the number of conditions exceeds the
number of unknowns, unless the universality assumption is partially
relaxed.

As a final comment, let us stress that, although the discussion in
this section has referred,  for the sake of simplicity, to
\emph{discrete} mixtures, all the dependence on the details of the
composition occurs through a finite number of moments, so that the
results remain meaningful even for continuous \emph{polydisperse}
mixtures \cite{L96}. In that case, instead of a set of mole
fractions $\{x_i\}$ and a set of diameters $\{\sigma_i\}$, one has
to deal with a distribution function $w(\sigma)$ such that
$w(\sigma)\D\sigma$ is the fraction of particles with a diameter
comprised between $\sigma$ and $\sigma+\D\sigma$. Therefore, the
moments \eqref{moments} are now defined as
\beq
\muM_n=\int_0^\infty \D\sigma\, \sigma^n w(\sigma),
\label{momentspoly}
\eeq
and with such a change the results we have derived for discrete
mixtures also hold for polydisperse systems.

\subsection{Non-Additive Systems}
\label{ss2.2}
Non-additive hard-core mixtures, where the distance of closest
approach between particles of different species is no longer the
arithmetic mean of the diameters of both particles, have received
much less attention than additive mixtures, in spite of their in
principle more versatility to deal with interesting aspects
occurring in real systems (such as fluid-fluid phase separation) and
of their potential use as reference systems in perturbation
calculations on the thermodynamic and structural properties of, say,
Lennard--Jones mixtures. Nevertheless, the study of non-additive
systems goes back fifty years \cite{PL54,AO54,K55} and is still a
rapidly developing and challenging problem.

As  mentioned in the paper by Ballone {et al.} \cite{BPGG86}, where
the relevant references may be found, experimental work on alloys,
aqueous electrolyte solutions, and molten salts suggests that
hetero-coordination and homo-coordination may be interpreted in
terms of excluded volume effects due to non-additivity of the
repulsive part of the intermolecular potential. In particular,
positive non-additivity leads naturally to demixing in HS mixtures,
so that some of the experimental findings of phase separation in the
above mentioned (real) systems may be accounted for by using a model
of a binary mixture of (positive) non-additive HS. On the other
hand, negative non-additivity seems to account well for chemical
short-range order in amorphous and liquid binary mixtures with
preferred hetero-coordination \cite{GPE89}.

\subsubsection{Some Preliminary Definitions}
 Let us  consider an $N$-component
mixture of {non-additive} HS in $d$ dimensions. In this case,
$\sigma_{ij}=\frac{1}{2}(\sigma_i+\sigma_j)(1+\Delta_{ij})$, where
$\Delta_{ij}\geq -1$ is a symmetric matrix with zero diagonal
elements ($\Delta_{ii}=0$) that characterizes the degree of
non-additivity of the interactions. If $\Delta_{ij}>0$ the
non-additivity character of the $ij$ interaction is said to be
\emph{positive}, while it is \emph{negative} if $\Delta_{ij}<0$. In
the case of a binary mixture ($N=2$), the only non-additivity
parameter is $\Delta\equiv \Delta_{12}=\Delta_{21}$. The virial EOS
\eqref{1} remains being valid in the non-additive case.

The contact values $g_{ij}(\sigma_{ij})$  can be expanded in a power
series in density as
\begin{equation}
g_{ij}(\sigma_{ij})=1+v_{d}\rho
\sum_{k=1}^{N}x_{k}c_{k;ij}+(v_{d}\rho)^2
\sum_{k,\ell=1}^{N}x_{k}x_\ell c_{k\ell;ij}+\mathcal{O}(\rho^3).
\label{1M}
\end{equation}
{The coefficients $c_{k;ij}$, $c_{k\ell;ij}$, \ldots are independent
of the composition of the mixture, but they are in general
complicated nonlinear functions of the diameters $\sigma_{ij}$,
$\sigma_{ik}$, $\sigma_{jk}$, $\sigma_{k\ell}$, \ldots . Insertion
of the expansion (\ref{1M})}  into Eq.\ (\ref{1}) yields the virial
expansion of $Z$, namely
\begin{eqnarray}
Z(\rho)&=&1+\sum_{n=2}^\infty  \overline{B}_{n} (v_d\rho)^{n-1}\nonumber\\
&=&1+v_{d}\rho
\sum_{i,j=1}^{N}\overline{B}_{ij}x_{i}x_{j}+(v_{d}\rho
)^{2}\sum_{i,j,k=1}^{N}\overline{B}_{ijk}x_{i}x_{j}x_{k}\nonumber\\
&&+(v_{d}\rho )^{3}\sum_{i,j,k,\ell=1 }^{N}\overline{B}_{ijk\ell
}x_{i}x_{j}x_{k}x_{\ell }+ \mathcal{O}(\rho^4). \label{2M}
\end{eqnarray}
Note that, for further convenience, we have introduced the
coefficients $\overline{B}_n\equiv v_d^{-(n-1)} B_n$, where $B_n$
are the usual virial coefficients [cf.\ Eq.\ \eqref{virial}]. The
composition-independent second, third, and fourth (barred) virial
coefficients are given by
\begin{equation}
\overline{B}_{ij}=2^{d-1}\sigma_{ij}^d,
\label{n1x}
\end{equation}
\begin{equation}
\overline{B}_{ijk}=\frac{2^{d-1}}{3}\left(c_{k;ij}\sigma_{ij}^d+
c_{j;ik}\sigma_{ik}^d+c_{i;jk}\sigma_{jk}^d\right),
\label{n2}
\end{equation}
\beqa
\overline{B}_{ijk\ell}&=&\frac{2^{d-1}}{6}\left(c_{k\ell;ij}\sigma_{ij}^d+
c_{j\ell;ik}\sigma_{ik}^d+c_{i\ell;jk}\sigma_{jk}^d+
c_{jk,i\ell}\sigma_{i\ell}^d+c_{ik,j\ell}\sigma_{j\ell}^d\right.\nn
&&\left. +c_{ij;k\ell}\sigma_{k\ell}^d\right).
 \label{n3}
\eeqa

\subsubsection{A Simple Proposal for the Equation of State of
$d$-Dimensional Non-Additive Mixtures}

Our goal now is to generalize the e1 proposal given by Eq.\
(\ref{gije1}) to the non-additive case \cite{SHY05}. We will not try
to extend the e2 and e3 proposals, Eqs.\ \eqref{gije2} and
\eqref{e3}, because of  two reasons. First, given the inherent
complexity of non-additive systems, we want to keep the approach as
simple as possible. Second, we are more interested in the EOS than
in the contact values themselves and, as mentioned earlier, the e1
proposal provides excellent EOS, at least in the additive case,
despite the simplicity of  the corresponding contact values.

As the simplest possible extension, we impose again the point
particle and equal size consistency conditions, Eqs.\ \eqref{2} and
\eqref{3}, and thus keep in this case also the ansatz \eqref{5} and
the linear structure of Eq.\ (\ref{9}). However, instead of using
Eq.\ \eqref{zij}, we determine the parameters $z_{ij}$ as to
reproduce Eq.\ (\ref{1M}) to first order in the density. The result
is {readily found to be} \cite{SHY05}
\beq
z_{ij}=\left(\frac{b_3}{b_2}-1\right)^{-1}\left(\frac{\sum_k
x_kc_{k;ij}}{\muM_d}-1\right). \label{new1bis} \eeq Here
$b_2=2^{d-1}$ and $b_3$ are the second and third virial
coefficients for the single component fluid, as defined by Eq.\
\eqref{virial_s}. {The proposal of Eq.\ (\ref{9}) supplemented by
Eq.\ (\ref{new1bis}) is, by construction, accurate for densities
low enough as to justify the truncated  approximation
$g_{ij}(\sigma_{ij})\approx 1+v_d\rho\sum_k x_k c_{k;ij}$. On the
other hand, the limitations of this truncated expansion for
moderate and large densities may be compensated by the use of
$g_{\pure}$. When Eqs.\ \eqref{5}, (\ref{9}), and (\ref{new1bis})
are inserted into Eq.\ (\ref{1}) one gets \beq
Z(\eta)=1+\frac{\eta}{1-\eta}\frac{b_3\muM_d \overline{B}_2-b_2
\overline{B}_3}{(b_3-b_2)\muM_d^2}+
\left[Z_{\pure}(\eta)-1\right]\frac{\overline{B}_3-\muM_d
\overline{B}_2}{(b_3-b_2)\muM_d^2}. \label{new2}
\eeq

Equation ({\ref{new2}) is the sought generalization of Eq.\
(\ref{Ze1}) to non-additive hard-core systems. As in the additive
case, the the density dependence in the EOS of the mixture   is
rather simple: $Z(\eta)-1$ is expressed as a linear combination of
$\eta/(1-\eta)$ and $Z_{\pure}(\eta)-1$, with coefficients such that
the second and third virial coefficients are reproduced. {Again,
Eq.\ (\ref{new2}) is bound to be accurate for sufficiently low
densities, while the limitations of the truncated expansion for
moderate and large densities are compensated by the use of the EOS
of the pure fluid.}

The exact second virial coefficient $\overline{B}_2$ is known from
Eq.\ \eqref{n1x}. In principle,  one should use the exact
coefficients $c_{k;ij}$ to compute $\overline{B}_3$. However,  to
the best of our knowledge they are only known for $d \leq 3$. Since
our objective is to have a proposal which is explicit for any $d$,
we can make use of a reasonable approximation for them \cite{SHY05},
as described below.

\subsubsection{An Approximate Proposal for $c_{k;ij}$}
 The values
of the coefficients $c_{k;ij}$ are exactly known for $d=1$ and $d=3$
and from these results one may approximate them in $d$ dimensions as
\cite{SHY05}
\begin{equation}
c_{k;ij}=\sigma_{k;ij}^d+\left(\frac{b_3}{b_2}-1\right)\frac{\sigma_{k;ij}^{d-1}}{\sigma_{ij}}
\sigma_{i;jk}\sigma_{j;ik},
\label{n7}
\end{equation}
where we have called
\begin{equation}
\sigma_{k;ij}\equiv\sigma_{ik}+\sigma_{jk}-\sigma_{ij}
\label{diamekk}
\end{equation}
and it is understood that $\sigma_{k;ij}\geq 0$ for all sets $ijk$.
Clearly, $\sigma_{i;ij}=\sigma_i$. For a binary mixture Eq.\
(\ref{n7}) yields
\beq
\begin{array}{l}
c_{1;11}=({b_3}/{b_2})\sigma_1^d,\\
c_{2;11}=(2\sigma_{12}-\sigma_1)^d+
\left({b_3}/{b_2}-1\right)\sigma_1(2\sigma_{12}-\sigma_1)^{d-1},\\
 c_{1;12}={\sigma}_1^d+
\left({b_3}/{b_2}-1\right){(2\sigma_{12}-\sigma_1)\sigma_1^d}/{\sigma_{12}}
.
\end{array}
\label{43c}
\eeq
Of course, Eqs.\ \eqref{n7} and \eqref{43c} reduce to the exact
results for $d=1$ ($b_2=b_3=1$) and for $d=3$ ($b_2=4$, $b_3=10$).

The quantities $\sigma_{k;ij}$ may be given a simple geometrical
interpretation. Assume that we have three spheres of species $i$,
$j$, and $k$ aligned in the sequence $ikj$. In such a case, the
distance of closest approach between the centers of spheres $i$ and
$j$ is $\sigma_{ik}+\sigma_{jk}$. If the sphere of species $k$ were
not there, that distance would of course be $\sigma_{ij}$. Therefore
$\sigma_{k;ij}$ as given by Eq.\ (\ref{diamekk}) represents a kind
of effective diameter of sphere $k$, as seen from the point of view
of the interaction between spheres $i$ and $j$.

Inserting Eq.\ (\ref{n7}) into Eq.\ \eqref{new1bis}, one gets
\beq
z_{ij} =\left(\frac{b_3}{b_2}-1\right)^{-1}\left(\frac{\sum_k
x_k\sigma_{k;ij}^d}{\muM_d}-1\right)+ \frac{\sum_k x_k
\sigma_{k;ij}^{d-1}\sigma_{i;jk}\sigma_{j;ik}}{\muM_d\sigma_{ij}} .
\label{new1}
\eeq
 It can be easily checked that in the additive case
($\sigma_{k;ij}\to \sigma_k$), Eq.\ (\ref{new1}) reduces to Eq.\
(\ref{zij}).

Equations\ \eqref{n7} and \eqref{43c} are restricted to the
situation $\sigma_{k;ij}\geq 0$ for any choice of $i$, $j$, and $k$,
{{i.e.}}, $2\sigma_{12}\geq \text{max}(\sigma_1,\sigma_2)$ in the
binary case. This excludes the possibility of dealing with mixtures
with extremely high negative non-additivity in which one sphere of
species $k$ might ``fit in'' between two spheres of species $i$ and
$j$ in contact. Since for $d=3$ and $N=2$ the coefficients
$c_{k;ij}$ are also known for such mixtures \cite{H96b}, we may
extend our proposal to deal with these cases:
\beq
\begin{array}{l}
 c_{1;11}=(b_3/{b_2})\sigma_1^d,\\
 c_{2;11}=\widehat{\sigma}_{2}^d+
\left({b_3}/{b_2}-1\right)\sigma_1\widehat{\sigma}_{2}^{d-1},\\
 c_{1;12}=(2\sigma_{12}-\widehat{\sigma}_2)^d+
\left({b_3}/{b_2}-1\right){\widehat{\sigma}_2\sigma_1^d}/{\sigma_{12}}
,
\end{array}
\label{negative}
\eeq
where we have defined
\begin{equation}
\widehat{\sigma}_2=\text{max}\left(2\sigma_{12}-\sigma_1,0\right).
\label{negative2}
\end{equation}
With such an extension, we recover the exact values of $c_{k;ij}$
for a binary mixture of hard spheres ($d=3$), even if $\sigma_1> 2
\sigma_{12}$ or $\sigma_2> 2 \sigma_{12}$.

The EOS \eqref{new2} becomes explicit when $\overline{B}_3$ is
obtained from Eq.\ \eqref{n2} by using the approximation \eqref{n7}.
The resulting virial coefficient is the exact one for $d=1$ and
$d=3$. For hard disks ($d=2$), it turns out that the approximate
third virial coefficient is practically indistinguishable from the
exact one \cite{SHY05}. When the approximate $\overline{B}_3$ is
used,  Eq.\ \eqref{new2} reduces to Eq.\ \eqref{Ze1} in the additive
case.

\begin{figure}
\centering
    \includegraphics[height=8cm]{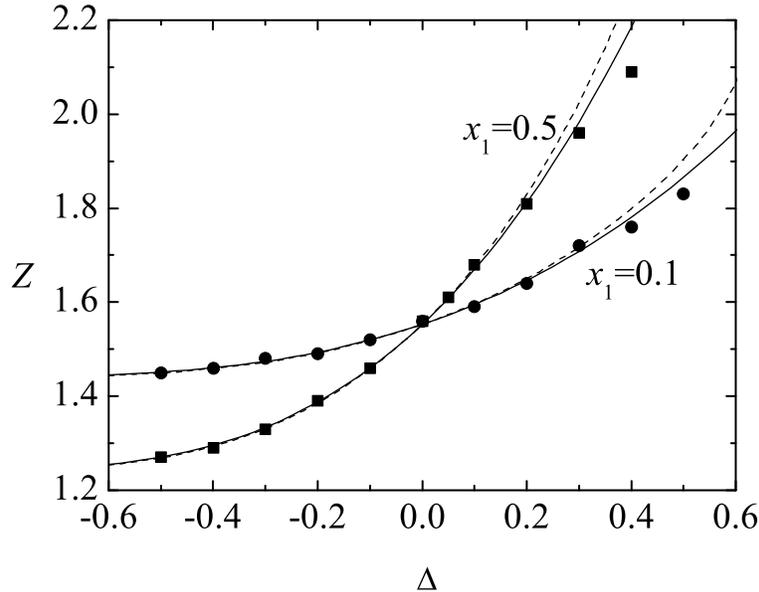}
%
%
\caption{Plot of the compressibility factor versus the
non-additivity parameter $\Delta$ for a symmetric binary mixture of
non-additive hard spheres ($d=3$) at $\eta=\pi/30$ and two different
compositions. The solid lines are our  proposal, Eq.\
(\protect\ref{new2}), with $Z_\pure=Z_\pure^\text{CS}$, while the
dashed lines are Hamad's proposal (Refs.\
\protect\cite{H96a,H96c,H99}). The symbols are results from Monte
Carlo simulations (Refs.\ \protect\cite{JJR94a,JJR94b}).}
\label{non-additive}       
\end{figure}

From the comparison with simulation results, both for the
compressibility factor and higher order virial coefficients, we find
that the EOS (\ref{new2}) does a  good job for non-additive
mixtures, thus representing a reasonable compromise between
simplicity and accuracy, provided that $Z_{\pure}$  is accurate
enough. This is illustrated in Fig.\ \ref{non-additive}, where the
proposal (\protect\ref{new2}) with $Z_\pure=Z_\pure^\text{CS}$ and a
similar proposal by Hamad \cite{H96a,H96c,H99} are compared with
simulation data \cite{JJR94a,JJR94b} for some three-dimensional
symmetric mixtures. A more extensive comparison \cite{SHY05} shows
that Eq.\ \eqref{new2}  seems to work better (especially as the
density is increased) in the case of positive non-additivities, at
least for $d=1$, $d=2$, and $d=3$, but its performance is also
reasonably good in highly asymmetric mixtures, even for negative
$\Delta$. Of course the full assessment of this proposal is still
pending since it involves many facets (non-additivity parameters,
size ratios, density, and composition). Without this full assessment
and given its rather satisfactory performance so far, going beyond
the approximation given by Eq.\ (\ref{9}) (taking similar steps to
the ones described in  Subsections \ref{ss2.1} and \ref{sss2.1.2}
for additive systems) does not seem to be necessary at this stage,
although it is in principle feasible.

\subsection{Demixing}
\label{ss2.3}

Demixing is a common phase transition in fluid mixtures usually
originated on the asymmetry of the interactions (e.g., their
strength and/or range) between the different components in the
mixture. In the case of athermal systems such as HS mixtures in $d$
dimensions, if fluid-fluid separation occurs, it would represent a
neat example of an entropy-driven phase transition, {i.e.}, a phase
separation based only on the size asymmetry of the components. The
existence of demixing in binary additive three dimensional HS
mixtures has been studied theoretically since decades, and the issue
is still controversial. In this subsection we will present our
results following different but related routes that attempt to
clarify some aspects of this problem.

\subsubsection{Binary Mixtures of Additive $d$-Dimensional Spheres ($d=3$, $d=4$ and $d=5$)}
\label{sss2.3.1}

Now we look at the possible instability of a binary fluid mixture of
HS of diameters $\sigma_1$ and $\sigma_2$ ($\sigma_1
>\sigma_2$) in $d$ dimensions by looking at the Helmholtz free
energy per unit volume, $f$, which is given by
\begin{equation}
\frac{f}{\rho k_{B}T}= -1+\sum_{i=1}^{2}x_{i}\ln \left( x_{i}\rho
\lambda_{i}^d\right) +\int_{0}^{\eta }\D\eta'  \frac{ Z(\eta'
)-1}{\eta '} , \label{FEN}
\end{equation}
where  $\lambda_{i}$ is the thermal de Broglie wavelength of species
$i$. We locate the spinodals through the condition
$f_{11}f_{22}-f_{12}^{2}=0$, with $f_{ij}\equiv \partial ^{2}{
f}/\partial \rho _{i}\partial \rho _{j} $. Due to the spinodal
instability, the mixture separates into two phases of different
composition. The coexistence conditions are determined through the
equality of the pressure $p$ and the two chemical potentials $\mu
_{1}$ and $\mu _{2}$ in both phases ($\mu_{i}=\partial {f}/\partial
\rho _{i}$), leading to binodal (or coexistence) curves.

We begin with the case $d=3$. It is well known that the BMCSL EOS,
Eq.\ \eqref{BMCSL}, does not lead to demixing. However, other EOS
for HS mixtures have been shown to predict demixing
\cite{RDA01,CB98b}, including the EOS that is obtained by truncating
the virial series after a certain number of terms \cite{VM03,HT04}.
In particular, it turns out that both $Z=Z_{\text{eCS1}}$, Eq.\
\eqref{eCS1},  and $Z=Z_{\text{eCS2}}$, Eq.\ \eqref{eCS2}, lead to
demixing for certain values of the parameter $\gamma \equiv
{\sigma_2}/{\sigma_1}$ that measures the size asymmetry. The
critical values of the pressure, the composition, and the packing
fraction are presented in Table \ref{TableCr} for a few values of
$\gamma$.

\begin{table}
\centering \caption{Critical constants ${p_c\sigma_1^3}/{k_B T}$,
$x_{1c}$, and $\eta_c$ for different $\gamma$-values as obtained
from the two extended CS equations (\ref{eCS1}) and (\ref{eCS2}).}
\label{TableCr}       %
\begin{tabular}{c|ccc|ccc}
\hline
&\multicolumn{3}{c|}{eCS1}&\multicolumn{3}{c}{eCS2}\\
\hline
 $\gamma$&$p_c\sigma_1^3/k_BT$&$x_{1c}$
&$\eta_c$&$p_c\sigma_1^3/k_BT$&$x_{1c}$ &$\eta_c$\\
\hline
0.05 &  3599 & 0.0093 & 0.822 & 1096 & 0.0004 & 0.204\\
0.1 & 1307 & 0.0203 & 0.757 & 832.0 & 0.0008 & 0.290 \\
0.2 & 653.4 & 0.0537 & 0.725 & --- & --- &  ---\\
0.3 & 581.9 & 0.0998 & 0.738 & --- &  ---&  ---\\
0.4 & 663.4 & 0.1532 & 0.766 & --- & ---& --- \\
\hline
\end{tabular}
\end{table}

As discussed earlier, the eCS1 EOS and, to a lesser extent, the eCS2
EOS are both in reasonably good agreement with the available
simulation results for the compressibility factor
\cite{YCH96,MV99,BMLS96} and lead to the exact second and third
virial coefficients but differ in the predictions for $B_n$ with
$n\geq 4$. The scatter in the values for the critical constants
shown in Table \ref{TableCr} is evident and so there is no
indication as to whether one should prefer one equation over the
other in connection with this problem. Notice, for instance, that
the eCS2 does not predict demixing for $\gamma\geq 0.2$, while both
the values of the critical pressures and packing fractions for which
it occurs according to the eCS1 EOS suggest that the transition
might be metastable with respect to a fluid-solid transition.

%
\begin{figure}
\centering
    \includegraphics[width=0.95\columnwidth]{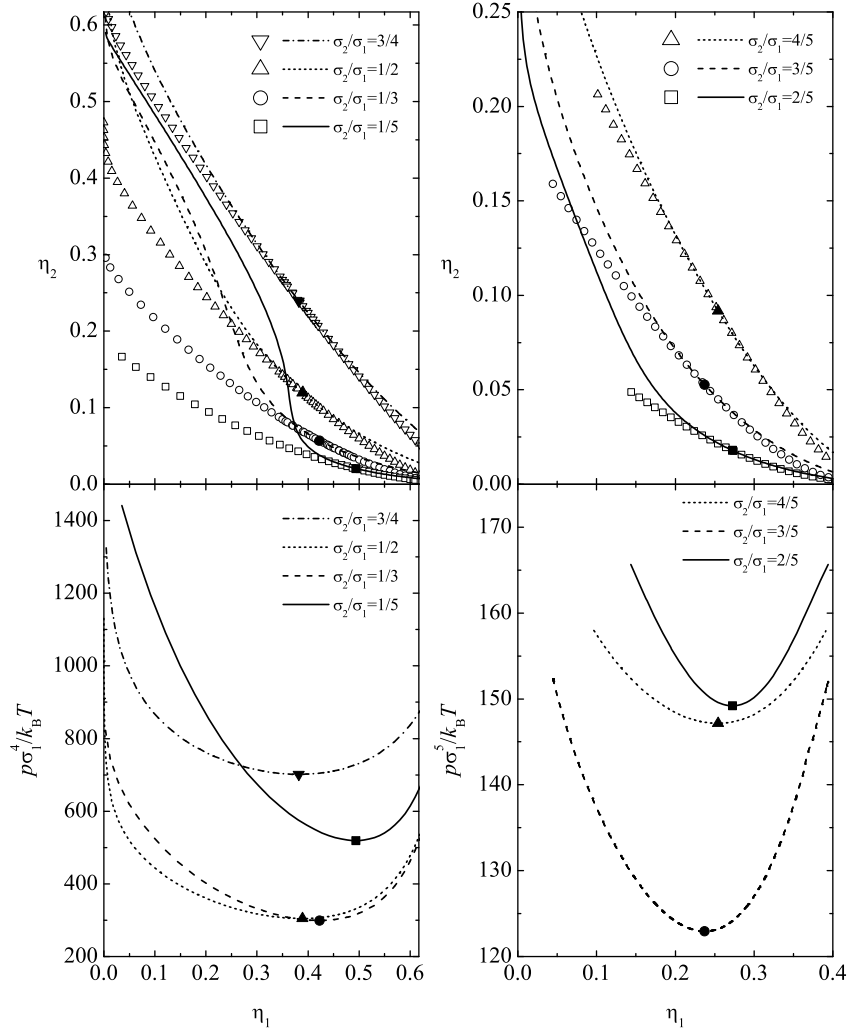}
%
%
\caption{Spinodal curves (upper panels: lines) and binodal curves
(upper panels: open symbols; lower panels: lines) in a 4D system
(left panels) and in a 5D system (right panels). The closed symbols
are the critical consolute points. }
\label{demix_4D5D}       
\end{figure}

Now we turn to the cases $d=4$ and $d=5$. Here we use the extended
Luban--Michels equation (eLM1) described in Subsection \ref{ss2.1}
[see Eq.\ \eqref{Ze1} and Table \ref{Tableg_s}].  As seen in Fig.\
\ref{demix_4D5D}, the location of the critical point tends to go
down and to the right in the $\eta_2$ vs $\eta_1$ plane as $\gamma$
decreases for $d=4$ \cite{YSH00a}. On the other hand, while it also
tends to go down as $\gamma$ decreases if $d=5$, its behavior in the
$\eta_2$ vs $\eta_1$ plane is rather more erratic in this case.
 Also, the value of the critical
pressure $p_{\text{c}}$ (in units of $k_BT/\sigma _{1}^{d}$) is not
a monotonic function of $\gamma $; its minimum value lies between
$\gamma =1/3$ and $\gamma =1/2$ when $d=4$, and it is around $\gamma
=3/5$ for $d=5$. This non-monotonic behavior is also observed for
three-dimensional HS \cite{CB98b,HT04}.

It is conceivable that the demixing transition in binary mixtures of
hard hyperspheres in four and five dimensions described above may be
metastable with respect to a fluid-solid transition, as it may also
be the case of 3D HS. In fact, the value of the pressure at the
freezing transition for the single component fluid is \cite{LM90}
$p_{\text{f}}\sigma ^{d}/k_BT\simeq 12.7$ ($d=3$), 11.5 ($d=4$), and
12.2 ($d=5$), {i.e.}, $p_{\text{f}}\sigma ^{d}/k_BT$ does not change
appreciably with the dimensionality but is clearly very small in
comparison with the critical pressures $p_{\text{c}}\sigma
_{1}^{d}/k_BT$ we obtain for the mixture; for instance,
$p_{\text{c}}\sigma_1^{d}/k_BT\simeq 600$ ($d=3$, $\gamma=3/10$),
300 ($d=4$, $\gamma=1/3$) and 123 ($d=5$, $\gamma=3/5$). However,
one should also bear in mind that, if the concentration $x_{1}$ of
the bigger spheres decreases, the value of the pressure at which the
solid-fluid transition in the mixture occurs in 3D is also
considerably increased with respect to $p_{\text{f}}$ [{cf.}\ Fig.\
6 of Ref.\ \cite{CB98b}]. Thus, for concentrations $x_{1}\simeq
0.01$ corresponding to the critical point of the fluid-fluid
transition, the maximum pressure of the fluid phase greatly exceeds
$p_{\text{f}}$. If a similar trend with composition also holds in 4D
and 5D, and given that the critical pressures become smaller as the
dimensionality $d$ is increased, it is not clear whether the
competition between the fluid-solid and the fluid-fluid transitions
in these dimensionalities will always be won by the former. The
point clearly deserves further investigation.

An interesting feature  must be mentioned. There is a remarkable
similarity between the binodal curves represented in the
$p\sigma_{i}^{d}$--$\eta _{1}$ and in the $\mu _{i}$--$\eta _{1}$
planes \cite{YSH00a}. By eliminating $\eta _{1}$ as if it were a
parameter, one can represent the binodal curves in a $\mu _{i}$ vs
$p\sigma_{i}^{d}$ plane.  Provided the origin of the chemical
potentials is such as to make $\lambda_i=\sigma_i$, the binodals  in
the $\mu _{i}$--$p\sigma _{i}^{d}$ plane practically collapse into a
single curve (which is in fact almost a straight line) for each
dimensionality ($d=3$, $d=4$, and $d=5$) \cite{YSH00a}. A closer
analysis of this phenomenon shows, however, that it is mainly due to
the influence on $\mu_i$ of terms which are quantitatively dominant
but otherwise irrelevant to the coexistence conditions.

\subsubsection{Binary Mixtures of Non-Additive Hard Hyperspheres in the Limit of High Dimensionality}
\label{sss2.3.2}

Let us now consider a binary mixture of non-additive HS of diameters
$\sigma_1$ and $\sigma_2$  in $d$ dimensions. Thus in this case
$\sigma_{12}\equiv \frac{1}{2}(\sigma_1+\sigma_2)(1+\Delta)$ where
as before $\Delta$ may be either positive or negative. Further
assume (something that will become exact in the limit $d\to\infty$
\cite{CFP91}) that the EOS of the mixture is described by the second
virial coefficient only, namely
\beq
p=\rho k_BT
\left[1+B_2(x_1)\rho\right],
\label{1dei}
\eeq
where, according to Eq.\ \eqref{n1x},
\beq
B_2(x_1)=v_d
2^{d-1}\left(x_1^2\sigma_{1}^d+x_2^2\sigma_{2}^d+2x_1x_2\sigma_{12}^d\right).
\label{2di}
\eeq
The Helmholtz free energy per unit volume is given by ${f}/{\rho
k_BT}=-1+\sum_{i=1}^2 x_i\ln\left(x_i\rho\lambda_i^d\right)
+B_2\rho$, where Eq.\ \eqref{FEN} has been used.
 The Gibbs free energy \emph{per particle} is
\beq
g=(f+p)/\rho=\sum_{i=1}^2 x_i\ln\left(x_i\rho\lambda_i^d\right)
+2B_2(x_1)\rho,
\label{4di}
\eeq
where without loss of generality we have set $k_B T=1$. { Given a
size ratio $\gamma$, a value of $\Delta$, and a dimensionality $d$,
the consolute critical point $(x_{1c},p_c)$ is the solution to
$\left({\partial^2 g}/{\partial x_1^2}\right)_p=\left({\partial^3
g}/{\partial x_1^3}\right)_p=0$, provided of course it exists}.
Then,  one can get the critical density $\rho_c$ from Eq.\
(\ref{1dei}).

We now introduce the scaled quantities \cite{SH05}
\beq
\ph\equiv 2^{d-1}v_d d^{-2}p\sigma_1^d/k_BT,\quad \yy \equiv
d^{-1}B_2\rho.
\label{new1di}
\eeq
Consequently, Eqs.\ (\ref{1dei}) and (\ref{4di}) can be rewritten as
\beq
\ph=\yy \left(\yy +d^{-1}\right)/{\widetilde{B}_2},
\label{new2di}
\eeq
\beq
g=\sum_{i=1}^2 x_i\ln\left(x_i\Lambda_i\right)+ \ln \left({A_d \yy
}/{\widetilde{B}_2}\right)+2d\yy ,
\label{new3di}
\eeq
where $\widetilde{B}_2\equiv B_2/2^{d-1}v_d\sigma_1^d$,
$\Lambda_i\equiv (\lambda_i/\sigma_1)^d$, and $A_d\equiv
d/2^{d-1}v_d$. Next we take the limit $d\to\infty$ and assume that
the volume ratio $\gh\equiv \gamma^d$ is kept fixed and that there
is a (slight) non-additivity $\Delta= d^{-2}\Deltah$ such that the
scaled non-additivity parameter $\Deltah$ is also kept fixed in this
limit. Thus, the second virial coefficient can be approximated by
\beq
\widetilde{B}_2=\widetilde{B}_2^{(0)}+\widetilde{B}_2^{(1)}d^{-1}+\mathcal{O}(d^{-2}),
\quad \widetilde{B}_2^{(0)}=\left(x_1+x_2\gh^{1/2}\right)^2,\quad
\widetilde{B}_2^{(1)}=x_1x_2\gh^{1/2}\KK,
\eeq
 with
\beq
\KK\equiv \frac{1}{4}\left(\ln\gh\right)^2+2 \Deltah .
\eeq
Let us remark that, { in order to find  a consolute critical point},
it is essential to keep the term of order $d^{-1}$ if $\Deltah \leq
0$. The EOS (\ref{new2di}) can then be inverted to yield
\beq
\yy =\yy ^{(0)}+\yy ^{(1)}d^{-1}+\mathcal{O}(d^{-2}) , \quad \yy
^{(0)}=\sqrt{\ph \widetilde{B}_2^{(0)}},\quad \yy ^{(1)}=
-\frac{1}{2} \left(1-\yy
^{(0)}\frac{\widetilde{B}_2^{(1)}}{\widetilde{B}_2^{(0)}}\right).
\eeq
In turn, the Gibbs free energy (\ref{new3di}) becomes
\beq
\begin{array}{ll}
&g=g^{(0)}d+g^{(1)}+\mathcal{O}(d^{-1}), \\ &g^{(0)}=2 \yy
^{(0)},\quad g^{(1)}=\sum_{i=1}^2
x_i\ln\left(x_i\Lambda_i\right)+\ln\left(A_d {\yy
^{(0)}}/{\widetilde{B}_2^{(0)}}\right)+2 \yy ^{(1)},
\end{array}
\eeq
while the chemical potentials $\mu_1=g+x_2\left(\partial g/\partial
x_1\right)_p$ and $\mu_2=g-x_1\left(\partial g/\partial
x_1\right)_p$ are given by
\beq
\begin{array}{ll}
&\mu_i=\mu_i^{(0)}d+\mu_i^{(1)}+\mathcal{O}(d^{-1}) ,\quad
\mu_1^{(0)}=2\ph^{1/2},\\& \mu_1^{(1)}=\ln\left(A_d x_1
\Lambda_1\sqrt{\ph/\widetilde{B}_2^{(0)}}\right)-1/\sqrt{\widetilde{B}_2^{(0)}}+(x_2/x_1)
(\gh\ph)^{1/2}\widetilde{B}_2^{(1)}/\widetilde{B}_2^{(0)},
\end{array}
\eeq
where $\mu_2$ is obtained from $\mu_1$ by the changes
$x_1\leftrightarrow x_2$, $\Lambda_1\to \Lambda_2/\gh$, $\gh\to
1/\gh$, $\ph\to \ph\gh$, $\widetilde{B}_2\to \widetilde{B}_2/\gh$.

{ The coordinates of the critical point are readily found to be}
\beq
x_{1c}=\frac{\gh^{3/4}}{1+\gh^{3/4}},\quad
\ph_c=\frac{\left(1+\gh^{1/4}\right)^4}{4\gh \KK^2}.
\label{8cp}
\eeq
Note that $x_{1c}$ is independent of $\Deltah$. The coexistence
curve, which has to be obtained numerically, follows from the
conditions $\mu_i^{(1)}(x_A,\ph)=\mu_i^{(1)}(x_B,\ph)$ ($i=1,2$)
where $x_1=x_A$ and $x_1=x_B$ are the mole fractions of the
coexisting phases. Once the critical consolute point has been
identified in the pressure/concentration plane, we can obtain the
critical density. The dominant behaviors of $\widetilde{B}_2$ and
$\yy $ at the critical point are
\beq
\widetilde{B}_2^{(0)}(x_{1c})=\frac{\gh}{\left(1-\gh^{1/4}+\gh^{1/2}\right)^2},
\quad \yy
_c^{(0)}=\frac{\left(1+\gh^{1/4}\right)^2}{2\left(1-\gh^{1/4}+\gh^{1/2}\right)\KK}.
\eeq
Hence, the critical density readily follows after substitution in
the scaling relation given in Eq.\ (\ref{new1di}). It is also
convenient to consider the scaled version $\etah\equiv d^{-1}
2^d\eta$ of the packing fraction
$\eta=v_d\rho\sigma_1^d\left(x_1+x_2\gh\right)$. At the critical
point, it takes the nice expression
\beq
\etah_c= \frac{\left(\gh^{1/8}+\gh^{-1/8}\right)^2}{\KK}.
\label{new9di}
\eeq

%
\begin{figure}[t]
\centering
    \includegraphics[height=8cm]{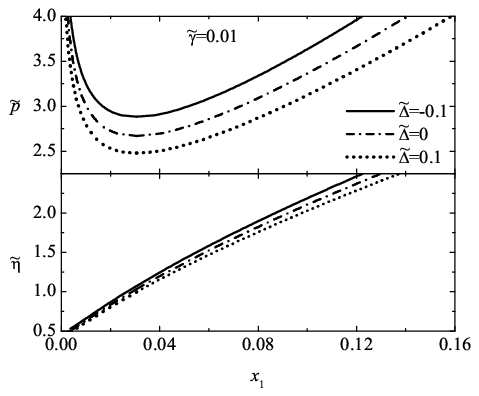}
%
%
\caption{Binodal curves in the planes $\ph$ vs $x_1$ and $\etah$ vs
$x_1$ corresponding to $\gh=0.01$ and $\Deltah=-0.1$, $\Deltah=0$,
and $\Deltah=0.1$.}
\label{coexistence_d_infty}       
\end{figure}
The previous results clearly indicate that a demixing transition is
possible {not only for additive or positively non-additive mixtures
but} even for negative non-additivities. The only requirement is
$\KK>0$, {i.e.}, $\Deltah
>-\frac{1}{8}\left(\ln\gh\right)^2$ {or, equivalently, $\Delta
>-\frac{1}{8}\left(\ln\gamma\right)^2$}. Figure
\ref{coexistence_d_infty} shows the binodal curves corresponding to
$\gh=0.01$ and $\Deltah=-0.1$ (negative non-additivity), $\Deltah=0$
(additivity), and $\Deltah=0.1$ (positive non-additivity).

While the high dimensionality limit has allowed us to address the
problem in a mathematically simple and clear-cut way, the
possibility of demixing with negative non-additivity is not an
artifact of that limit. As said before, demixing is known to occur
for positive non-additive binary mixtures of HS in three dimensions
and there is compelling evidence on the existence of this phenomenon
in the additive case, at least in the metastable fluid region. Even
though in a three-dimensional mixture the EOS is certainly more
complicated than Eq.\ (\ref{1dei}) and the demixing transition that
we have just discussed for negative non-additivity is possibly
metastable with respect to the freezing transition, the main effects
at work (namely the competition between depletion due to size
asymmetry and hetero-coordination due to negative non-additivity)
are also present. In fact, it is interesting to point out that Roth
et al.\ \cite{REL01}, using the approximation of an effective single
component fluid with pair interactions to describe a binary mixture
of non-additive 3D HS and employing an empirical rule based on the
effective second virial coefficient, have also suggested that
demixing is possible for small negative non-additivity  and high
size asymmetry. Our exact results lend support to this suggestion
 and confirm  that, in some cases, the
limit $d\to\infty$ highlights features already present in real
systems.

\section{The Rational Function Approximation (RFA) Method for the Structure of Hard-Sphere Fluids}
\label{sec3}

 The RDF $g(r)$ and its close relative the (static) structure
factor $S(q)$ are the basic quantities used to discuss the structure
of a single component fluid \cite{BH76,M76,F85,HM86}. The latter
quantity is defined as
\beq
S({q})=1+\rho \widetilde{h}(q),
\label{S(q)bis}
\eeq
where
\beq
\widetilde{h}(q)=\int\D\mathbf{r}\, \ee^{-\I
\mathbf{q}\cdot\mathbf{r} }h(r)
\label{d20bis}
\eeq
is the Fourier transform of the total correlation function
$h(r)\equiv g(r)-1$, $\I$ being the imaginary unit. An important
related quantity is the direct correlation function $c(r)$, which is
defined in Fourier space through the Ornstein--Zernike (OZ) relation
\cite{BH76,M76,F85,HM86}
\beq
\widetilde{c}(q)=\frac{\widetilde{h}(q)} {1+\rho\widetilde{h}(q)},
\label{d19}
\eeq
where $\widetilde{c}(q)$ is the Fourier transform of $c(r)$

 The usual approach to obtain $g(r)$ is through one of
the integral equation theories, where the OZ equation is
complemented by a closure relation between $c(r)$ and $h(r)$
\cite{BH76}. However, apart from requiring in general hard numerical
labor, a disappointing aspect is that the substitution of the
(necessarily) approximate values of $g(r)$ obtained from them in the
(exact) statistical mechanical formulae may lead to  the
thermodynamic inconsistency problem.

The two basic routes to obtain the EOS of a single component fluid
of HS are the virial route, Eq.\ \eqref{Z_s}, and the
compressibility route
\beqa
\chi_\pure\equiv k_BT \left(\frac{\partial \rho}{\partial
p}\right)_T&=&\left[1-\rho \widetilde{c}(0)\right]^{-1}= S(0)\nn
 &=&1+2^{d}d\eta
\sigma^{-d}\int_0^\infty \D r\, r^{d-1} h(r).
\label{chi}
\eeqa
Thermodynamic consistency implies that
\beq
\chi_\pure^{-1}(\eta)=\frac{\D }{\D \eta}[\eta Z_\pure(\eta)],
\label{consistent}
\eeq
but, in general, this condition is not satisfied by an approximate
RDF.
 In the case of a HS mixture,
the virial route is given by Eq.\ \eqref{1}, while the
compressibility route is indicated below [{cf.}\ Eq.\
\eqref{1.1bb}].

In this section we describe the RFA method, which is an alternative
to the integral equation approach and in particular leads by
construction to thermodynamic consistency.

\subsection{The Single Component HS Fluid}
\label{ss_3.1}

We begin with the case of a single component  fluid of HS of
diameter $\sigma$. The following presentation is equivalent to the
one given in Refs.\ \cite{YS91,YHS96}, where all details can be
found, but more suitable than the former for direct generalization
to the case of mixtures.

The starting point will be the Laplace transform
\beq
\label{2.1}
G(s)=\int_0^\infty \D r\, \ee^{-sr}r g(r)
\eeq
and the auxiliary
function $\Psi(s)$ defined through
\beq
\label{2.2}
G(s)=\frac{s}{2\pi}\left[\rho+\ee^{s\sigma}\Psi(s)\right]^{-1}.
\eeq
The choice of $G(s)$ as the Laplace transform of $r g(r)$ and the
definition of $\Psi(s)$ from Eq.\ \eqref{2.2} are suggested by the
exact form of $g(r)$ to first order in density \cite{YS91}.

Since $g(r)=0$ for $r<\sigma$ while $g(\sigma^+)=\text{finite}$, one
has
\beq
\label{3.2s}
g(r)=\Theta(r-\sigma)\left[g(\sigma^+)+
g'(\sigma^+)(r-\sigma)+\cdots\right],
\eeq
where $g'(r)\equiv \D g(r)/\D r$. This property imposes a constraint
on the large $s$ behavior of $G(s)$, namely
\beq
\label{3.3s}
 \ee^{\sigma s}s G(s)=\sigma g(\sigma^+ )
+\left[g(\sigma^+ )+\sigma g'(\sigma^+)\right] s^{-1}+{\cal
O}(s^{-2}).
\eeq
Therefore, $\lim_{s\to\infty} \ee^{s\sigma}sG(s)=\sigma
g(\sigma^+)=\text{finite}$ or, equivalently,
\beq
\label{2.3}
\lim_{s\to\infty}s^{-2}\Psi(s)=\frac{1}{2\pi\sigma
g(\sigma^+)}=\text{finite}.
\eeq
On the other hand,  according to Eq.\ \eqref{chi} with $d=3$,
\beqa
\chi_\pure&=&1-24\eta\sigma^{-3}\lim_{s\to 0}\frac{\D}{\D
s}\int_0^\infty \D r\, \ee^{-s r}r\left[g(r)-1\right]\nn
&=&1-24\eta\sigma^{-3}\lim_{s\to 0}\frac{\D}{\D
s}\left[G(s)-s^{-2}\right].
\label{GG}
\eeqa
Since the (reduced) isothermal compressibility $\chi_\pure$ is also
finite, one has $\int_0^\infty \D
r\,r^2\left[g(r)-1\right]=\text{finite}$, so that the weaker
condition
 $\int_0^\infty \D
r\,r\left[g(r)-1\right]=\lim_{s\to 0}[G(s)-s^{-2}]=\text{finite}$
must hold. This in turn implies
\beq
\label{2.4}
\Psi(s)=-\rho+\rho\sigma s-\frac{1}{2}\rho \sigma^2
s^2 +\left(\frac{1}{6}\rho\sigma^3+\frac{1}{2\pi}\right)s^3-
\left(\frac{1}{24}\rho\sigma^3+ \frac{1}{2\pi}\right)\sigma
s^4+{\cal O}(s^5).
\eeq

\subsubsection{First-Order Approximation (PY Solution)}

An interesting aspect to be remarked is that the minimal input we
have just described on the physical requirements related to the
structure and thermodynamics of the system is enough to  determine
the small and large $s$ limits of $\Psi(s)$, Eqs.\ \eqref{2.3} and
\eqref{2.4}, respectively. While infinite choices for $\Psi(s)$
would comply with such limits, a particularly simple form is a
\emph{rational function}. In particular, the rational function
having the least number of coefficients to be determined is
\beq
\label{2.5}
\Psi(s)=\frac{\SE^\zero+\SE^\one s+\SE^\two s^2+\SE^\three
s^3}{L^\zero+L^\one s},
\eeq
where one of the coefficients can be given an arbitrary non-zero
value. We choose $\SE^\three=1$.   With such a choice and in view of
Eq.\ (\ref{2.4}), one finds $\SE^\zero=-\rho L^\zero$,
$\SE^\one=-\rho(L^\one-\sigma L^\zero)$, $\SE^\two=\rho(\sigma
L^\one-\frac{1}{2} \sigma^2 L^\zero)$, and
\beq
\label{2.6}
L^\zero=2\pi \frac{1+2\eta}{(1-\eta)^2},
\eeq
\beq
\label{2.7}
L^\one=2\pi \sigma\frac{1+\eta/2}{(1-\eta)^2}.
\eeq
Upon substitution of these results into Eqs.\ (\ref{2.2}) and
(\ref{2.5}), we get
\beq
\label{2.8}
G(s)=\frac{\ee^{-\sigma s}}{2\pi s^2} \frac{L^\zero+L^\one s}{
1-\rho\left[\varphi_2(\sigma s)\sigma^3 L^\zero +\varphi_1(\sigma
s)\sigma^2 L^\one\right]},
\eeq
where
\beq
\label{2.9}
\varphi_n(x)\equiv x^{-(n+1)}\left(\sum_{m=0}^n \frac{(-x)^m}{m!}-
\ee^{-x}\right).
\eeq
In particular,
\beq
\varphi_0(x)=\frac{1-\ee^{-x}}{x},\quad
\varphi_1(x)=\frac{1-x-\ee^{-x}}{x^2},\quad
\varphi_2(x)=\frac{1-x+x^2/2-\ee^{-x}}{x^3}.
\eeq
Note that $\lim_{x\to 0}\varphi_n(x)=(-1)^n/(n+1)!$.

It is remarkable that  Eq.\ (\ref{2.8}), which has been derived here
as the simplest rational form for $\Psi(s)$ complying with the
requirements \eqref{2.3} and \eqref{2.4}, coincides with the
solution to the PY closure, $c(r)=0$ for $r>\sigma$, of the OZ
equation \cite{W63}. Application of Eq.\ \eqref{2.3} yields the PY
contact value $g_\pure^\text{PY}$ and compressibility factor
$Z_\pure^\text{PY}$ shown in Table \ref{Tableg_s}. Analogously, Eq.\
\eqref{GG} yields
\beq
\chi_\pure^{\text{PY}}=\frac{(1-\eta)^4}{(1+2\eta)^2}.
\label{chiPY}
\eeq
It can be easily checked that the thermodynamic relation
\eqref{consistent} is not satisfied by the PY theory.

\subsubsection{Second-Order Approximation}

In the  spirit of the RFA, the simplest extension of the rational
approximation (\ref{2.5}) involves two new terms, namely $\alpha
s^4$ in the numerator and $L^\two s^2$ in the denominator, both of
them necessary in order to satisfy Eq.\ (\ref{2.3}). Such an
addition leads to
\beq
\label{2.5bis}
\Psi(s)=\frac{\SE^\zero+\SE^\one s+\SE^\two s^2+\SE^\three
s^3+\alpha s^4}{L^\zero+L^\one s+L^\two s^2}.
\eeq
Applying Eq.\ \eqref{2.4}, it is possible to express $\SE^\zero$,
$\SE^\one$, $\SE^\two$, $\SE^\three$, $L^\zero$, and $L^\one$ in
terms of $\alpha$ and  $L^\two$. This leads to
\beq
\label{2.10}
G(s)=\frac{\ee^{-\sigma s}}{2\pi s^2} \frac{L^\zero+L^\one
s+{L^\two} s^2}{ 1+\alpha s-\rho\left[\varphi_2(\sigma s)\sigma^3
L^\zero +\varphi_1(\sigma s)\sigma^2 L^\one +\varphi_0(\sigma
s)\sigma L^\two\right]},
\eeq
where
\beq
\label{2.11}
L^\zero=2\pi \frac{1+2\eta}{(1-\eta)^2}
+\frac{12\eta}{1-\eta}\left(
\frac{\pi}{1-\eta}\frac{\alpha}{\sigma}-\frac{L^\two}{\sigma^2}\right),
\eeq
\beq
\label{2.12}
L^\one=2\pi
\sigma\frac{1+\frac{1}{2}\eta}{(1-\eta)^2} +\frac{2}{1-\eta}\left(
\pi\frac{1+2\eta}{1-\eta}\alpha-3\eta\frac{L^\two}{\sigma}\right).
\eeq
Thus far, irrespective of the values of the coefficients $L^\two$
and $\alpha$, the conditions $\lim_{s\to\infty}
\ee^{s\sigma}sG(s)=\text{finite}$ and $\lim_{s\to
0}[G(s)-s^{-2}]=\text{finite}$ are satisfied. Of course, if
$L^\two=\alpha=0$, one recovers the PY approximation. More
generally, we may determine these coefficients by prescribing the
compressibility factor $Z_\pure$ (or equivalently the contact value
$g_\pure$) and then, in order to ensure thermodynamic consistency,
compute from it the isothermal compressibility $\chi_\pure$ by means
of Eq.\ \eqref{consistent}. {}From Eqs.\ \eqref{2.3} and \eqref{GG}
one gets
\beq
\label{2.13}
{L^\two} ={2\pi \alpha \sigma}g_\pure,
\eeq
\beq
\label{2.14}
\chi_\pure=\left(\frac{2\pi}{L^\zero}\right)^2
\left[1-\frac{12\eta}{1-\eta}\frac{\alpha}{\sigma}\left(1+2
\frac{\alpha}{\sigma}\right)+\frac{12\eta}{\pi}\frac{\alpha
L^\two}{\sigma^3} \right].
\eeq
Clearly, upon substitution of Eqs.\
(\ref{2.11}) and (\ref{2.13}) into Eq.\ (\ref{2.14}) a quadratic
algebraic equation for $\alpha$ is obtained. The physical root is
\beq
\alpha=-\frac{12\eta(1+2\eta)\SE_4}{(1-\eta)^2+36\eta\left[1+\eta-Z_\pure(1-\eta)\right]\SE_4},
\label{Ralfa}
\eeq
where
\begin{equation}
\SE_4=\frac{1-\eta }{36\eta \left( Z_\pure-\frac{1}{3}\right)
}\left\{ 1-\left[
1+\frac{Z_\pure-\frac{1}{3}}{Z_\pure-Z_\pure^\text{PY}}\left(
\frac{\chi_\pure}{\chi_\pure^\text{PY}}-1\right) \right]
^{1/2}\right\} .
\label{S4}
\end{equation}
The other root must be discarded because it corresponds to a
negative value of $\alpha$, which, according to Eq.\ \eqref{2.13},
yields a negative value of $L^\two$. This would imply the existence
of a positive real value of $s$ at which $G(s)=0$
\cite{YS91,YHS96}}, which is not compatible with a positive definite
RDF. However, according to the form of Eq.\ (\ref{S4}) it may well
happen that, once $Z_\pure$ has been chosen, there exists a certain
packing fraction $\eta_\text{g}$ above which $\alpha$ is no longer
positive. This may be interpreted as an indication that, at the
packing fraction $\eta_\text{g}$ where $\alpha$ vanishes, the system
ceases to be a fluid and a glass transition in the HS fluid occurs
\cite{YHS96,RHSY98,RH03}.

Expanding \eqref{2.10} in powers of $s$ and using  Eq.\ (\ref{3.3s})
one can obtain the derivatives of the RDF at $r=\sigma^+$
\cite{RH97}. In particular, the first derivative is
\beq
g'(\sigma^+)=\frac{1}{2\pi\alpha\sigma}\left[L^\one-
L^\two\left(\frac{1}{\alpha}+\frac{1}{\sigma}\right)\right],
\label{3.19s}
\eeq
which may have some use in connection with perturbation theory
\cite{LL73}.

It is worthwhile to point out that the structure implied by Eq.\
(\ref{2.10}) coincides in this single component case with the
solution of the Generalized Mean Spherical Approximation (GMSA)
\cite{W73}, where the OZ relation is solved under the ansatz that
the direct correlation function has a Yukawa form outside the core.

For a given $Z_\pure$, once $G(s)$ has been determined, inverse
Laplace transformation yields $r g(r)$. First, note that Eq.\
\eqref{2.2} can be formally rewritten as
\beq
G(s)=-\frac{s}{2\pi}\sum_{n=1}^\infty
\rho^{n-1}\left[-\Psi(s)\right]^{-n} \ee^{-ns\sigma}.
\eeq
Thus, the RDF is then given by
\begin{equation}
g\left({r}\right) =\frac{1}{2\pi r} \sum_{n=1}^{\infty
}\rho^{n-1}\psi _{n}\left(r-n\sigma\right) \Theta
\left(r-n\sigma\right) ,
\label{g(r)}
\end{equation}
with $\Theta \left(x\right)$ denoting the Heaviside step function
and
\begin{equation}
\psi_{n}\left( r\right) =-\mathcal{L}^{-1}\left\{ s\left[ -\Psi
\left( s\right) \right] ^{-n}\right\} ,
\label{varphi}
\end{equation}
$\mathcal{L}^{-1}$ denoting the inverse Laplace transform.
Explicitly, using the residue theorem,
\begin{equation}
\psi _{n}\left( r\right) =- \sum_{i=1}^{4} \ee^{s_{i}r}
\sum_{m=1}^{n} \frac{a_{mn}^{(i)}}{\left(n - m\right)!(m-1)! }
r^{n-m} ,
\label{z8bis}
\end{equation}
where
\begin{equation}
a_{mn}^{(i)} = \lim_{s \rightarrow s_{i}} \left(\frac{\D}{\D
s}\right)^{m-1} s\left[-\Psi \left( s\right)/(s-s_i) \right] ^{-n},
\label{z8bisa}
\end{equation}
$s_{i}$ ($i=1,\ldots,4$) being the poles of $1/\Psi(s)$, {i.e.}, the
roots of $\SE^\zero+\SE^\one s+\SE^\two s^2+\SE^\three s^3+\alpha
s^{4}=0$. Explicit expressions of $g(r)$ up to the second
coordination shell $\sigma\leq r\leq 3\sigma$ can be found in Ref.\
\cite{DLS06}.

On the other hand, the  static structure factor $S(q)$ [{cf}. Eq.\
\eqref{S(q)bis}] and the Fourier transform $\widetilde{h}(q)$
 may be related to $G(s)$ by noting that
\beq
\widetilde{h}(q)=\frac{4\pi}{q}\int_0^\infty \D r \, r \sin(qr)
h(r)=-2\pi \left.\frac{G(s)-G(-s)}{s} \right|_{s=\I q}.
 \label{S(q)}
\end{equation}
 Therefore, the basic structural quantities of the single component HS
fluid, namely the RDF and the static structure factor, may be
analytically determined within the RFA method once the
compressibility factor $Z_\pure$, or equivalently the contact value
$g_\pure$, is specified. In Fig.\ \ref{g(r)mono} we compare
simulation data of $g(r)$ for a density $\rho \sigma^3=0.9$
\cite{KLM04} with the RFA prediction and a recent approach by
Trokhymchuk {et al.} \cite{TNJH05}, where
$Z_\pure=Z_\pure^{\text{CS}}$ [{cf}.\ Table \ref{Tableg_s}] and the
associated compressibility
\beq
\chi_\pure^\text{CS}=\frac{(1-\eta)^4}{1+4\eta+4\eta^2-4\eta^3+\eta^4}
\eeq
are taken in both cases. Both theories are rather accurate, but the
RFA captures better the maxima and minima of $g(r)$ \cite{HSY06}.

\begin{figure}[b]
\centering
    \includegraphics[height=8cm]{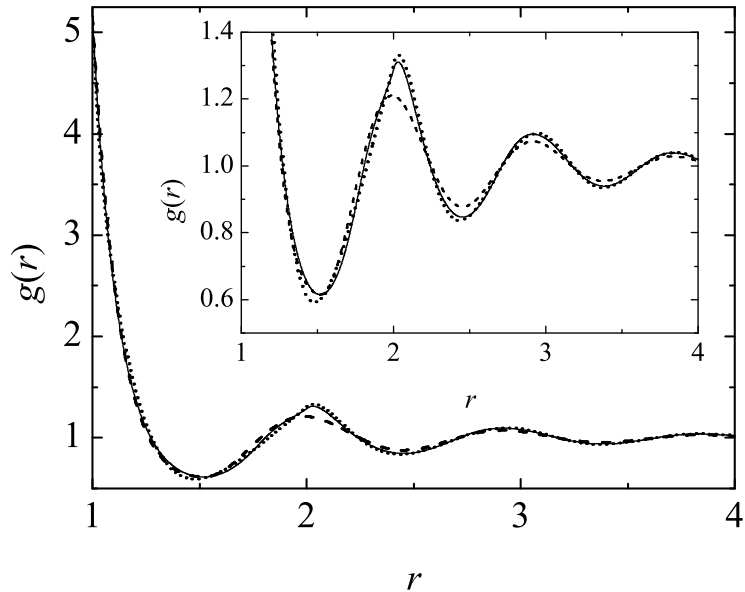}
%
%
\caption{Radial distribution function of a single component HS fluid
for $\rho \sigma^3=0.9$. The solid lines represent simulation data
\protect\cite{KLM04}. The dashed lines represent the results of the
approach of Ref.\ \protect\cite{TNJH05}, while the dotted lines
refer to those of the RFA method. The inset shows the oscillations
of $g(r)$ in more detail.}
\label{g(r)mono}       
\end{figure}

It is also possible to obtain within the RFA method the direct
correlation function $c(r)$. Using Eqs.\ \eqref{d19} and
\eqref{S(q)}, and applying the residue theorem, one gets, after some
algebra,
\beq
\label{c(r)}
c(r)=\left(\aK_+\frac{\ee^{\kappa r}}{r}+\aK_-\frac{\ee^{-\kappa
r}}{r}+\frac{\aK_{-1}}{r}+\aK_0+\aK_1 r+\aK_3 r^3\right)
\Theta(1-r)+\aK\frac{\ee^{-\kappa r}}{r},
\eeq
where
\beq
\kappa=\frac{1}{\alpha}\sqrt{12\alpha\eta L^\two/\pi+1-12\alpha(1+2
\alpha)\eta/(1- \eta)},
\label{kappa}
\eeq
\beqa
 \aK_\pm&=&\frac{\ee^{\mp\kappa}}{4 \alpha^2 (1 - \eta)^4
\kappa^6} \Bigl\{ 2\left[1+2(1+3 \alpha)\eta\right]\pm \left[2 +
\eta + 2\alpha (1 + 2 \eta )\right] \kappa \nn
 && \left.+
(1 - \eta )\left[\kappa^2 - \eta \left(12 +(\kappa \pm
6)\kappa\right)\right]L^\two/\pi\right\}\Bigl\{ 12 \eta
\left[1+2(1+3 \alpha)\eta\right] \nn
 &&\pm 6 \eta \left[3\eta
- 2 \alpha (1-4 \eta)\right] \kappa - 6 \eta(1 + 2 \alpha)(1 -\eta)
\kappa^2 - (1 - \eta)^2 \kappa^3 (\alpha \kappa\mp 1)\nn
 &&\left.+ 6\eta
(1 - \eta)  \left[\kappa^2 - \eta \left(12 +(\kappa \pm
6)\kappa\right)\right]L^\two/\pi \right\},
\label{eca2}
\eeqa
\beq
\aK_{-1}=-\left(\frac{L^\two}{2 \pi \alpha} + \aK_+ \ee^{\kappa} +
\aK_- \ee^{-\kappa}+ \aK_0 + \aK_1 + \aK_3\right),
\label{eca3}
\eeq
\beq
\aK_0=-\left[\frac{1+2\left(1+3 \alpha\right)\eta-6 \eta  \left( 1 -
\eta\right)L^\two/\pi }{\alpha \kappa \left(1 -\eta
\right)^2}\right]^2 ,
\label{eca4}
\eeq
\beqa
\aK_1&=&\frac{6\eta}{\kappa^2}\aK_0+\frac{3 \eta}{2 \alpha^2
\kappa^2 \left(1 -\eta \right)^4}\left\{\left[2+\eta+2\alpha(1+
2\eta)\right]^2 - 4 \left(1 - \eta\right)\left[1 +\eta \right.
\right.\nn
 &&\left. \left.  \times (7
+ \eta + 6 \alpha \left(2 + \eta \right))\right]L^\two /\pi +12 \eta
\left(2 +\eta\right)(1-\eta)^2{L^\two}^2/\pi^2\right\},
\label{eca5}
\eeqa
\beq
\aK_3=\frac{ \eta}{2}\aK_0,
\label{eca6}
\eeq
\beq
\aK=-\left(\aK_++\aK_-+\aK_{-1}\right).
\label{aprima}
\eeq
In Eqs.\ \eqref{kappa}--\eqref{aprima} we have taken $\sigma=1$ as
the length unit. Note that Eq.\ \eqref{aprima} guarantees that
$c(0)=\text{finite}$, while Eq.\ \eqref{eca3} yields
$c(\sigma^+)-c(\sigma^-)=L^\two/2\pi\alpha=g(\sigma^+)$. The latter
equation proves the continuity of the  indirect correlation function
$\gamma(r)\equiv h(r)- c(r)$ at $r=\sigma$. With the above results,
Eqs.\ \eqref{g(r)} and \eqref{c(r)},  one may immediately write the
function $\gamma(r)$. Finally, we note that the bridge function
$B(r)$ is linked to $\gamma(r)$ and to the cavity (or background)
function $y(r)\equiv \ee^{\phi(r)/k_BT}g(r)$, where $\phi(r)$ is the
interaction potential,  through
\begin{equation}
 B(r)= \ln y(r)-\gamma(r),
\label{b(r)}
\end{equation}
and so, within the RFA method, the bridge function is also
completely specified analytically for $r>\sigma$ once $Z_\pure$ is
prescribed.

If one wants to have $B(r)$ also for $0\leq r \leq \sigma$, then an
expression for the cavity function is required in that region. Here
we propose such an expression using a limited number of constraints.
First, since the cavity function and its first derivative are
continuous at $r=\sigma$, we have
\beq
y(1)=g_\pure,\quad
\frac{y'(1)}{y(1)}=\frac{L^\one}{L^\two}-\frac{1}{\alpha}-1,
\label{yp1}
\eeq
where Eqs.\ \eqref{2.13} and \eqref{3.19s} have been used and again
$\sigma=1$ has been taken. Next, we consider the following exact
zero-separation theorems \cite{L95}:
\beq
\ln y(0)=Z_\pure(\eta)-1+\int_0^\eta
d\eta'\frac{Z_\pure(\eta')-1}{\eta'},
\label{36zst}
\eeq
\beq
\frac{y'(0)}{y(0)}=-6\eta y(1).
\label{37zst}
\eeq
\begin{figure}[t]
\centering
    \includegraphics[height=8cm]{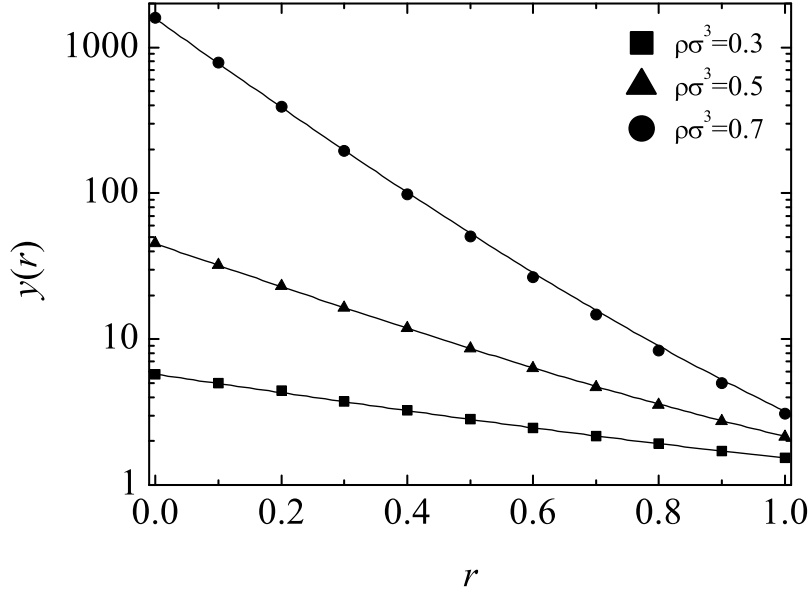}
%
%
\caption{Cavity function  of a single component HS fluid in the
overlap region for $\rho \sigma^3=0.3$, 0.5, and 0.7. The solid
lines represent our proposal \protect\eqref{lny} with
$Z_\pure=Z_\pure^{\text{CS}}$, while the symbols represent Monte
Carlo simulation results \protect\cite{LM84}.}
\label{y(r)_hs}       
\end{figure}
\begin{figure}[t]
\centering
    \includegraphics[height=8cm]{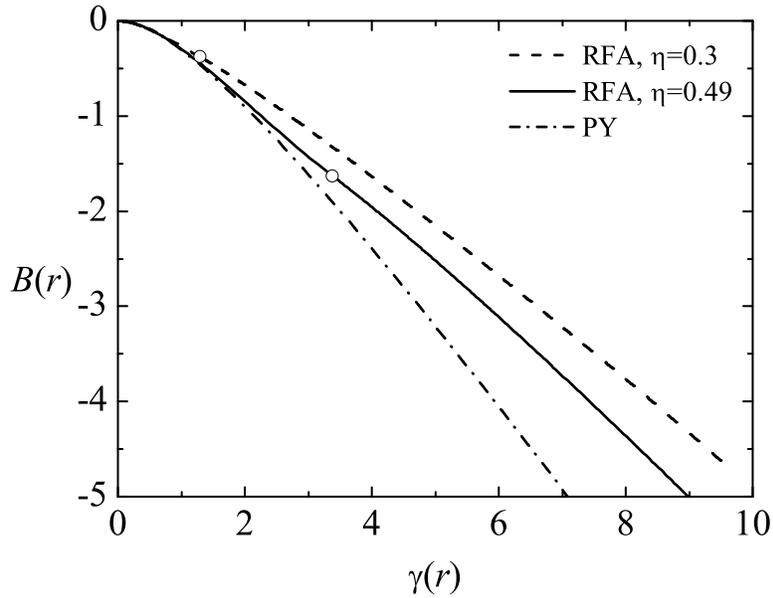}
%
%
\caption{Parametric plot of the bridge function $B(r)$ versus the
indirect correlation function $\gamma(r)$. The dashed line refers to
the RFA for $\eta=0.3$, while the solid line refers to the RFA for
$\eta=0.49$. In each case, the branch of the curve to the right of
the circle corresponds to $r\leq 1$, while that to the left
corresponds to $r\geq 1$. For comparison, the PY closure
$B(r)=\ln[1+\gamma(r)]-\gamma(r)$ is also plotted (dash-dotted
line).}
\label{B_vs_gamma}       
\end{figure}
The four conditions \eqref{yp1}--\eqref{37zst} can be enforced by
assuming a cubic polynomial form for $\ln y(r)$ inside the core,
namely
\beq
y(r)=\exp\left(Y_0+Y_1 r+Y_2 r^2+Y_3 r^3\right),\quad (0\leq r\leq
1),
\label{lny}
\eeq
where
\beq
Y_0=Z_\pure(\eta)-1+\int_0^\eta
d\eta'\frac{Z_\pure(\eta')-1}{\eta'},
\eeq
\beq
Y_1=-6\eta y(1),
\eeq
\beq
Y_2=3\ln y(1)-\frac{y'(1)}{y(1)}-3Y_0-2Y_1,
\eeq
\beq
Y_3=-2\ln y(1)+\frac{y'(1)}{y(1)}+2Y_0+Y_1.
\eeq
The proposal \eqref{lny} is compared with available Monte Carlo data
\cite{LM84} in Fig.\ \ref{y(r)_hs}, where an excellent agreement can
be observed.

Once the cavity function $y(r)$ provided by the RFA method is
complemented by \eqref{lny}, the bridge function $B(r)$ can be
obtained at any distance. Figure \ref{B_vs_gamma} presents a
parametric plot of the bridge function versus the indirect
correlation function as given by the RFA method for two different
packing fractions, as well as the result associated with the PY
closure. The fact that one gets a smooth curve means that within the
RFA the oscillations in $\gamma(r)$ are highly correlated to those
of $B(r)$. Further, the effective closure relation in the RFA turns
out to be density dependent, in contrast with what occurs for the PY
theory. Note that the absolute value $|B(r)|$  for a given value of
$\gamma(r)$ is smaller in the RFA than the PY value and that the RFA
and PY curves become paradoxically closer for larger densities.
Since the PY theory is known to yield rather poor values of the
cavity function inside the core \cite{MS06,SM06}, it seems likely
that the present differences may represent yet another manifestation
of the superiority of the RFA method, a point that certainly
deserves to be further explored.

\subsection{The Multicomponent HS Fluid} \label{ss_3.2}
The method outlined in the preceding subsection will be now extended
to an $N$-component mixture of additive HS. Note that in a
multicomponent system the isothermal compressibility $\chi$ is given
by
\beqa
\label{1.1bb}
\chi^{-1}&=&\frac{1}{k_BT}\left(\frac{\partial p}{\partial
\rho}\right)_{T,\{x_j\}}=\frac{1}{k_BT} \sum_{i=1}^N
x_i\left(\frac{\partial p}{\partial
\rho_i}\right)_{T,\{x_j\}}\nonumber\\
&=&1-\rho\sum_{i,j=1}^N x_ix_j \widetilde{c}_{ij}(0),
\eeqa
where  $\widetilde{c}_{ij}(q)$ is the Fourier transform of the
direct correlation function $c_{ij}(r)$, which is defined by the OZ
equation
\beq
\label{1.3bb}
\widetilde{h}_{ij}(q)=\widetilde{c}_{ij}(q)+ \sum_{k=1}^N
\rho_k\widetilde{h}_{ik}(q)\widetilde{c}_{kj}(q),
\eeq
where $h_{ij}(r)\equiv g_{ij}(r)-1$. Equations \eqref{1.1bb} and
\eqref{1.3bb} are the multicomponent extensions of Eqs.\ \eqref{chi}
and \eqref{d19}, respectively. Introducing the quantities
$\widehat{h}_{ij}(q)\equiv \sqrt{\rho _{i}\rho
_{j}}\widetilde{h}_{ij}(q)$ and
$\widehat{c}_{ij}(q)\equiv\sqrt{\rho_{i}\rho_{j}}\,\widetilde{c}_{ij}(q)$,
the OZ relation \eqref{1.3bb} becomes, in matrix notation,
\begin{equation}
 \widehat{\sf c}(q)= \widehat{\sf h}(q)\cdot[{\sf
I}+\widehat{\sf h}(q)]^{-1} ,
\label{C(q)}
\end{equation}
where $\openone$ is the $N\times N$ identity matrix. Thus, Eq.\
\eqref{1.1bb} can be rewritten as
\beq
\label{3.18}
\chi^{-1}=\sum_{i,j=1}^N
\sqrt{x_ix_j}\left[\delta_{ij}-\widehat{c}_{ij}(0)\right]=
\sum_{i,j=1}^N \sqrt{x_ix_j} {\left[\openone+\widehat{\sf
h}(0)\right]_{ij}^{-1}}.
\eeq

Similarly to what we did in the single component case,  we introduce
the Laplace transforms of $r g_{ij}(r)$:
\beq
\label{3.1}
G_{ij}(s)=\int_0^\infty \D r\, \ee^{-sr}r g_{ij}(r).
\eeq
The counterparts of Eqs.\ \eqref{3.2s} and \eqref{3.3s} are
\beq
\label{3.2}
g_{ij}(r)=\Theta(r-\sigma_{ij})\left[g_{ij}(\sigma_{ij}^+)+
g_{ij}'(\sigma_{ij}^+)(r-\sigma_{ij})+\cdots\right],
\eeq
\beq
\label{3.3}
 \ee^{\sigma_{ij}s}s G_{ij}(s)=\sigma_{ij}g_{ij}(\sigma_{ij}^+ )
+\left[g_{ij}(\sigma_{ij}^+
)+\sigma_{ij}g_{ij}'(\sigma_{ij}^+)\right] s^{-1}+{\cal O}(s^{-2}).
\eeq
Moreover, the condition of a finite compressibility implies that
$\widetilde{h}_{ij}(0)=\text{finite}$. As a consequence, for small
$s$,
\beq
\label{3.4}
s^2
G_{ij}(s)=1+H_{ij}^{(0)}s^2+H_{ij}^{(1)}s^3+\cdots
\eeq
with $H_{ij}^\zero=\text{finite}$ and
$H_{ij}^\one=-\widetilde{h}_{ij}(0)/4\pi=\text{finite}$, where
\beq
\label{3.5}
H_{ij}^\n\equiv \frac{1}{n!}\int_0^\infty \D r\, (-r)^n r h_{ij}(r).
\eeq

 We are now in the position to generalize the approximation
(\ref{2.10}) to the $N$-component case \cite{YSH98a}. While such a
generalization may be approached in a variety of ways,  two
motivations are apparent. On the one hand, we want to recover the PY
result \cite{L64} as a particular case in much the same fashion as
in the single component system. On the other hand, we want to
maintain the development as simple as possible. Taking all of this
into account, we propose
\beq
\label{3.6}
G_{ij}(s)=\frac{\ee^{-\sigma_{ij} s}}{2\pi s^2}
\left({\sf L}(s)\cdot \left[(1+\alpha s)\openone-{ \sf
A}(s)\right]^{-1}\right)_{ij},
\eeq
where $\mathsf{L}(s)$ and $\mathsf{A}(s)$ are the matrices
\beq
\label{3.7}
L_{ij}(s)=L_{ij}^\zero+L_{ij}^\one
s+L_{ij}^\two s^2,
\eeq
\beq
\label{3.8}
A_{ij}(s)=\rho_i\left[\varphi_2(\sigma_i s)\sigma_i^3 L_{ij}^\zero
+\varphi_1(\sigma_i s)\sigma_{i}^2 L_{ij}^\one +\varphi_0(\sigma_{i}
s)\sigma_{i} L_{ij}^\two\right],
\eeq
the functions $\varphi_n(x)$ being defined by Eq.\ \eqref{2.9}. We
note that, by construction, Eq.\ (\ref{3.6}) complies with the
requirement $\lim_{s\rightarrow\infty} \ee^{\sigma_{ij}s}s
G_{ij}(s)=\text{finite}$. Further, in view of Eq.\ (\ref{3.4}), the
coefficients of $s^0$ and $s$ in the power series expansion of $s^2
G_{ij}(s)$ must be 1 and 0, respectively. This yields $2N^2$
conditions that allow us to express ${\sf L}^\zero$ and ${\sf
L}^\one$ in terms of ${\sf L}^\two$ and $\alpha$. The solution is
\cite{YSH98a}
\beq
\label{3.13}
L_{ij}^\zero=\lambdak+\lambdakk\sigma_j+2\lambdakk\alpha-
\lambdak\sum_{k=1}^N \rho_k\sigma_k L_{kj}^\two,
\eeq
\beq
\label{3.14}
L_{ij}^\one=\lambdak\sigma_{ij}+\frac{1}{2}\lambdakk\sigma_i\sigma_j
+(\lambdak+\lambdakk\sigma_i)\alpha-\frac{1}{2}\lambdak\sigma_i
\sum_{k=1}^N \rho_k\sigma_k L_{kj}^\two,
\eeq
where $\lambdak\equiv 2\pi/(1-\eta)$ and $\lambdakk\equiv
6\pi(\mt/\mth)\eta/(1-\eta)^2$.

In parallel with the development of the single component case, ${\sf
L}^\two$ and $\alpha$ can be chosen arbitrarily. Again, the choice
$L_{ij}^\two=\alpha=0$ gives the PY solution \cite{L64,BH77}. Since
we want to go beyond this approximation, we will determine those
coefficients by taking prescribed values for $g_{ij}(\sigma_{ij} )$,
which in turn, via Eq.\ \eqref{1}, give the EOS of the mixture. This
also leads to the required value of $\chi^{-1}=\partial(\rho
Z)/\partial \rho$, thus making the theory thermodynamically
consistent. In particular, according to Eq.\ (\ref{3.3}),
\beq
\label{3.17}
{L_{ij}^\two}={2\pi\alpha\sigma_{ij}} g_{ij}(\sigma_{ij}^+ ).
\eeq
The condition related to $\chi$ is more involved. Making use of Eq.\
(\ref{3.4}), one can get $\widetilde{h}_{ij}(0)=-4\pi H_{ij}^\one$
in terms of ${\sf L}^\two$ and $\alpha$ and then insert it into Eq.\
\eqref{3.18}. Finally, elimination of $L_{ij}^\two$ in favor of
$\alpha$ from Eq.\ (\ref{3.17}) produces an algebraic equation of
degree $2N$, whose physical root is determined by the requirement
that $G_{ij}(s)$ is positive definite for  positive real $s$. It
turns out that the physical solution corresponds to the smallest of
the real roots. Once $\alpha$ is known, upon substitution into Eqs.\
(\ref{3.6}), (\ref{3.13}), (\ref{3.14}), and (\ref{3.17}), the
scheme is complete. Also, using Eq.\ (\ref{3.3}), one can easily
derive the result
\beq
\label{3.19}
g_{ij}'(\sigma_{ij}^+)=\frac{1}{2\pi\alpha\sigma_{ij}}\left[L_{ij}^\one-
L_{ij}^\two\left(\frac{1}{\alpha}+\frac{1}{\sigma_{ij}}\right)\right].
\eeq
It is straightforward to check that the results of the preceding
subsection are recovered  by setting $\sigma_i=\sigma$, regardless
of the values of the mole factions.

Once $G_{ij}(s)$ has been determined, inverse Laplace transformation
directly yields $rg_{ij}(r)$. Although in principle this can be done
analytically, it is more practical to use one of the efficient
methods discussed by  Abate and Whitt \cite{AW92} to numerically
invert Laplace transforms \cite{notebook}.

\begin{figure}
\centering
        \includegraphics[width=0.95\columnwidth]{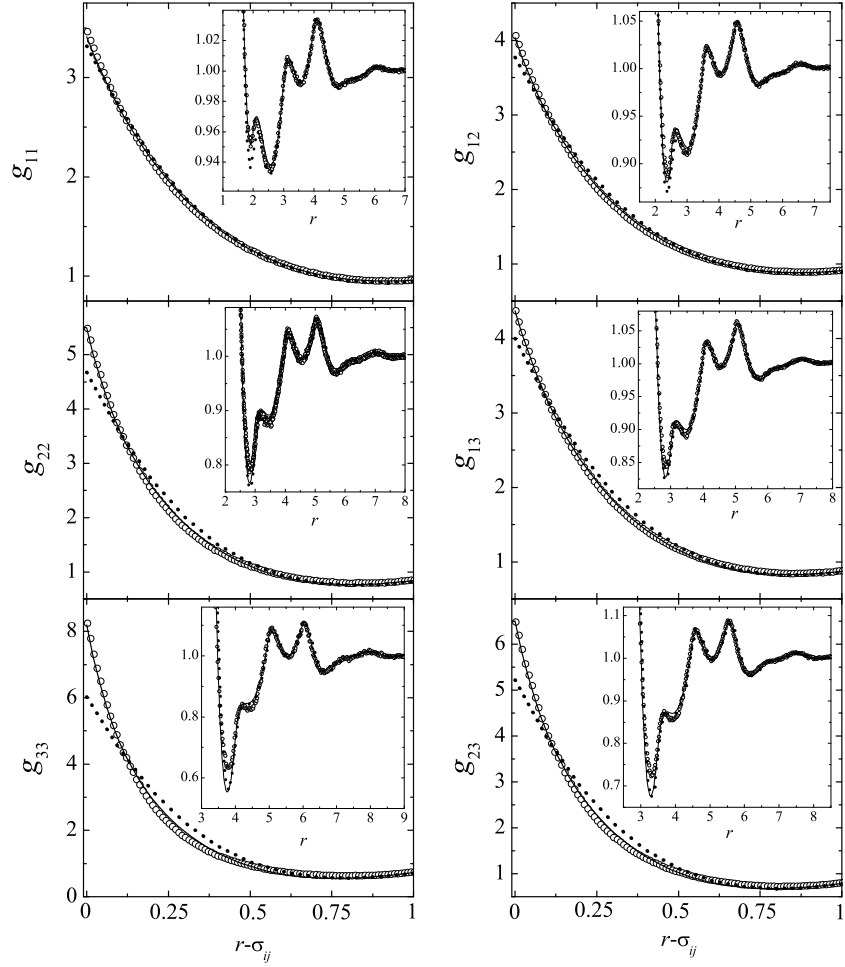}
%
%
\caption{Radial distribution functions $g_{ij}(r)$ for a ternary
mixture with diameters $\sigma_1=1$, $\sigma_2=2$, and $\sigma_3=3$
at a packing fraction $\eta=0.49$ with mole fractions $x_1=0.7$,
$x_2=0.2$, and $x_3=0.1$. The circles are simulation results
\protect\cite{MMYSH02}, the solid lines are the RFA predictions, and
the dotted lines are the PY predictions.}
\label{ternary}       
\end{figure}

In Fig.\ \ref{ternary} we present a comparison between the results
of the RFA method  with the PY theory and simulation data
\cite{MMYSH02} for the RDF of a ternary mixture. In the case of the
RFA, we have used the eCS2 contact values and the corresponding
isothermal compressibility. The improvement of the RFA over the PY
prediction, particularly in the region near contact, is noticeable.
Although the RFA accounts nicely for the observed oscillations, it
seems to somewhat overestimate the depth of the first minimum.

Explicit knowledge of $G_{ij}(s)$ also allows us to determine
 the Fourier transform $\widetilde{h}_{ij}(q)$ through
the relation
\beq
\widetilde{h}_{ij}(q)=-2\pi \left.\frac{G_{ij}(s)-G_{ij}(-s)}{s}
\right|_{s=\I q}.
\label{1.7}
\end{equation}
The structure factor $S_{ij}(q)$ may be expressed in terms of
$\widetilde{h}_{ij}(q)$ as \cite{HM86}
\beq
\label{1.6}
S_{ij}(q)=x_i \delta_{ij}+\rho x_i x_j
\widetilde{h}_{ij}(q).
\eeq
In the particular case of a binary mixture, rather than the
individual structure factors $S_{ij}(q)$, it is some combination of
them which may be easily associated with fluctuations of the
thermodynamic variables \cite{AL67,BT70}. Specifically, the
quantities \cite{HM86}
\beq
\label{1.8}
S_{nn}(q)=S_{11}(q)+S_{22}(q)+2S_{12}(q),
\eeq
\beq
\label{1.9}
S_{nc}(q)=x_2 S_{11}(q)-x_1 S_{22}(q)+(x_2-x_1)S_{12}(q),
\eeq
\beq
\label{1.10}
S_{cc}(q)=x_2^2S_{11}(q)+x_1^2S_{22}(q)-2x_1
x_2S_{12}(q)
\eeq
are sometimes required.

After replacement of
$\widehat{h}_{ij}(q)=\sqrt{\rho_i\rho_j}\widetilde{h}_{ij}(q)$ in
Eq.\ \eqref{C(q)}, one easily gets $\widetilde{c}_{ij}(q)$.
Subsequent inverse Fourier transformation yields $c_{ij}(r)$. The
result gives $c_{ij}(r)$ for $r>\sigma_{ij}$ as the superposition of
$N$ Yukawas \cite{YSH00}, namely
\begin{equation}
 c_{ij}(r)= \sum_{\ell=1}^N K_{ij}^{(\ell)}
 \frac{\ee^{-\kappa_\ell r}}{r} ,
\label{cij(r)}
\end{equation}
where $q=\pm \I \kappa_\ell$ with $\ell=1,\ldots,N$ are the zeros of
$\det\left[{\sf I}+\widehat{\sf h}(q)\right]$ and the
 amplitudes $K_{ij}^{(\ell)}$ are obtained by applying the residue theorem
 as
\begin{equation}
 K_{ij}^{(\ell)}=\frac{\I \kappa_\ell}{2\pi}
 \lim_{q\rightarrow \I \kappa_\ell} \widetilde{c}_{ij}(q)(q-\I \kappa_\ell).
\label{kij}
\end{equation}

The indirect correlation functions $\gamma_{ij}(r) \equiv h_{ij}(r)
- c_{ij}(r)$ readily follow from the previous results for the RDF
and direct correlation functions. Finally, in this case the bridge
functions $B_{ij}(r)$ for $r>\sigma_{ij}$ are linked to $g_{ij}(r)$
and $c_{ij}(r)$ through
\begin{equation}
 B_{ij}(r)= \ln g_{ij}(r)-\gamma_{ij}(r)
\end{equation}
 and so once more we have a full set of analytical results for the structural properties of a
 multicomponent fluid mixture of HS once the contact
 values $g_{ij}(\sigma_{ij})$ are specified.

\section{Other Related Systems\label{sec3bis}}
The philosophy behind the RFA method to derive the structural
properties of three-dimensional HS systems can be adapted to deal
with other related systems. The main common features of the RFA can
be summarized as follows. First, one chooses to represent the RDF in
Laplace space. Next, using as a guide the low-density form of the
Laplace transform, an auxiliary function is defined which is
approximated by a rational  or a rational-like form. Finally, the
coefficients are determined by  imposing some basic consistency
conditions. In this section we consider the cases of
sticky-hard-sphere,
 square-well, and hard-disk fluids. In the two former cases the RFA
 program is followed quite literally, while in the latter case it is
done more indirectly through the RFA method as applied to hard rods
($d=1$) and hard spheres ($d=3$).

\subsection{Sticky Hard Spheres\label{sec3bis.1}}

The  sticky-hard-sphere (SHS) fluid model has received a lot of
attention since it was first introduced by Baxter in 1968 \cite{B68}
and later extended to multicomponent mixtures by Perram and Smith
\cite{PS75} and, independently, by Barboy \cite{BT79}. In this
model, the molecular interaction may be defined via square-well (SW)
potentials of infinite depth and vanishing width, thus embodying the
two essential characteristics of real molecular interactions, namely
a harsh repulsion and an attractive part. In spite of their known
shortcomings \cite{BJG97}, an important feature of SHS systems is
that they allow for an exact solution of the OZ equation in the PY
approximation \cite{B68,PS75}. Furthermore, they are thought to be
appropriate for describing structural properties of colloidal
systems, micelles, and microemulsions, as well as some aspects of
gas-liquid equilibrium, ionic fluids and mixtures, solvent mediated
forces, adsorption phenomena, polydisperse systems, and fluids
containing chainlike molecules
\cite{BB74,varios_SHS,SG87,MF04a,MF04b}.

Let us consider an $N$-component mixture of spherical particles
interacting according to the SW potential
\begin{equation}
\phi _{ij}(r)=\left\{
\begin{array}{ll}
\infty , & r<\sigma _{ij}, \\
-\epsilon _{ij}, & \sigma _{ij}<r<R_{ij}, \\
0, & r>R_{ij}.
\label{SW}
\end{array}
\right.
\end{equation}
As in the case of additive HS,  $\sigma _{ij}=(\sigma _{i}+\sigma
_{j})/2$ is the distance between the centers of a sphere of species
$i$ and a sphere of species $j$ at contact. In addition, $ \epsilon
_{ij}$ is the well depth and $R_{ij}-\sigma _{ij}$ indicates the
well width. We now take the SHS limit \cite{B68}, namely
\begin{equation}
R_{ij}\to \sigma _{ij},\quad \epsilon _{ij}\to \infty ,\quad \tau
_{ij}\equiv \frac{1}{12 }\frac{\sigma _{ij}}{R_{ij}-\sigma
_{ij}}\ee^{-\epsilon _{ij}/k_{B}T}=\text{ finite},
\label{169kk}
\end{equation}
where the $\tau _{ij}$ are monotonically increasing functions of the
temperature $T$ and their inverses measure the degree of
``adhesiveness'' of the interacting spheres $i$ and $j$. Even
without strictly taking the mathematical limits \eqref{169kk},
short-range SW fluids can be well described  in practice by the SHS
model \cite{MYS06}.

The virial EOS for the SHS mixture is given by
\begin{eqnarray}
Z &=&1+\frac{1}{6}\rho \sum_{i,j=1}^N x_{i}x_{j}\int \D{\bf
r}\,ry_{ij}(r)\frac{\D}{\D r}
\ee^{-\phi _{ij}(r)/k_{B}T}  \nonumber \\
&=&1+\frac{2\pi }{3}\rho \sum_{i,j=1}^N x_{i}x_{j}\sigma
_{ij}^{3}y_{ij}(\sigma _{ij})\left[ 1-\frac{1}{12\tau _{ij}}\left(
3+\frac{ y_{ij}'(\sigma_{ij})}{y_{ij}(\sigma _{ij})}\right) \right]
,
\label{vir}
\end{eqnarray}
where $y_{ij}(r)\equiv g_{ij}(r)\ee^{\phi _{ij}(r)/k_{B}T}$ is the
cavity function and $y_{ij}'(r) = \D y_{ij}(r)/\D r$. Since
$y_{ij}(r)$ must be continuous, it follows that
\begin{equation}
g_{ij}(r)=y_{ij}(r)\left[ \Theta (r-\sigma _{ij})+\frac{\sigma
_{ij}}{12\tau _{ij}}\delta(r-\sigma _{ij})\right] .
\end{equation}
The case of a  HS system is recovered by taking the limit of
vanishing adhesiveness $\tau_{ij}^{-1}\to 0$, in which case Eq.\
\eqref{vir} reduces to the three-dimensional version of Eq.\
\eqref{1}. On the other hand, the compressibility EOS, Eq.\
\eqref{1.1bb}, is valid for any interaction potential, including
SHS.

As in the case of HS, it is convenient to  define the Laplace
transform \eqref{3.1}. The condition $y_{ij}(\sigma
_{ij})=\text{finite}$ translates into the following large $s$
behavior of $G_{ij}(s)$:
\begin{equation}
\ee^{\sigma _{ij}s}G_{ij}(s)=\sigma _{ij}^{2}y_{ij}(\sigma
_{ij})\left( \frac{1 }{12\tau _{ij}}+\sigma _{ij}^{-1}s^{-1}\right)
+{\cal O}(s^{-2}),
\label{shs10}
\end{equation}
which differs from \eqref{3.3}: while  $\ee^{\sigma
_{ij}s}G_{ij}(s)\sim s^{-1}$ for HS, $\ee^{\sigma
_{ij}s}G_{ij}(s)\sim s^{0}$ for SHS. However,  the small $s$
behavior is still given by Eq.\ \eqref{3.4}, as a consequence of the
condition $\chi ^{-1}= \text{finite}$.

The RFA proposal for SHS mixtures \cite{SYH98} keeps the form
\eqref{3.6}, except that now
\begin{equation}
L_{ij}(s)=L_{ij}^{(0)}+L_{ij}^{(1)}s+L_{ij}^{(2)}s^{2}+L_{ij}^{(3)}s^{3},
\label{shs14}
\end{equation}
\begin{equation}
A_{ij}(s)=\rho _{i}\left[ \varphi _{2}(\sigma _{i}s)\sigma
_{i}^{3}L_{ij}^{(0)}+\varphi _{1}(\sigma _{i}s)\sigma
_{i}^{2}L_{ij}^{(1)}+\varphi _{0}(\sigma _{i}s)\sigma
_{i}L_{ij}^{(2)}-\ee^{-\sigma _{i}s}L_{ij}^{(3)}\right] ,
\label{shs15}
\end{equation}
instead of Eqs.\ \eqref{3.7} and \eqref{3.8}. By construction, Eqs.\
(\ref{3.6}), \eqref{shs14}, and \eqref{shs15} comply with the
requirement $\lim_{s\rightarrow \infty }\ee^{\sigma
_{ij}s}G_{ij}(s)=\text{finite}$. Further, in view of Eq.\
(\ref{3.4}), the coefficients of $s^{0}$ and $s$ in the power series
expansion of $ s^{2}G_{ij}(s)$ must be 1 and 0, respectively. This
yields $2N^{2}$ conditions that allow us to express ${\sf L}^{(0)}$
and ${\sf L}^{(1)}$ in terms of ${\sf L}^{(2)}$, ${\sf L}^{(3)}$,
and $\alpha $ as \cite{SYH98}
\begin{equation}
L_{ij}^{(0)}=\lambdak +\lambdakk\sigma _{j}+2\lambdakk\alpha
-\lambdak \sum_{k=1}^{N}\rho _{k}\left( \sigma
_{k}L_{kj}^{(2)}-L_{kj}^{(3)}\right) -\lambdakk\sum_{k=1}^{N}\rho
_{k}\sigma _{k}L_{kj}^{(3)},
\label{shs20}
\end{equation}
\beqa
L_{ij}^{(1)}&=&\lambdak \sigma _{ij}+\frac{1}{2}\lambdakk\sigma
_{i}\sigma _{j}+(\lambdak +\lambdakk\sigma _{i})\alpha -\frac{1}{2}
\lambdak \sigma _{i}\sum_{k=1}^{N}\rho _{k}\left( \sigma
_{k}L_{kj}^{(2)}-L_{kj}^{(3)}\right)\nn && -\frac{1}{2}\left(
\lambdak +\lambdakk\sigma _{i}\right) \sum_{k=1}^{N}\rho _{k}\sigma
_{k}L_{kj}^{(3)},
\label{shs21}
\eeqa
where $\lambdak$ and $\lambdakk$ are defined below Eq.\
\eqref{3.14}.
 We have the freedom to choose ${\sf
L}^{(3)}$ and $\alpha $, but ${\sf L} ^{(2)}$ is constrained by the
condition (\ref{shs10}), {{ i.e.}}, the ratio between the first and
second  terms in the expansion of $\ee^{\sigma _{ij}s}G_{ij}(s)$ for
large $s$ must be exactly equal to $\sigma _{ij}/12\tau _{ij}$.

\subsubsection{First-Order Approximation (PY Solution)}

The simplest approximation consists of making $\alpha =0$. In view
of the condition $\ee^{\sigma _{ij}s}G_{ij}(s)\sim s^{0}$ for large
$s$, this implies $L_{ij}^{(3)}=0$. In that case, the large $s$
behavior that follows from Eq.\ (\ref{3.6}) is
\begin{equation}
2\pi \ee^{\sigma _{ij}s}G_{ij}(s)=L_{ij}^{(2)}+\left[
L_{ij}^{(1)}+\left( {\sf L}^{(2)}\cdot {\sf D}\right) _{ij}\right]
s^{-1}+{\cal O}(s^{-2}),
\label{shs22}
\end{equation}
where
\begin{equation}
D_{ij}\equiv \rho _{i}\left( \frac{1}{2}\sigma
_{i}^{2}L_{ij}^{(0)}-\sigma _{i}L_{ij}^{(1)}+L_{ij}^{(2)}\right) .
\label{shs23}
\end{equation}
Comparison with Eq.\ (\ref{shs10}) yields
\begin{equation}
y_{ij}(\sigma _{ij})=\frac{6\tau _{ij}}{\pi \sigma
_{ij}^{2}}L_{ij}^{(2)},
\label{shs24}
\end{equation}
\begin{equation}
\frac{12\tau _{ij}L_{ij}^{(2)}}{\sigma _{ij}}=L_{ij}^{(1)}+
\sum_{k=1}^{N}L_{ik}^{(2)}D_{kj}.
\label{shs25}
\end{equation}
Taking into account Eqs.\ (\ref{shs20}) and (\ref{shs21}) (with
$L_{ij}^{(2)}=L_{ji}^{(2)}$ and of course also with $\alpha =0$ and
${\sf L} ^{(3)}=0$), Eq.\ (\ref{shs25}) becomes a closed equation
for ${\sf L} ^{(2)}$:
\begin{equation}
\frac{12\tau _{ij}L_{ij}^{(2)}}{\sigma _{ij}}=\lambdak \sigma
_{ij}+\frac{1}{2 }\lambdakk\sigma _{i}\sigma
_{j}-\frac{1}{2}\lambdak \sum_{k=1}^{N}\rho _{k}\sigma _{k}\left(
L_{ki}^{(2)}\sigma _{j}+L_{kj}^{(2)}\sigma _{i}\right)
+\sum_{k=1}^{N}\rho _{k}L_{ki}^{(2)}L_{kj}^{(2)}.
\label{shs26}
\end{equation}
 The physical root $\mathsf{L}^{(2)}$ of
Eq.\ \eqref{shs26} is the one vanishing in the HS limit
$\tau_{ij}\to\infty$. Once known, Eq.\ (\ref{shs24}) gives the
contact values.

This first-order approximation obtained from the RFA method turns
out to coincide with the exact solution of the PY theory for SHS
\cite{PS75}.

\subsubsection{Second-Order Approximation}

As in the case of HS mixtures, a more flexible proposal is obtained
by keeping $\alpha $ (and, consequently, $L_{ij}^{(3)}$) different
from zero. In that case, instead of Eq.\ \eqref{shs22}, one has
\begin{equation}
2\pi \ee^{\sigma _{ij}s}G_{ij}(s)=\frac{L_{ij}^{(3)}}{\alpha }\left[
1+\left( \frac{L_{ij}^{(2)}}{L_{ij}^{(3)}}-\frac{1}{\alpha }\right)
s^{-1}\right] + {\cal O}(s^{-2}).
\label{shs27}
\end{equation}
This implies
\begin{equation}
L_{ij}^{(3)} =\frac{\pi \sigma _{ij}^{2}}{6\tau _{ij}}\alpha
y_{ij}(\sigma _{ij}),
\label{shs28}
\end{equation}
\begin{equation}
\frac{12\tau _{ij}L_{ij}^{(3)}}{\sigma
_{ij}}=L_{ij}^{(2)}-\frac{L_{ij}^{(3)} }{\alpha }.
\label{shs29}
\end{equation}
If we fix $y_{ij}(\sigma _{ij})$, Eqs.\ (\ref{shs20}),
(\ref{shs21}), (\ref{shs28}), and (\ref{shs29}) allow one to express
${\sf L}^{(0)}$, ${\sf L}^{(1)}$, ${\sf L }^{(2)}$, and ${\sf
L}^{(3)}$ as {\em linear\/} functions of $\alpha $. Thus, only the
scalar parameter $\alpha $ remains to be fixed, analogously to what
happens in the HS case. As done in the latter case, one possibility
is to choose $\alpha$ in order to reproduce  the isothermal
compressibility $\chi $ given by Eq.\ \eqref{3.18}. To do so, one
needs to find the coefficients $H_{ij}^{(1)}$ appearing in Eq.\
(\ref{3.4}). The result is \cite{SYH98}
\begin{equation}
{\sf H}^{(0)}={\sf C}^{(0)}\cdot \left( \openone-{\sf
A}^{(0)}\right) ^{-1},
\label{shsA1}
\end{equation}
\begin{equation}
{\sf H}^{(1)}={\sf C}^{(1)}\cdot \left( \openone-{\sf
A}^{(0)}\right) ^{-1},
\label{shsA3}
\end{equation}
where
\begin{equation}
C_{ij}^{(0)}=\frac{1}{2\pi }L_{ij}^{(2)}+\sum_{k=1}^{N}A_{kj}^{(2)}-
\sum_{k=1}^{N}\sigma _{ik}\left( \alpha \delta
_{kj}-A_{kj}^{(1)}\right) -\sum_{k=1}^{N}\frac{1}{2}\sigma
_{ik}^{2}\left( \delta _{kj}-A_{kj}^{(0)}\right) ,
\label{shsA2}
\end{equation}
\begin{eqnarray}
\label{shsA4}
C_{ij}^{(1)} &=&\frac{1}{2\pi
}L_{ij}^{(3)}+\sum_{k=1}^{N}A_{kj}^{(3)}+ \sum_{k=1}^{N}\sigma
_{ik}A_{kj}^{(2)}-\sum_{k=1}^{N}\left( \frac{1}{2} \sigma
_{ik}^{2}+H_{ik}^{(0)}\right) \left( \alpha \delta
_{kj}-A_{kj}^{(1)}\right)  \nonumber   \\
&&-\sum_{k=1}^{N}\left( \frac{1}{6}\sigma _{ik}^{3}+\sigma
_{ik}H_{ik}^{(0)}\right) \left( \delta _{kj}-A_{kj}^{(0)}\right) ,
\end{eqnarray}
\begin{equation}
A_{ij}^{\n}=(-1)^{n}\rho _{i}\left[ \frac{\sigma _{i}^{n+3}}{(n+3)!}
L_{ij}^{(0)}-\frac{\sigma
_{i}^{n+2}}{(n+2)!}L_{ij}^{(1)}+\frac{\sigma
_{i}^{n+1}}{(n+1)!}L_{ij}^{(2)}-\frac{\sigma _{i}^{n}}{n!}
L_{ij}^{(3)}\right] .
\label{shs17}
\end{equation}
Equation (\ref{shsA3}) gives ${\sf H}^{(1)}$ in terms of $\alpha $:
$ H_{ij}^{(1)}=P_{ij}(\alpha )/[Q(\alpha )]^{2}$, where
$P_{ij}(\alpha )$ denotes a polynomial in $\alpha $ of degree $2N$
and $Q(\alpha )$ denotes a polynomial of degree $N$.  It turns out
then that, seen as a function of $\alpha $, $\chi $ is the ratio of
two polynomials of degree $2N$. Given a value of $\chi $, one may
solve for $\alpha $. The physical solution, which has to fulfill the
requirement that $G_{ij}(s)$ is positive definite for positive real
$s$, corresponds to the smallest positive real root.

Once $\alpha $ is known, the scheme is complete: Eq.\ (\ref{shs28})
gives ${\sf L}^\three$, then ${\sf L}^\two$ is obtained from Eq.\
(\ref{shs29}), and finally ${\sf L}^\one$ and ${\sf L}^\zero$ are
given by Eqs.\ (\ref{shs20}) and  (\ref{shs21}), respectively.
Explicit knowledge of $ G_{ij}(s)$ through Eqs.\ (\ref{3.6}),
\eqref{shs14}, and \eqref{shs15} allows one to determine the Fourier
transform $\widetilde{h}_{ij}(q)$ and the structure factor
$S_{ij}(q)$ through Eqs.\  \eqref{1.7} and \eqref{1.6},
respectively. Finally, inverse Laplace transformation of $G_{ij}(s)$
yields $g_{ij}(r)$ \cite{notebook}.

\subsubsection{Single Component SHS Fluids}
The special case of single component SHS fluids \cite{YS93b,YS93c}
can be obtained from the multicomponent one by taking
$\sigma_{ij}=\sigma$ and $\tau_{ij}=\tau$. Thus, the Laplace
transform of $rg(r)$ in the RFA is
\beq
\label{shs2.10}
G(s)=\frac{\ee^{-s}}{2\pi s^2} \frac{L^\zero+L^\one s+{L^\two}
s^2+L^\three s^3}{ 1+\alpha s-\rho\left[\varphi_2(s) L^\zero
+\varphi_1(s) L^\one +\varphi_0(s) L^\two-\ee^{-s}L^\three\right]},
\eeq
where we have taken $\sigma=1$. Equations \eqref{shs20} and
\eqref{shs21} become
\beq
\label{shs2.11}
L^\zero=2\pi \frac{1+2\eta}{(1-\eta)^2} +\frac{12\eta}{1-\eta}\left(
\frac{\pi
\alpha}{1-\eta}-{L^\two}\right)+\frac{12\eta}{(1-\eta)^2}(1-4\eta)L^\three,
\eeq
\beq
\label{shs2.12}
L^\one=2\pi \frac{1+\frac{1}{2}\eta}{(1-\eta)^2}
+\frac{2}{1-\eta}\left(
\pi\frac{1+2\eta}{1-\eta}\alpha-3\eta{L^\two}\right)-\frac{18\eta^2}{(1-\eta)^2}L^\three.
\eeq

The choice $\alpha=L^\three=0$ makes Eq.\ \eqref{shs2.10} coincide
with the exact solution to the PY approximation for SHS \cite{B68},
where $L^\two$ is the physical root ({i.e.}, the one vanishing in
the limit $\tau\to\infty$) of the quadratic equation [see Eq.\
\eqref{shs26}]
\beq
12\tau
L^\two=2\pi\frac{1+2\eta}{(1-\eta)^2}-\frac{12\eta}{1-\eta}L^\two+\frac{6}{\pi}\eta
{L^\two}^2.
\eeq

 \begin{figure}[b]
\centering
    \includegraphics[height=8cm]{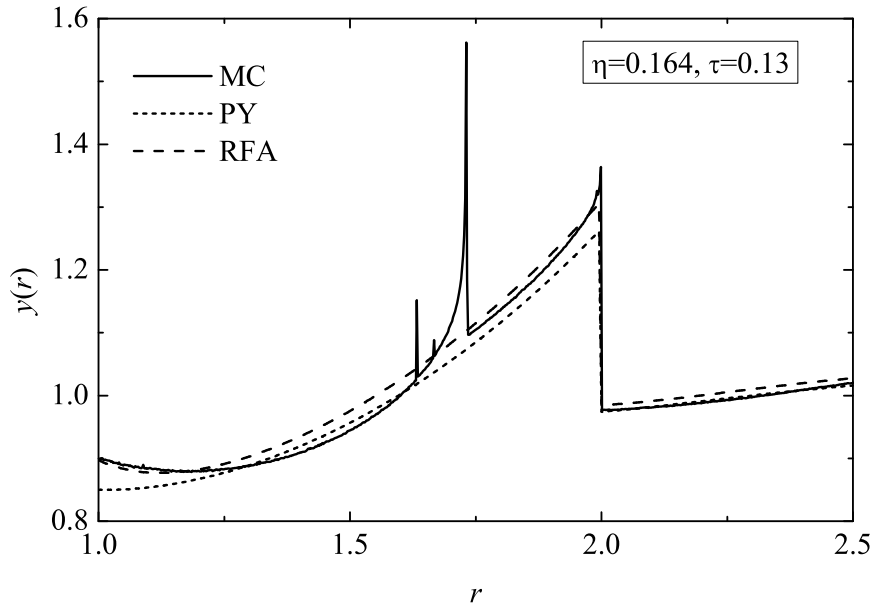}
%
%
\caption{Cavity function of a single component SHS fluid for
$\eta=0.164$ and $\tau=0.13$. The solid line represents simulation
data \protect\cite{MF04a}. The dotted and dashed lines represent the
PY and RFA approaches, respectively.}
\label{SHS}       
\end{figure}

We can go beyond the PY approximation by prescribing a contact value
$y(1)$, so that, according to Eqs.\ \eqref{shs28} and \eqref{shs29},
\begin{equation}
L^\three=\frac{\pi}{6}\frac{\alpha}{\tau} y(1),
\label{monoshs28}
\end{equation}
\begin{equation}
L^\two=\left(12\tau+\frac{1}{\alpha}\right)L^\three.
\label{monoshs29}
\end{equation}
By prescribing the isothermal compressibility $\chi$, the parameter
$\alpha$ can be obtained as the physical solution (namely, the one
remaining finite in the limit $\tau\to\infty$) of a quadratic
equation \cite{YS93c}. Thus, given an EOS for the SHS fluid, one can
get the thermodynamically consistent values of $y(1)$ and $\chi$ and
 determine from them all the coefficients appearing in Eq.\ \eqref{shs2.10}.

 Figure \ref{SHS} shows the cavity function for $\eta=0.164$ and
$\tau=0.13$ as obtained from Monte Carlo simulations \cite{MF04a}
and as predicted by the PY and RFA theories, the latter making use
of the EOS recently proposed by Miller and Frenkel \cite{MF04b}. It
can be observed that the RFA is not only more accurate than the PY
approximation near $r=1$ but also near $r=2$. On the other hand,
none of these two approximations account for the singularities
(delta-peaks and/or discontinuities)  of $y(r)$ at
$r=\sqrt{{8}/{3}}, {5}/{3}, \sqrt{3}, 2,\ldots$ \cite{SG87,MF04a}.

\subsection{Single Component Square-Well Fluids\label{sec3bis.2}}
Now we consider again the SW interaction potential \eqref{SW} but
for a single fluid, {i.e.}, $\sigma_{ij}=\sigma$,
$\epsilon_{ij}=\epsilon$, $R_{ij}=R$. Since no exact solution of the
PY theory for the SW potential is known, the application of the RFA
method is more challenging in this case than for HS and SHS fluids.

As in the cases of HS and SHS, the key quantity is  the Laplace
transform of $rg(r)$ defined by Eq.\ \eqref{2.1}. It is again
convenient to introduce the auxiliary function $\Psi(s)$ through
Eq.\ \eqref{2.2}. As before, the conditions $g(r)=\text{finite}$ and
$\chi=\text{finite}$ imply Eqs.\ \eqref{2.3} and \eqref{2.4},
respectively. However, the important difference between  HS and SHS
fluids is that in the latter case $G(s)$ must reflect the fact that
$g(r)$ is discontinuous at $r=R$ as a consequence of the
discontinuity of the potential $\phi(r)$ and the continuity of the
cavity function $y(r)$. This implies that $G(s)$, and hence
$\Psi(s)$, must contain the exponential term $\ee^{-(R-\sigma)s}$.
This manifests itself in the low-density limit, where the condition
$\lim_{\rho\to 0}y(r)=1$ yields
\beq
\lim_{\rho\to
0}\Psi(s)=\frac{1}{2\pi}\frac{s^3}{\ee^{1/T^*}(1+s)-\ee^{-(R-1)s}
(\ee^{1/T^*}-1)(1+R s )},
\label{SW1}
\eeq
where  $ T^*\equiv k_BT/\epsilon$ and we have taken $\sigma=1$.

In the spirit of the RFA method, the simplest form that complies
with Eq.\ \eqref{2.3} and is consistent with Eq.\ \eqref{SW1} is
\cite{YS94}
\begin{equation}
\Psi (s)=\frac{1}{2\pi}\frac{-12\eta+\SE_1 s+\SE_2 s^2+\SE_3
s^3}{1+\aQ+\aKQ_1 s-\ee^{-(R-1)s}\left(\aQ+\aKQ_2 s\right)},
\label{eq:F(t)}
\end{equation}
where the coefficients $\aQ$, $\aKQ_1$, $\aKQ_2$, $\SE_1$, $\SE_2$,
and $\SE_3$ are functions of $\eta$, $T^*$, and $R$. The condition
\eqref{2.4} allows one to express the parameters $\aKQ_1$, $\SE_1$,
$\SE_2$, and $\SE_3$ as linear functions of $\aQ$ and $\aKQ_2$
\cite{YS94,AS01}:
\beqa
\label{c5}
\aKQ_1&=&\frac{1}{1+2\eta}\left[1+\frac{\eta}{2}+2\eta(R^3-1)
\aKQ_2-\frac{\eta}{2}(R-1)^2(R^2+2R+3) {\aQ}\right]\nn &&
 +\aKQ_2-(R-1)\aQ,
\eeqa
\beq
\label{c6}
\SE_1=\frac{6\eta^2}{1+2\eta}\left[{3}-4(R^3-1) \aKQ_2+
(R-1)^2(R^2+2R+3) {\aQ}\right],
\eeq
\beqa
\label{c7}
\SE_2&=&\frac{6\eta}{1+2\eta}\left\{1-\eta-2(R-1)\left[1-2\eta
R(R+1)\right]\aKQ_2 \right.\nn
&&\left.+(R-1)^2\left[(1-\eta(R+1)^2\right]{\aQ}\right\},
\eeqa
\beqa
\label{c7bis}
\SE_3&=&\frac{1}{1+2\eta}\left\{(1-\eta)^2+6\eta(R-1)
\left(R+1-2\eta R^2\right)\aKQ_2\right.\nn &&\left. -\eta(
R-1)^2[4+2R-\eta(3R^2+2 R+1)]{\aQ}\right\}.
\eeqa
{}From Eq.\ (\ref{2.3}), we have
\beq
g(1^+)=\frac{\aKQ_1}{ \SE_3}.
\label{SWB0}
\eeq
The complete RDF is given by Eq.\ \eqref{g(r)}, where now Eq.\
\eqref{eq:F(t)} must be used in Eq.\ \eqref{varphi}. In particular,
$\psi_1(r)$ and $\psi_2(r)$ are
\beq
\psi_1(r)=\psi_{10}(r)\Theta(r)+\psi_{11}(r+1- R)\Theta(r+1- R),
\label{SWB1}
\eeq
\beq
\psi_2(r)=\psi_{20}(r)\Theta(r)+\psi_{21}(r+1- R)\Theta(r+1-
R)+\psi_{22}(r+2-2 R)\Theta(r+2-2 R),
\label{SWB4}
\eeq
where
\beq
\psi_{1k}(r)={2\pi}\sum_{i=1}^3 \frac{\aC_{1k}(s_i)}{\SE'(s_i)}s_i
\ee^{s_i x},
\label{SWB2}
\eeq
\beq
\psi_{2k}(r)=-4\pi^2\sum_{i=1}^3 \left[r
\aC_{2k}(s_i)+\aC_{2k}'(s_i)-\aC_{2k}(s_i)\frac{\SE''(s_i)}{\SE'(s_i)}\right]\frac{\ee^{s_i
r}}{[\SE'(s_i)]^2}.
\label{SWB5}
\eeq
Here, $s_i$ are the three distinct roots of $\SE(s)\equiv
-12\eta+\SE_1 s+\SE_2 s^2+\SE_3s^3$  and
\beq
\aC_{10}(s)\equiv 1+\aQ+\aKQ_1 s,\quad \aC_{11}(s)\equiv
-(\aQ+\aKQ_2 s).
\label{SWB3}
\eeq
\beq
\aC_{20}(s)\equiv s[\aC_{10}(s)]^2,\quad \aC_{21}(s)\equiv 2s
\aC_{10}(s)\aC_{11}(s), \quad \aC_{22}(s)\equiv s[\aC_{11}(s)]^2.
\label{SWB6}
\eeq

\begin{figure}
\centering
    \includegraphics[width=0.95\columnwidth]{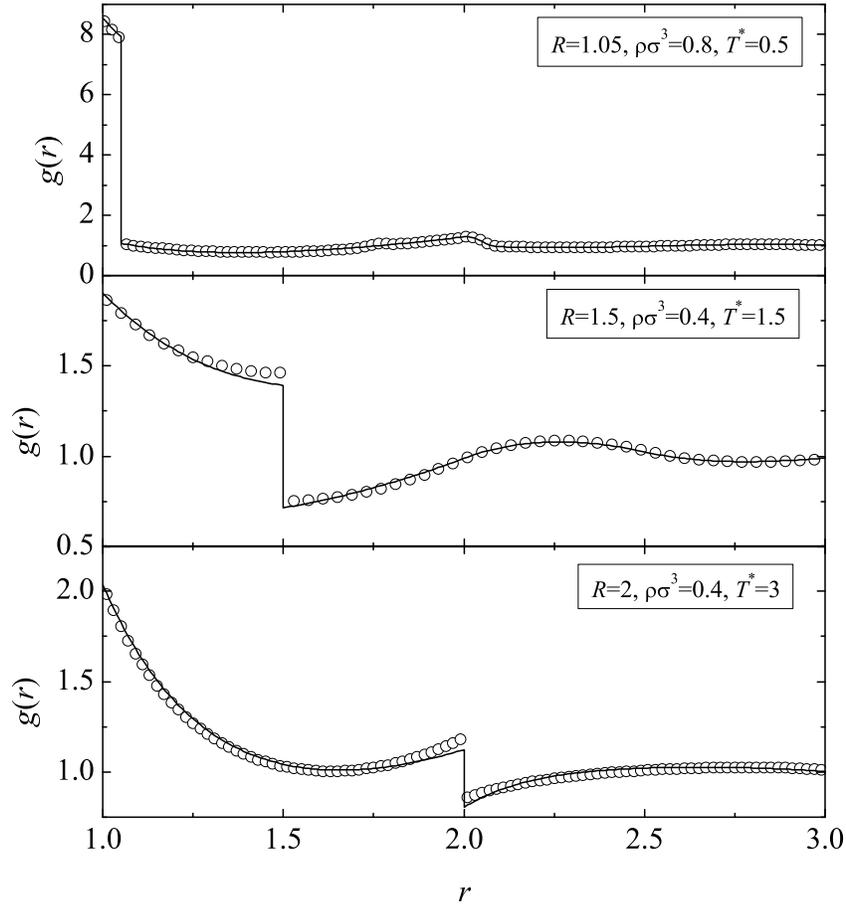}
%
%
\caption{Radial distribution function of a single component SW fluid
for $R=1.05$, $\rho \sigma^3=0.8$, and $T^*=0.5$ (top panel),  for
$R=1.5$, $\rho \sigma^3=0.4$, and $T^*=1.5$ (middle panel), and for
$R=2.0$, $\rho \sigma^3=0.4$, and $T^*=3.0$ (bottom panel). The
circles represent simulation data \protect\cite{LSYS05} and  the
solid lines refer to the results obtained from the RFA method.}
\label{figSW}       
\end{figure}

To close the proposal, we need to determine the parameters $\aQ$ and
$\aKQ_2$ by imposing two new conditions. An obvious condition is the
continuity of the cavity function at $r=R$, what implies
\beq
g( R^+)=e^{1/T^*}g( R^-).
\label{SWB8}
\eeq
 This yields
\beq
\left(1-e^{-1/T^*}\right)\psi_{10}(
R-1)=-\psi_{11}(0)=2\pi\frac{\aKQ_2}{ \SE_3}.
\label{SWB7}
\eeq
As an extra condition, we could enforce the continuity of the first
derivative $y'(r)$ at $r=R$ \cite{A00}. However, this complicates
the problem too much without any relevant gain in accuracy. In
principle, it might be possible to impose consistency with a given
EOS,  via either the virial route, the compressibility route, or the
energy route. But this is not practical since no simple EOS for SW
fluids is at our disposal for wide values of density, temperature,
and range. As a compromise between simplicity and accuracy, we fix
the parameter $\aQ$ at its exact zero-density limit value, namely
$\aQ=e^{1/T^*}-1$ \cite{YS94}. Therefore, Eq.\ \eqref{SWB7} becomes
a {transcendental} equation for $\aKQ_2$ that needs to be solved
numerically. For narrow SW potentials, however, it is possible to
replace the exact condition (\ref{SWB8})  by a simpler one allowing
$\aKQ_2$ to be obtained analytically \cite{AS01}, which is
especially useful {for determining} the thermodynamic properties
\cite{AS01,LSAS03}.

It can be proven that the RFA proposal \eqref{eq:F(t)} reduces to
the exact solutions of the PY equation \cite{W63,B68} in the HS
limit, i.e., $\epsilon\to 0$ or $R\to 1$, and in the SHS limit,
i.e., $\epsilon\to\infty$ and $R\to 1$ with
$(R-1)\ee^{1/T^*}=\text{finite}$ \cite{YS94,AS01}.

Comparison with computer simulations \cite{YS94,AS01,LSAS03,LSYS05}
shows that the RFA for SW fluids is rather accurate at any fluid
density if the potential well is sufficiently narrow (say $R\leq
1.2$), as well as for any width if the density is small enough (say
$\rho\sigma^3\leq 0.4$). However, as the width and/or the density
increase, the RFA predictions  worsen, especially at low
temperatures. As an illustration, Fig.\ \ref{figSW} compares the RDF
provided by the RFA with Monte Carlo data \cite{LSYS05} for three
representative cases.

\subsection{Hard Disks}\label{sec3bis.3}}
As is well known, the PY theory is exactly solvable for HS fluids
with an odd number of dimensions \cite{FI81,L84,RHS04}. In
particular, in the case of hard rods ($d=1$), the PY theory provides
the exact RDF $g(r)$ or, equivalently, the exact cavity function
$y(r)$ outside the hard core ({i.e.}, for $r>\sigma$). However, it
does not reproduce the exact $y(r)$ in the overlapping region
({i.e.}, for $r<\sigma$) \cite{MS06}. The full exact one-dimensional
cavity function is \cite{MS06}
\beq
y_\hr(r|\eta)=\frac{\ee^{-(r-1)\eta/(1-\eta)}}{1-\eta}+
\sum_{n=2}^\infty
\frac{\eta^{n-1}\ee^{-(r-n)\eta/(1-\eta)}}{(1-\eta)^n
(n-1)!}(r-n)^{n-1}\Theta(r-n),
\label{2.46}
\eeq
where the subscript $\hr$ stands for hard rods and, as usual,
$\sigma=1$ has been taken. Consequently, one has
\beq
g_\hr(1^+|\eta)=\frac{1}{1-\eta},\quad \int_0^\infty dr\, r
h_\hr(r|\eta)\equiv
H_\hr^\zero(\eta)=-\frac{1}{2}+\frac{2}{3}\eta-\frac{1}{4}\eta^2.
\label{HS2}
\eeq

When $d$ is even, the PY equation is not analytically solvable for
the HS interaction. In particular, in the important case of hard
disks ($d=2$), one must resort to numerical solutions of the PY
equation \cite{BH76,CRR76}. Alternatively, a simple heuristic
approach has proven to yield reasonably good results \cite{YS93a}.
Such an approach is based on the na\"{\i}ve assumption that the
structure and spatial correlations of a hard-disk fluid share some
features with those of a hard-rod and a hard-sphere fluid. This
fuzzy idea becomes a more specific one by means of the following
simple model \cite{YS93a}:
\beq
g_\hd(r|\eta)=\nu(\eta)g_\hr(r|\omega_1(\eta)\eta)+[1-\nu(\eta)]g_\hs(r|\omega_3(\eta)\eta).
\label{HS1}
\eeq
Here, the subscript $\hd$ stands for hard disks ($d=2$) and the
subscript $\hs$ stands for hard spheres ($d=3$). The parameter
$\nu(\eta)$ is a density-dependent mixing parameter, while
$\omega_1(\eta)\eta$ and $\omega_3(\eta)\eta$ are the packing
fractions in one and three dimensions, respectively, which are
``equivalent'' to the packing fraction $\eta$ in two dimensions. In
Eq.\ \eqref{HS1}, it is natural to take for $g_\hr(r|\eta)$ the
exact solution, Eq.\ \eqref{2.46}. As for $g_\hr(r|\eta)$, one might
use the RFA recipe described in Section \ref{sec3}. However, in
order to keep the model \eqref{HS1} as simple as possible, it is
sufficient for practical purposes to take the PY solution, Eq.\
\eqref{2.8}. In the latter approximation,
\beq
g_\hs(1^+|\eta)=\frac{1+\eta/2}{(1-\eta)^2},\quad \int_0^\infty \D
r\, r h_\hs(r|\eta)\equiv
H_\hs^\zero(\eta)=-\frac{10-2\eta+\eta^2}{20(1+2\eta)}.
\label{HS3}
\eeq

In order to close the model \eqref{HS1}, we still need to determine
the parameters $\nu(\eta)$, $\omega_1(\eta)$, and $\omega_3(\eta)$.
To that end, we first impose the condition that Eq.\ \eqref{HS1}
must be consistent with a {prescribed} contact value
$g_\hd(1^+|\eta)$ or, equivalently, with a prescribed
compressibility factor $Z_\hd(\eta)=1+2\eta g_\hd(1^+|\eta)$, with
independence of the choice of the mixing parameter $\nu(\eta)$. In
other words,
\beq
g_\hd(1^+|\eta)=g_\hr(1^+|\omega_1(\eta)\eta)=g_\hs(1^+|\omega_3(\eta)\eta).
\label{HS4}
\eeq
Making use of Eqs.\ \eqref{HS2} and \eqref{HS3}, this yields
\beq
\omega_1(\eta)=\frac{g_\hd(1^+|\eta)-1}{\eta g_\hd(1^+|\eta)},\quad
\omega_3(\eta)=\frac{4g_\hd(1^+|\eta)+1-\sqrt{24g_\hd(1^+|\eta)+1}}{4\eta
g_\hd(1^+|\eta)}.
\label{HS5}
\eeq
Once $\omega_1(\eta)$ and $\omega_3(\eta)$ are known, we can
determine $\nu(\eta)$ by imposing that the model \eqref{HS1}
reproduces the isothermal compressibility $\chi_\hd(\eta)$
thermodynamically consistent with the prescribed $Z_\hd(\eta)$ [cf.
Eq.\ \eqref{consistent}]. {}From Eqs.\ \eqref{chi} and \eqref{HS1}
one has
\beq
\chi_\hd(\eta)=1+8\eta\int_0^\infty \D r\,
r\left\{\nu(\eta)h_\hr(r|\omega_1(\eta)\eta)+\left[1-\nu(\eta)\right]
h_\hs(r|\omega_3(\eta)\eta)\right\},
\label{HS7}
\eeq
so that
\beq
\nu(\eta)=\frac{\left[\chi_\hd(\eta)-1\right]/8\eta
-H_\hs^\zero(\omega_3(\eta)\eta)}{H_\hr^\zero(\omega_1(\eta)\eta)-H_\hs^\zero(\omega_3(\eta)\eta)},
\label{HS8}
\eeq
where $H_\hr^\zero(\eta)$ and $H_\hs^\zero(\eta)$ are given by Eqs.\
\eqref{HS2} and \eqref{HS3}, respectively.

Once a sensible EOS for hard disks is chosen [see, for instance,
Table \ref{Tableg_s}], Eqs.\ \eqref{HS5} and \eqref{HS8} provide the
parameters of the model \eqref{HS1}. The results show that the
scaling factor $\omega_1(\eta)$ is a decreasing function, while
$\omega_3(\eta)$ is an increasing function \cite{YS93a}. As for the
mixing parameter $\nu(\eta)$, it is hardly dependent of density and
takes values around $\nu(\eta)\simeq 0.35$--0.40.

Comparison of the interpolation model \eqref{HS1} with computer
simulation results shows a surprisingly good agreement, despite the
crudeness of the model and the absence of empirical fitting
parameters, especially at low and moderate densities \cite{YS93a}.
The discrepancies become important only for distances beyond the
location of the second peak and for densities close to the stability
threshold.

\vspace{1cm}

\section{Perturbation Theory\label{sec4}}

When one wants to deal with realistic intermolecular interactions,
the problem of deriving the thermodynamic and structural properties
of the system becomes rather formidable.  Thus,  perturbation
theories of liquids have been devised since the mid twentieth
century. In the case of single component fluids, the use of an
accurate and well characterized RDF  for the HS fluid in a
perturbation theory opens up the possibility of deriving a closed
theoretical scheme for the determination of the thermodynamic and
structural properties of more realistic models, such as the
Lennard--Jones (LJ) fluid. In this section, we will consider this
model system, which captures the basic physical properties of real
non-polar fluids, to illustrate the procedure.

In the application of the perturbation theory of liquids, the
stepping stone has been the use of the HS RDF obtained from the
solution to the PY equation. Unfortunately, the absence of
thermodynamic consistency present in the PY approximation (as well
as in other integral equation theories) may clearly contaminate the
results derived from its use within a perturbative treatment. In
what follows we will reanalyze the different theoretical schemes for
the thermodynamics of LJ fluids that have been constructed with
perturbation theory, taking as the reference system the HS fluid.
This includes the consideration of the RDF as obtained with the RFA
method, which embodies thermodynamic consistency,  as well as the
proposal of a unifying framework in which all schemes fit in. With
our development, we will be able to present a formulation which
lends itself to relatively easy numerical calculations while
retaining the merits that analytical results provide, namely a
detailed knowledge and control of all the approximations involved.

Let us consider a three-dimensional fluid system defined by a pair
interaction potential $\phi( r) $. The virial and energy EOS express
the compressibility factor $Z$ and the excess part of the Helmholtz
free energy per unit volume $f^{\text{ex}}$, respectively, in terms
of the RDF of the system as
\beq
Z=1-\frac{2}{3}\pi\rho \beta\int_0^\infty \D r\,
\frac{\partial\phi(r)}{\partial r} g(r)r^3,
\label{z1}
\eeq
\begin{equation}
\frac{f^{\text{ex}}}{\NN k_BT}=2\pi \rho \beta \int_{0}^{\infty }\D
r \,\phi(r) g( r) r^{2} ,
\label{z2}
\end{equation}
where $\beta \equiv 1/k_BT$. Let us now assume that $\phi(r)$ is
 split into a known (reference) part $\phi _{0}( r) $
and a perturbation part $\phi _{1}( r)$. The usual perturbative
expansion for the Helmholtz free energy to first order in $\beta $
leads to \cite{M76}
\begin{equation}
\frac{f}{\NN k_BT}=\frac{f_{0}}{\NN k_BT}+2\pi \rho \beta
\int_{0}^{\infty }\D r \,\phi_{1}(r) g_{0}( r)
r^{2}+\mathcal{O}\left( \beta ^{2}\right) ,
\label{AO1}
\end{equation}
where $f_{0}$ and $g_{0}( r) $ are the free energy and the RDF of
the reference system, respectively.

The LJ potential is
\beq
\phi _{\LJ}( r) =4\epsilon \left( r^{-12}-r^{-6}\right),
\eeq
 where $\epsilon $ is the depth of the well
and, for simplicity, we have taken the distance at which the
potential vanishes as the length unit, {i.e.}, $\phi _{\LJ}( r=1)
=0$. For this potential the reference system may be forced to be a
HS system, {i.e.}, one can set
\begin{equation}
\phi _{0}( r) =\phi _{\hs}( r)=\left\{
\begin{array}{ll}
\infty , & r\leq \diam , \\
0, & r>\diam ,
\end{array}
\right.
\end{equation}
where $\diam$ is a conveniently chosen effective HS diameter. In
this case the Helmholtz free energy to this order is approximated by
\begin{equation}
\frac{f_{\LJ}}{\NN k_BT}\approx \frac{f_{\hs}}{\NN k_BT}+2\pi \rho
\beta \int_{\diam }^{\infty }\D r\,\phi _{\LJ}( r) g_\hs(
{r}/{\diam}) r^{2}.  \label{AO1b}
\end{equation}
 Note that Eq.\ (\ref{AO1b}) may be rewritten in terms of the
Laplace transform $G(s)$ of $(r/\diam) g_\hs({r}/{\diam}) $ as
\begin{equation}
\frac{f_{\LJ}}{\NN k_BT}\approx \frac{f_{\hs}}{\NN k_BT}+2\pi \rho
\beta \diam^{3}\int_{0}^{\infty }\D s \,\Phi_{\LJ}(s) G(s) ,
\label{ALT}
\end{equation}
where $\Phi _{\LJ}( s) $ satisfies
\begin{equation}
r \phi_{\LJ}( r) =\diam \int_{0}^{\infty }\D s\,
\ee^{-rs/\diam}\Phi_{\LJ}(s) ,
\label{FLJ}
\end{equation}
so that
\begin{equation}
\Phi _{\LJ}(s) =4\epsilon
\diam^{-2}\left[\frac{(s/\diam)^{10}}{10!}-\frac{(s/\diam)^{4}}{4!}\right]
.  \label{FLJT}
\end{equation}

Irrespective of the value of the diameter $\diam$ of the reference
system, the right hand side of Eq.\ (\ref{ALT}) represents {\it
always} an upper bound for the value of the free energy of the real
system. Therefore, it is natural to determine $\diam$ so as to
provide the least upper bound. This is precisely the variational
scheme of Mansoori and Canfield \cite{MC69,MPC69} and Rasaiah and
Stell  \cite{RS70}, usually referred to as MC/RS, and originally
implemented with the PY theory for $G(s)$, Eq.\ \eqref{2.8}. In our
case, however, we will consider $G(s)$ as given by the RFA method,
Eq.\ (\ref{2.10}). Therefore, at fixed $\rho$ and $\beta$, the
effective diameter $\diam$ in the MC/RS scheme is obtained from the
conditions
\beqa
\frac{\partial }{\partial \diam}\left\{ \int_{0}^{\eta_0}\D \eta\,
\frac{Z_\hs(\eta)-1}{\eta }\right. &+&48\beta\epsilon
\diam^{-2}\int_{0}^{\infty }\D s\, G(s|\eta_0)\nn
&&\left.\times\left[\frac{(s/\diam)^{10}}{10!}-\frac{(s/\diam)^{4}}{4!}\right]\right\}
=0,
\eeqa
\beqa
\frac{\partial^2 }{\partial \diam^2}\left\{ \int_{0}^{\eta_0}\D
\eta\, \frac{Z_\hs(\eta)-1}{\eta }\right. &+&48\beta\epsilon
\diam^{-2}\int_{0}^{\infty }\D s\, G(s|\eta_0)\nn
&&\left.\times\left[\frac{(s/\diam)^{10}}{10!}-\frac{(s/\diam)^{4}}{4!}\right]\right\}
>0.
\eeqa
In these equations, use has been made of the thermodynamic
relationship between the free energy and the compressibility factor,
Eq.\ \eqref{FEN}. Moreover, we have called
$\eta_0\equiv(\pi/6)\rho\diam^3$ and have made explicit with the
notation $G(s|\eta_0)$ the fact that the HS RDF depends on the
packing fraction $\eta_0$.

Even if the reference system is not forced to be a HS fluid, one can
still use Eq.\ (\ref{ALT}) provided an adequate choice for $\diam $
is made such that the expansion involved in the right hand side of
Eq.\ (\ref{AO1}) yields the right hand side of  Eq.\ (\ref{ALT}) to
order $\beta ^{2}$. This is the idea of the Barker and Henderson
 \cite{BH67} first order perturbation scheme (BH$_{1}$), where the
effective HS diameter is
\begin{equation}
\diam=\int_{0}^{\infty }\D r\,\left[ 1-\ee^{-\beta \phi _{\LJ}\left(
r\right) }\right] .
\label{dBH}
\end{equation}

The same ideas may be carried out to higher order in the
perturbation expansion. The inclusion of the second order term in
the expansion yields the so-called macroscopic compressibility
approximation \cite{M76} for the free energy, namely
\begin{eqnarray}
\frac{f_{\LJ}}{\NN k_BT} &=&\frac{f_{0}}{\NN k_BT}+2\pi \rho \beta
\int_{0}^{\infty
}\D r\,\phi _{1}( r) g_{0}( r) r^{2} \nonumber \\
&&-\pi \rho \beta ^{2}\chi _{0}\int_{0}^{\infty }\D r\,\phi
_{1}^{2}( r) g_{0}( r) r^{2}+\mathcal{O}\left( \beta ^{3}\right) ,
\label{9p}
\end{eqnarray}
where $\chi _{0}$ is the (reduced) isothermal compressibility of the
reference system \cite{note}.

To implement a particular perturbation scheme in this approximation
under a unifying framework that eventually leads to easy numerical
evaluation, two further assumptions may prove convenient. First, the
perturbation potential $\phi _{1}( r) \equiv \phi _{\LJ}(r) -\phi
_{0}( r) $  may be split into two parts using some ``molecular
size'' parameter $\xi\geq \diam $ such that
\begin{equation}
\phi _{1}( r) =\left\{
\begin{array}{ll}
\phi _{1a}( r),  &   0\leq r\leq \xi,  \\
\phi _{1b}( r),  &   r>\xi.
\end{array}
\right.
\label{10p}
\end{equation}
Next, a choice for the RDF for the reference system is done in the
form
\begin{equation}
g_{0}\left( r\right) \approx \theta( r) y_\hs( r/\diam) ,
\label{11pp}
\end{equation}
where $y_\hs$ is the cavity (background) correlation function of the
HS system and $\theta( r) $ is a step function defined by
\begin{equation}
\theta ( r) =\left\{
\begin{array}{ll}
\theta _{a}( r),  &   0\leq r\leq \xi,  \\
\theta _{b}( r),   & r>\xi,
\end{array}
\right.
\label{12p}
\end{equation}
in which the functions $\theta _{a}( r) $ and $\theta _b( r) $
depend on the scheme.

With these assumptions the integrals involved in Eq.\ (\ref{9p}) may
be rewritten as
\begin{eqnarray}
I_{n} &\equiv&\int_{0}^{\infty }\D r\,\phi_{1}^n( r) g_{0}( r)
r^{2}\nn &=&\int_{0}^{\diam}\D r\,\phi_{1a}^n( r) \theta _{a}( r)
y_\hs( r/\diam) r^{2} +\int_{\diam}^{\xi }\D r\,\phi_{1a}^n( r)
\theta _{a}( r) g_\hs( r/\diam) r^{2} \nn &&+\int_{\xi }^{\infty }\D
r\,\phi _{1b}^n( r) \theta _{b}( r) g_\hs( r/\diam) r^{2},
\label{13p}
\end{eqnarray}
with $n=1,2$ and  where the fact that $y_\hs( r/\diam)
=g_\hs(r/\diam) $ when $r>\diam$ has been used. Decomposing the last
integral as $\int_\xi^\infty=\int_{\diam}^\infty-\int_{\diam}^\xi$
and applying the same step as in Eq.\ \eqref{ALT}, Eq.\ \eqref{13p}
becomes
\begin{eqnarray}
I_{n} &=&\diam^3\int_{0}^{\infty }\D s\,\Phi_{nb}( s) G(
s)+\int_{0}^{\diam}\D r\,\phi _{1a}^n( r) \theta _{a}( r) y_\hs(
r/\diam) r^{2} \nonumber \\
&&+\int_{\diam}^{\xi }\D r\,\left[ \phi _{1a}^n( r) \theta _{a}( r)
-\phi _{1b}^n( r) \theta _{b}( r) \right] g_\hs( r/\diam) r^{2},
\label{17p}
\end{eqnarray}
where the functions $\Phi_{1b}(s)$ and $\Phi_{2b}(s)$ are defined by
the relation
\begin{equation}
{r}\phi _{1b}^n( r)\theta_b(r) =\diam\int_{0}^{\infty }\D
s\,\ee^{-rs/\diam}\Phi _{nb}( s).
\end{equation}

In the Barker--Henderson  second order perturbation scheme
(BH$_{2}$), one takes
\beq
\theta _{a}( r) =0 ,\quad \theta _{b}( r) =1 ,\quad \xi =\diam,\quad
\phi _{1a}( r) =0,\quad \phi _{1b}( r) =4\epsilon\left( {r^{-12}}-
{r^{-6}}\right),
\eeq
and $\diam $ is computed according to Eq.\ (\ref{dBH}). This choice
ensures  that
\begin{eqnarray}
\frac{f_{\LJ}}{\NN k_BT} &=&\frac{f_{\hs}}{\NN k_BT}+2\pi \rho \beta
\int_{\diam}^{\infty}\D r\,\phi _{1}( r) g_\hs( r/\diam) r^{2}  \nonumber \\
&&-\pi \rho \beta ^{2}\chi _{\hs}\int_{\diam}^{\infty }\D r\, \phi
_{1}^2( r) g_\hs( r/\diam) r^{2}+O\left( \beta ^{3}\right) .
\end{eqnarray}

On the other hand,  if one chooses
\beq
\theta _{a}( r)  =\exp \left[ -\beta \left( \phi _{\LJ}(r) +\epsilon
\right) \right],\quad \theta _{b}( r)  =1 ,\quad \xi =2^{1/6},
\label{WCAF1}
\eeq
\beq
\phi _{1a}( r)  =-\epsilon ,\quad \phi _{1b}( r)  =4\epsilon\left(
{r^{-12}}- {r^{-6}}\right),
\label{WCAF2}
\eeq
the scheme leads to the Weeks--Chandler--Andersen (WCA) theory
\cite{WCA71} if one determines the HS diameter through the condition
$\chi _{0}=\chi _{\hs}$ \cite{VW72}, which in turn implies
\begin{equation}
\int_{0}^{\diam}\D r\, r^{2}\ee^{-\beta \phi _{0}( r)
}y_\hs(r/\diam)=\int_{\diam}^{2^{1/6} }\D r\,r^{2}g_\hs( r/\diam)
\left[ 1-\ee^{-\beta \phi _{0}( r) }\right].
\label{WCAd}
\end{equation}
To close the scheme, the HS cavity function has to be provided in
the range $0\leq r\leq \diam$. Fortunately, relatively simple
expressions for $y_\hs(r/\diam)$ are available in the literature
\cite{HG75,BS85,ZS88}, apart from our own proposal, Eq.\
\eqref{lny}.

Note that $\theta_b(r)$ and $\phi _{1b}( r) $, and thus also
$\Phi_{nb}( s) $,
 are the same functions in the BH$_{2}$ and WCA schemes. It is convenient, in order to have
all the quantities needed to evaluate $f_{\LJ}$ in these schemes, to
provide explicit expressions for $\Phi_{1b}( s)$ and $\Phi_{2b}(
s)$. These are given by [cf.\ Eq.\ (\ref{FLJT})]
\beq
\Phi_{1b}( s) =\Phi _{\LJ}( s) ,  \label{Fi1b}
\eeq
\beq
\Phi _{2b}(s)=16\epsilon ^{2}\diam^{-2}\left[
\frac{(s/\diam)^{22}}{22!}
-2\frac{(s/\diam)^{16}}{16!}+\frac{(s/\diam)^{10}}{10!}\right] .
\label{fi2b}
\eeq

Up to this point, we have embodied the most popular perturbation
schemes within a unified framework that requires as input {\it only}
the EOS of the HS fluid in order to compute the Helmholtz free
energy of the LJ system and leads to relatively easy numerical
computations. It should be clear that a variety of other possible
schemes, requiring the same little input, fit in our unified
framework, which is based on the RFA method for $g_\hs(r/\diam)$ and
$G(s)$. Once $f_{\LJ}$ has been determined, the compressibility
factor of the LJ fluid at a given order of the perturbation
expansion readily follows from Eqs.\ (\ref {AO1}) or (\ref{9p})
through the thermodynamic relation
\begin{equation}
Z_{\LJ}=\rho \left(\frac{\partial }{\partial \rho }
\frac{f_{\LJ}}{\NN k_BT}\right) _{T}.
\label{ZLJ}
\end{equation}

Taking into account that the HS fluid presents a fluid-solid
transition at a freezing packing fraction $\eta_\text{f}\simeq
0.494$ \cite{HV69} and a solid-fluid transition at a melting packing
fraction $\eta _\text{m}\simeq 0.54$ \cite{HV69}, the fluid-solid
and solid-fluid coexistence lines for the LJ system may be computed
from the values $(\rho,T)$ determined from the conditions $({\pi
}/{6})\rho\diam^3(\rho,T)=\eta_\text{f}$  and $({\pi
}/{6})\rho\diam^3(\rho,T)=\eta_\text{m}$, respectively, with the
effective diameter $\diam(\rho,T)$ obtained using any of the
perturbative schemes. Similarly, admitting that there is a glass
transition in the HS fluid at the packing fraction
$\eta_\text{g}\simeq 0.56$ \cite{S94}, one can now determine the
location of the liquid-glass transition line for the LJ fluid in the
$(\rho,T)$ plane from the simple relationship $({\pi
}/{6})\rho\diam^3(\rho,T)=\eta_\text{g}$. With a proper choice for
$Z_\hs$, it has been shown \cite{RH03,RH01,HR04} that the critical
point, the structure, and the phase diagram (including a glass
transition) of the LJ fluid may be adequately  described with this
approach.

\section{Perspectives}

In this chapter we have given a   self-contained account of
 a simple
(mostly analytical) framework for the study of the thermodynamic and
structural properties of hard-core systems. Whenever possible, the
developments have attempted to cater for mixtures with an arbitrary
number of components (including polydisperse systems) and arbitrary
dimensionality. We started considering the contact values of the RDF
because they enter directly into the EOS and are required as input
in the RFA method to compute the structural properties. With the aid
of consistency conditions, we were able to devise various
approximate proposals which, when used in conjunction with a
sensible choice for the contact value of the RDF of the single
component fluid (required in the formulation but otherwise chosen at
will), have been shown to be in reasonably good agreement with
simulation results and lead to accurate EOS both for additive and
non-additive mixtures. Some aspects of the results that follow from
the use of these EOS were illustrated by looking at demixing
problems in these mixtures, including the far from intuitive case of
a binary mixture of non-additive hard spheres in infinite
dimensionality.

After that, restricting ourselves to three-dimensional systems, we
described the RFA method as applied to a single component
hard-sphere fluid and to a multicomponent mixture of HS. Using this
approach, we have been able to obtain explicit analytical results
for the RDF, the  direct  correlation function, the static structure
factor, and the bridge function, in the end requiring as input
\emph{only} the contact value of the RDF of the single component HS
fluid (or equivalently its compressibility factor).  One of the nice
assets of the RFA approach is that it eliminates the thermodynamic
consistency problem which is present in most of the integral
equation formulations for the computation of structural quantities.
Once again, when a sensible choice for the single component EOS is
made, we have shown, through the comparison between the results of
the RFA approach and simulation data for some illustrative cases,
the very good performance of our development. Also, the use of the
RFA approach in connection with some other related systems (sticky
hard spheres, square-well fluids, and hard disks) has been
addressed.

The final part of the chapter concerns the use of HS results for
more realistic intermolecular potentials in the perturbation theory
of liquids. In this instance we have been able to provide a unifying
scheme  in which the most popular perturbation theory formulations
may be expressed and which was devised to allow for easy
computations. We illustrated this for a LJ fluid but it should be
clear that a similar approach might be followed for other fluids and
in fact it has recently been done in connection with the glass
transition of hard-core Yukawa fluids \cite{HR06}.

Finally, it should be clear that there are many facets of the
equilibrium and structural properties of hard-core systems that may
be studied with a similar approach but that up to now have not been
considered. For instance, the generalizations of the RFA approach
for systems such as hard hyperspheres, non-additive hard spheres,
square-well mixtures, penetrable spheres \cite{L01b}, or the Jagla
potential \cite{J99} appear as  interesting challenges. Similarly,
the extension of the perturbation theory scheme to the case of LJ
mixtures seems a worthwhile task. We hope to address some of these
problems in the future and would be very much rewarded if some
others were taken up by researchers who might find these
developments also a valuable tool for their work.

%

%

%
%

%
%

\begin{thebibliography}{100}
\expandafter\ifx\csname
natexlab\endcsname\relax\def\natexlab#1{#1}\fi
\expandafter\ifx\csname bibnamefont\endcsname\relax
  \def\bibnamefont#1{#1}\fi
\expandafter\ifx\csname bibfnamefont\endcsname\relax
  \def\bibfnamefont#1{#1}\fi
\expandafter\ifx\csname citenamefont\endcsname\relax
  \def\citenamefont#1{#1}\fi
\expandafter\ifx\csname url\endcsname\relax
  \def\url#1{\texttt{#1}}\fi
\expandafter\ifx\csname urlprefix\endcsname\relax\def\urlprefix{URL
}\fi \providecommand{\bibinfo}[2]{#2}
\providecommand{\eprint}[2][]{\url{#2}}

\bibitem{BH76}
J. A. Barker and D. Henderson, Rev. Mod. Phys. \textbf{48}, 587
(1976).

\bibitem{M76}
D. A. McQuarrie, {\emph{ Statistical Mechanics}} (Harper \& Row, N.
Y., 1976).

\bibitem{F85}
H. L. Friedman, {\em A Course in Statistical Mechanics} (Prentice
Hall, Englewood Cliffs, 1985).

\bibitem{HM86}
J.-P.~Hansen and I. R. McDonald, \textit{Theory of Simple Liquids},
(Academic Press, London, 1986).

\bibitem{LZ71}
J. L. Lebowitz and D. Zomick, J. Chem. Phys. \textbf{54}, 3335
(1971).

\bibitem{JM87}
\bibinfo{author}{\bibfnamefont{J.~T.} \bibnamefont{Jenkins}} \bibnamefont{and}
  \bibinfo{author}{\bibfnamefont{F.}~\bibnamefont{Mancini}},
  \bibinfo{journal}{J. Appl. Mech.} \textbf{\bibinfo{volume}{54}},
  \bibinfo{pages}{27} (\bibinfo{year}{1987}).

  \bibitem{BS01}
C. Barrio and J. R. Solana, J. Chem. Phys. \textbf{115}, 7123
(2001); \textbf{117}, 2451(E) (2002).

\bibitem{L64}
\bibinfo{author}{\bibfnamefont{J.~L.} \bibnamefont{Lebowitz}},
  \bibinfo{journal}{Phys. Rev. A} \textbf{\bibinfo{volume}{133}},
  \bibinfo{pages}{895} (\bibinfo{year}{1964}).

\bibitem{RFL59}
H. Reiss, H. L. Frisch, and J. L. Lebowitz, J. Chem. Phys.
\textbf{31}, 369 (1959); E. Helfand, H. L. Frisch, and J. L.
Lebowitz, J. Chem. Phys. \textbf{34}, 1037 (1961); J. L. Lebowitz,
E. Helfand, and E. Praestgaard, J. Chem. Phys. \textbf{43}, 774
(1965).

\bibitem{MR75}
 M. J. Mandell  and H. Reiss, J. Stat. Phys.  \textbf{13}, 113 (1975).

\bibitem{R88}
\bibinfo{author}{\bibfnamefont{Y.}~\bibnamefont{Rosenfeld}},
  \bibinfo{journal}{J. Chem. Phys.} \textbf{\bibinfo{volume}{89}},
  \bibinfo{pages}{4272} (\bibinfo{year}{1988}).


\bibitem{HC04}
M. Heying and D. S. Corti, J. Phys. Chem. B \textbf{108}, 19756
(2004).

\bibitem{B70}
\bibinfo{author}{\bibfnamefont{T.}~\bibnamefont{Boubl\'{\i}k}},
  \bibinfo{journal}{J. Chem. Phys.} \textbf{\bibinfo{volume}{53}},
  \bibinfo{pages}{471} (\bibinfo{year}{1970}).

\bibitem{GH72}
\bibinfo{author}{\bibfnamefont{E.~W.} \bibnamefont{Grundke}} \bibnamefont{and}
  \bibinfo{author}{\bibfnamefont{D.}~\bibnamefont{Henderson}},
  \bibinfo{journal}{Mol. Phys.} \textbf{\bibinfo{volume}{24}},
  \bibinfo{pages}{269} (\bibinfo{year}{1972}).

\bibitem{LL73}
\bibinfo{author}{\bibfnamefont{L.~L.} \bibnamefont{Lee}} \bibnamefont{and}
  \bibinfo{author}{\bibfnamefont{D.}~\bibnamefont{Levesque}},
  \bibinfo{journal}{Mol. Phys.} \textbf{\bibinfo{volume}{26}},
  \bibinfo{pages}{1351} (\bibinfo{year}{1973}).

\bibitem{MCSL71}
\bibinfo{author}{\bibfnamefont{G.~A.} \bibnamefont{Mansoori}},
  \bibinfo{author}{\bibfnamefont{N.~F.} \bibnamefont{Carnahan}},
  \bibinfo{author}{\bibfnamefont{K.~E.} \bibnamefont{Starling}},
  \bibnamefont{and}
  \bibinfo{author}{\bibfnamefont{J.}~\bibnamefont{T.~W.~Leland}},
  \bibinfo{journal}{J. Chem. Phys.} \textbf{\bibinfo{volume}{54}},
  \bibinfo{pages}{1523} (\bibinfo{year}{1971}).

\bibitem{HMLC96}
\bibinfo{author}{\bibfnamefont{D.}~\bibnamefont{Henderson}},
  \bibinfo{author}{\bibfnamefont{A.}~\bibnamefont{Malijevsk\'{y}}},
  \bibinfo{author}{\bibfnamefont{S.}~\bibnamefont{Lab\'{\i}k}},
  \bibnamefont{and} \bibinfo{author}{\bibfnamefont{K.~Y.} \bibnamefont{Chan}},
  \bibinfo{journal}{Mol. Phys.} \textbf{\bibinfo{volume}{87}},
  \bibinfo{pages}{273} (\bibinfo{year}{1996}).

\bibitem{YCH96}
\bibinfo{author}{\bibfnamefont{D.~H.~L.} \bibnamefont{Yau}},
  \bibinfo{author}{\bibfnamefont{K.-Y.} \bibnamefont{Chan}}, \bibnamefont{and}
  \bibinfo{author}{\bibfnamefont{D.}~\bibnamefont{Henderson}},
  \bibinfo{journal}{Mol. Phys.} \textbf{\bibinfo{volume}{88}},
  \bibinfo{pages}{1237} (\bibinfo{year}{1996}); \textbf{91}, 1137 (1997).


\bibitem{HC98}
\bibinfo{author}{\bibfnamefont{D.}~\bibnamefont{Henderson}} \bibnamefont{and}
  \bibinfo{author}{\bibfnamefont{K.~Y.} \bibnamefont{Chan}},
  \bibinfo{journal}{J. Chem. Phys.} \textbf{\bibinfo{volume}{108}},
  \bibinfo{pages}{9946} (\bibinfo{year}{1998});   Mol. Phys. \textbf{94}, 253
  (1998); \textbf{98}, 1005 (2000).

\bibitem{HBCW98}
\bibinfo{author}{\bibfnamefont{D.}~\bibnamefont{Henderson}},
  \bibinfo{author}{\bibfnamefont{D.}~\bibnamefont{Boda}},
  \bibinfo{author}{\bibfnamefont{K.~Y.} \bibnamefont{Chan}}, \bibnamefont{and}
  \bibinfo{author}{\bibfnamefont{D.~T.} \bibnamefont{Wasan}},
  \bibinfo{journal}{Mol. Phys.} \textbf{\bibinfo{volume}{95}},
  \bibinfo{pages}{131} (\bibinfo{year}{1998}).

  \bibitem{MHC99}
D. Matyushov, D. Henderson, and K.-Y. Chan, Mol. Phys. \textbf{96},
1813 (1999).

\bibitem{CCHW00}
\bibinfo{author}{\bibfnamefont{D.}~\bibnamefont{Cao}},
  \bibinfo{author}{\bibfnamefont{K.-Y.} \bibnamefont{Chan}},
  \bibinfo{author}{\bibfnamefont{D.}~\bibnamefont{Henderson}},
  \bibnamefont{and} \bibinfo{author}{\bibfnamefont{W.}~\bibnamefont{Wang}},
  \bibinfo{journal}{Mol. Phys.} \textbf{\bibinfo{volume}{98}},
  \bibinfo{pages}{619} (\bibinfo{year}{2000}).


\bibitem{ML97}
D. V. Matyushov and B. M. Ladanyi, J. Chem. Phys. \textbf{107}, 5815
(1997).

\bibitem{BS00}
\bibinfo{author}{\bibfnamefont{C.}~\bibnamefont{Barrio}} \bibnamefont{and}
  \bibinfo{author}{\bibfnamefont{J.~R.} \bibnamefont{Solana}},
  \bibinfo{journal}{J. Chem. Phys.} \textbf{\bibinfo{volume}{113}},
  \bibinfo{pages}{10180} (\bibinfo{year}{2000}).

\bibitem{VS02}
D. Viduna and W. R. Smith, Mol. Phys. \textbf{100}, 2903
(2002); J. Chem. Phys. \textbf{117}, 1214 (2002).

\bibitem{H75}
\bibinfo{author}{\bibfnamefont{D.}~\bibnamefont{Henderson}},
  \bibinfo{journal}{Mol. Phys.} \textbf{\bibinfo{volume}{30}},
  \bibinfo{pages}{971} (\bibinfo{year}{1975}).

\bibitem{SHY95}
A. Santos, M. L\'opez de Haro, and S. B. Yuste, J. Chem. Phys. \textbf{103},
 4622 (1995);
 M. L\'opez de Haro, A. Santos, and S. B. Yuste,  Eur. J. Phys. \textbf{19}, 281 (1998).

 \bibitem{L01}
S. Luding, Phys. Rev. E \textbf{63}, 042201 (2001); S.~Luding, Adv.
Compl. Syst. {\bf 4}, 379 (2002); S.~Luding and O.~Strau\ss{},
 in {\em
Granular Gases}, T.~P\"oschel and S.~Luding, eds. (LNP 564,
Springer-Verlag, Berlin, 2001),  pp.\ 389--409.

\bibitem{W63}
M. S. Wertheim, {Phys. Rev. Lett.} {\bf 10}, 321 (1963);
E. Thiele, J. Chem. Phys. \textbf{39}, 474 (1963).


\bibitem{CS69}
\bibinfo{author}{\bibfnamefont{N.~F.} \bibnamefont{Carnahan}} \bibnamefont{and}
  \bibinfo{author}{\bibfnamefont{K.~E.} \bibnamefont{Starling}},
  \bibinfo{journal}{J. Chem. Phys.} \textbf{\bibinfo{volume}{51}},
  \bibinfo{pages}{635} (\bibinfo{year}{1969}).

\bibitem{LM90}
\bibinfo{author}{\bibfnamefont{M.}~\bibnamefont{Luban}} \bibnamefont{and}
  \bibinfo{author}{\bibfnamefont{J.~P.~J.} \bibnamefont{Michels}},
  \bibinfo{journal}{Phys. Rev. A} \textbf{\bibinfo{volume}{41}},
  \bibinfo{pages}{6796} (\bibinfo{year}{1990}).

  \bibitem{H94}
\bibinfo{author}{\bibfnamefont{E.}~\bibnamefont{Hamad}}, \bibinfo{journal}{J.
  Chem. Phys.} \textbf{\bibinfo{volume}{101}}, \bibinfo{pages}{10195}
  (\bibinfo{year}{1994}).

\bibitem{V98}
\bibinfo{author}{\bibfnamefont{C.}~\bibnamefont{Vega}}, \bibinfo{journal}{J.
  Chem. Phys.} \textbf{\bibinfo{volume}{108}}, \bibinfo{pages}{3074}
  (\bibinfo{year}{1998}).

\bibitem{THM99}
\bibinfo{author}{\bibfnamefont{N.~M.} \bibnamefont{Tukur}},
  \bibinfo{author}{\bibfnamefont{E.~Z.} \bibnamefont{Hamad}}, \bibnamefont{and}
  \bibinfo{author}{\bibfnamefont{G.~A.} \bibnamefont{Mansoori}},
  \bibinfo{journal}{J. Chem. Phys.} \textbf{\bibinfo{volume}{110}},
  \bibinfo{pages}{3463} (\bibinfo{year}{1999}).

\bibitem{SYH99}
\bibinfo{author}{\bibfnamefont{A.}~\bibnamefont{Santos}},
  \bibinfo{author}{\bibfnamefont{S.~B.} \bibnamefont{Yuste}}, \bibnamefont{and}
  \bibinfo{author}{\bibfnamefont{M.}~\bibnamefont{{L\'opez de Haro}}},
  \bibinfo{journal}{Mol. Phys.} \textbf{\bibinfo{volume}{96}},
  \bibinfo{pages}{1} (\bibinfo{year}{1999}).

\bibitem{MV99}
\bibinfo{author}{\bibfnamefont{A.}~\bibnamefont{Malijevsk\'{y}}}
  \bibnamefont{and} \bibinfo{author}{\bibfnamefont{J.}~\bibnamefont{Veverka}},
  \bibinfo{journal}{Phys. Chem. Chem. Phys.} \textbf{\bibinfo{volume}{1}},
  \bibinfo{pages}{4267} (\bibinfo{year}{1999}).

\bibitem{SYH01}
\bibinfo{author}{\bibfnamefont{A.}~\bibnamefont{Santos}},
  \bibinfo{author}{\bibfnamefont{S.~B.} \bibnamefont{Yuste}}, \bibnamefont{and}
  \bibinfo{author}{\bibfnamefont{M.}~\bibnamefont{{L\'opez de Haro}}},
  \bibinfo{journal}{Mol. Phys.} \textbf{\bibinfo{volume}{99}},
  \bibinfo{pages}{1959} (\bibinfo{year}{2001}).


\bibitem{GAH01}
M. Gonz\'alez-Melchor, J. Alejandre, and M. L\'opez de Haro, J.
Chem. Phys. \textbf{114}, 4905 (2001).

\bibitem{HYS02}
\bibinfo{author}{\bibfnamefont{M.}~\bibnamefont{{L\'opez de Haro}}},
  \bibinfo{author}{\bibfnamefont{S.~B.} \bibnamefont{Yuste}}, \bibnamefont{and}
  \bibinfo{author}{\bibfnamefont{A.}~\bibnamefont{Santos}},
  \bibinfo{howpublished}{ Phys. Rev. E \textbf{66}, 031202} (\bibinfo{year}{2002}).

\bibitem{S99}
 A. Santos,  Mol.
Phys. \textbf{96}, 1185 (1999); \textbf{99}, 617(E) (2001).


\bibitem{RDA01}
C. Regnaut, A. Dyan, and S. Amokrane, Mol. Phys. \textbf{99}, 2055
(2001); \textbf{100}, 2907(E) (2002).

\bibitem{SYH02}
A. Santos, S. B. Yuste, and M. L\'opez de Haro, J. Chem. Phys.
\textbf{117}, 5785 (2002).

\bibitem{SYH05}
\bibinfo{author}{\bibfnamefont{A.}~\bibnamefont{Santos}},
  \bibinfo{author}{\bibfnamefont{S.~B.} \bibnamefont{Yuste}} \bibnamefont{and}
  \bibinfo{author}{\bibfnamefont{M.}~\bibnamefont{{L\'opez de Haro}}},
  \bibinfo{journal}{J. Chem. Phys.} \textbf{\bibinfo{volume}{123}},
  \bibinfo{pages}{234512} (\bibinfo{year}{2005});
M. L\'opez de Haro, S. B. Yuste, and A. Santos, Mol. Phys.
\textbf{104}, 3461 (2006).

\bibitem{LS04}
S. Luding and A. Santos,  J. Chem. Phys. \textbf{121}, 8458 (2004).

\bibitem{BMLS96}
\bibinfo{author}{\bibfnamefont{M.}~\bibnamefont{Baro\v{s}ov\'a}},
  \bibinfo{author}{\bibfnamefont{A.}~\bibnamefont{Malijevsk\'y}},
  \bibinfo{author}{\bibfnamefont{S.}~\bibnamefont{Lab\'{\i}k}},
  \bibnamefont{and} \bibinfo{author}{\bibfnamefont{W.~R.} \bibnamefont{Smith}},
  \bibinfo{journal}{Mol. Phys.} \textbf{\bibinfo{volume}{87}},
  \bibinfo{pages}{423} (\bibinfo{year}{1996}).

\bibitem{H-GR06}
H. Hansen-Goos and R. Roth,  J. Chem. Phys. \textbf{124}, 154506
(2006).

\bibitem{E90}
R. Evans, in \textit{Liquids and Interfaces}, edited by J.
Charvolin, J. F. Joanny, and J. Zinn-Justin (North-Holland,
Amsterdam, 1990).


\bibitem{R89}
Y. Rosenfeld, Phys. Rev. Lett. \textbf{63}, 980
(1989).

\bibitem{MBS97}
\bibinfo{author}{\bibfnamefont{A.}~\bibnamefont{Malijevsk\'y}},
  \bibinfo{author}{\bibfnamefont{M.}~\bibnamefont{Baro\v{s}ov\'a}},
  \bibnamefont{and} \bibinfo{author}{\bibfnamefont{W.~R.} \bibnamefont{Smith}},
  \bibinfo{journal}{Mol. Phys.} \textbf{\bibinfo{volume}{91}},
  \bibinfo{pages}{65} (\bibinfo{year}{1997}).

\bibitem{MMYSH02}
Al. Malijevsk\'y, A. Malijevsk\'y, S. B. Yuste, A. Santos, and M.
L\'opez de Haro, Phys. Rev. E \textbf{66}, 061203 (2002).

\bibitem{BPW04}
M. Buzzacchi, I. Pagonabarraga, and N. B. Wilding, J. Chem. Phys.
\textbf{121}, 11362 (2004).

\bibitem{Alexander}
Al. Malijevsk\'y, S. B. Yuste, A. Santos, and M. L\'opez de Haro,
preprint arXiv: cond-mat/0702284.


\bibitem{L96}
F. Lado, Phys. Rev. E \textbf{54}, 4411 (1996).


\bibitem{PL54}
I. Prigogine and S. Lafleur, Bull. Classe Sci. Acad. Roy.
Belg. \textbf{40}, 484, 497 (1954).

\bibitem{AO54}
S. Asakura and F. Oosawa, J. Chem. Phys. \textbf{22}, 1255 (1954);
J. Polym. Sci. \textbf{33}, 183 (1958).

\bibitem{K55}
R. Kikuchi, J. Chem. Phys. \textbf{23}, 2327 (1955).

\bibitem{BPGG86}P. Ballone, G. Pastore, G. Galli, and D. Gazzillo, Mol. Phys. \textbf{59},
275 (1986).

\bibitem{GPE89}
D. Gazzillo, G. Pastore, and S. Enzo, J. Phys.: Condens. Matter
\textbf{1}, 3469 (1989);
D. Gazzillo, G. Pastore, and R. Frattini, J. Phys.: Condens. Matter
\textbf{2},8465 (1990).

\bibitem{SHY05}
A. Santos, M. L\'opez de Haro, and S. B. Yuste,  J. Chem. Phys.
\textbf{122}, 024514 (2005).

\bibitem{H96b}
E. Z. Hamad, J. Chem. Phys. \textbf{105}, 3229 (1996).

\bibitem{H96a}
E. Z. Hamad, J. Chem. Phys. \textbf{105}, 3222 (1996).

\bibitem{H96c}
H. Hammawa and E. Z. Hamad, J. Chem. Soc. Faraday
Trans. \textbf{92}, 4943 (1996).

\bibitem{H99}
M. Al-Naafa, J. B. El-Yakubu, and E. Z. Hamad, Fluid Phase Equilibria \textbf{154},
33 (1999).



\bibitem{JJR94a}
J. Jung, M. S. Jhon, and F. H. Ree, J. Chem. Phys. \textbf{100}, 528 (1994).

\bibitem{JJR94b}
J. Jung, M. S. Jhon, and F. H. Ree, J. Chem. Phys. \textbf{100}, 9064 (1994).

\bibitem{CB98b}
T. Coussaert and M. Baus, J. Chem. Phys. \textbf{109}, 6012
(1998).


\bibitem{VM03}
A. Yu. Vlasov and A. J. Masters, Fluid Phase Equilibria
\textbf{212}, 183 (2003).

\bibitem{HT04}
M. L\'opez de Haro and C. F. Tejero,   J. Chem. Phys. \textbf{121},
6918 (2004).

\bibitem{YSH00a}
\bibinfo{author}{\bibfnamefont{S.~B.} \bibnamefont{Yuste}},
  \bibinfo{author}{\bibfnamefont{A.}~\bibnamefont{Santos}}, \bibnamefont{and}
  \bibinfo{author}{\bibfnamefont{M.}~\bibnamefont{{L\'opez de Haro}}},
  \bibinfo{journal}{Europhys. Lett.} \textbf{\bibinfo{volume}{52}},
  \bibinfo{pages}{158} (\bibinfo{year}{2000}).



\bibitem{CFP91}
H.-O. Carmesin, H. L. Frisch, and J. K. Percus, J.
Stat. Phys. \textbf{63}, 791 (1991).

\bibitem{SH05}
A. Santos and M. L\'opez de Haro, Phys. Rev. E \textbf{72},
010501(R) (2005).

\bibitem{REL01}
{R. Roth, R. Evans, and A. A. Louis, Phys. Rev. E \textbf{64},
051202 (2001).}

\bibitem{YS91}
S. B. Yuste and A. Santos, {Phys. Rev. A}
 {\bf 43}, 5418 (1991).

\bibitem{YHS96}
S. B. Yuste, M. L\'{o}pez de Haro, and A. Santos, {Phys. Rev. E}
{\bf 53}, 4820 (1996).

\bibitem{RHSY98}
M. Robles, M. L\'opez de Haro, A. Santos, and S. B. Yuste,  J. Chem.
Phys. \textbf{108}, 1290 (1998).

\bibitem{RH03}
M. Robles and M. L\'opez de Haro, Europhys. Lett. \textbf{62}, 56
(2003).

\bibitem{RH97}
M. Robles and M. L\'{o}pez de Haro, {J. Chem. Phys.}
 {\bf 107}, 4648 (1997).


\bibitem{W73}
E. Waisman, Mol. Phys. \textbf{25}, 45 (1973); D. Henderson and L.
Blum, Mol. Phys. {\bf 32}, 1627 (1976); J. S. H{\o}ye and L. Blum,
J. Stat. Phys. {\bf 16}, 399 (1977).


\bibitem{DLS06}
A. D\'{\i}ez, J. Largo, and J. R. Solana, J. Chem. Phys.
\textbf{125}, 074509 (2006).

\bibitem{KLM04}
J. Kolafa, S. Lab\'ik, and A. Malijevsk\'y, {Phys. Chem. Chem.
Phys.} {\bf 6}, 2335 (2004). {See also
http://www.vscht.cz/fch/software/hsmd/ for molecular dynamics
results of $g(r)$.}

\bibitem{TNJH05}
A. Trokhymchuk, I. Nezbeda, J. Jirs\'ak, and D. Henderson, {J. Chem.
Phys.} {\bf 123}, 024501 (2005).


\bibitem{HSY06}
M. L\'opez de Haro, A. Santos, and S. B. Yuste,  J. Chem. Phys.
\textbf{124}, 236102 (2006).

\bibitem{L95}
L. L. Lee, J. Chem. Phys. \textbf{103}, 9388 (1995); L. L. Lee, D.
Ghonasgi, and E. Lomba, J. Chem. Phys. \textbf{104}, 8058 (1996); L.
L. Lee and A. Malijevsk\'y, J. Chem. Phys. \textbf{114}, 7109
(2001).

\bibitem{LM84}
S. Lab\'{\i}k and A. Malijevsk\'y, Mol. Phys. \textbf{53}, 381
(1984).


\bibitem{MS06}
Al. Malijevsk\'y and A. Santos,  J. Chem. Phys. \textbf{124}, 074508
(2006).

\bibitem{SM06}
A. Santos and Al. Malijevsk\'y, Phys. Rev. E \textbf{75}, 021201
(2007).

\bibitem{YSH98a}
S. B. Yuste, A. Santos, and M. L\'opez de Haro,  J. Chem. Phys.
\textbf{108}, 3683 (1998).

\bibitem{BH77}
L. Blum and J. S. H{\o}ye, J. Phys. Chem. \textbf{81}, 1311 (1977).

\bibitem{AW92}
{J. Abate and W. Whitt},  {Queuing Systems} {\bf 10}, 5 (1992).

\bibitem{notebook}
A code using the Mathematica computer algebra system to obtain
$G_{ij}(s)$ and $g_{ij}(r)$ with the present method is available
from the web page
http://www.unex.es/eweb/fisteor/santos/filesRFA.html.

\bibitem{AL67}
N. W. Ashcroft and D. C. Langreth, Phys. Rev.  \textbf{156}, 685
(1967).

\bibitem{BT70}
A. B. Bathia and D. E. Thornton, Phys. Rev. B \textbf{8}, 3004
(1970).

\bibitem{YSH00}
S. B. Yuste, A. Santos, and M. L\'opez de Haro, Mol. Phys.
\textbf{98}, 439 (2000).

\bibitem{B68}
R. J. Baxter, J. Chem. Phys. {\bf 49}, 2270 (1968).

\bibitem{PS75}
J. W. Perram and E. R. Smith, {Chem. Phys. Lett.} \textbf{35}, 138
(1975).

\bibitem{BT79}  B. Barboy, Chem. Phys. {\bf 11}, 357 (1975); B. Barboy and
R. Tenne, Chem. Phys. {\bf 38}, 369 (1979).

\bibitem{BJG97}
G. Stell, J. Stat. Phys. {\bf 63}, 1203 (1991);
B. Bor\v{s}tnik, C. G. Jesudason, and G. Stell, J. Chem. Phys. {\bf
106}, 9762 (1997).


\bibitem{BB74}  B. Barboy, J. Chem. Phys. {\bf 61}, 3194 (1974).

\bibitem{varios_SHS}
J. W. Perram and E. R. Smith, Chem. Phys. Lett. {\bf 39}, 328
(1975);
P. T. Cummings, J. W. Perram, and E. R. Smith, Mol. Phys. {\bf 31},
535 (1976);
E. R. Smith and J. W. Perram, J. Stat. Phys.{\bf \ 17}, 47 (1977);
J. W. Perram and E. R. Smith, Proc. R. Soc. London A{\bf 353}, 193
(1977);
W. G. T. Kranendonk and D. Frenkel, { Mol. Phys.} \textbf{64}, 403
(1988);
C. Regnaut and J. C. Ravey, J. Chem. Phys. {\bf 91}, 1211 (1989);
G. Stell and Y. Zhou, J. Chem. Phys. {\bf 91}, 3618 (1989);
J. N. Herrera and L. Blum, J. Chem. Phys. {\bf 94}, 6190 (1991);
A. Jamnik, D. Bratko, and D. J. Henderson, J. Chem. Phys. {\bf 94},
8210 (1991);
S. V. G. Menon, C. Manohar, and K. S. Rao, {J. Chem. Phys.};
\textbf{95}, 9186 (1991);
Y. Zhou and G. Stell, J. Chem. Phys. {\bf 96}, 1504 (1992);
E. Dickinson, J. Chem. Soc. Faraday Trans. {\bf 88}, 3561 (1992);
C. F. Tejero and M. Baus, { Phys. Rev. E}, {\bfseries 48}, 3793
(1993);
K. Shukla and R. Rajagopalan, Mol. Phys. {\bf 81}, 1093 (1994);
C. Regnaut, S. Amokrane, and Y. Heno, J. Chem. Phys. {\bf  102},
6230 (1995);
C. Regnaut, S. Amokrane, and P. Bobola, Prog. Colloid Polym. Sci.
{\bf 98}, 151 (1995);
Y. Zhou, C. K. Hall, and G. Stell, Mol. Phys. {\bf 86}, 1485 (1995);
J. N. Herrera-Pacheco and J. F. Rojas-Rodr\'{\i}guez, Mol. Phys.
{\bf 86}, 837 (1995);
Y. Hu, H. Liu, and J. M. Prausnitz, J. Chem. Phys. {\bf 104}, 396
(1996);
O. Bernard and L. Blum, J. Chem. Phys. {\bf 104}, 4746 (1996);
L. Blum, M. F. Holovko, and I. A. Protsykevych, J. Stat. Phys. {\bf
84}, 191 (1996);
S. Amokrane, P. Bobola and C. Regnaut, Prog. Colloid Polym. Sci.
{\bf 100}, 186 (1996);
S. Amokrane and C. Regnaut, J. Chem. Phys. {\bf 106}, 376 (1997);
C. Tutschka, G. Kahl, and E. Riegler, {Mol. Phys.} \textbf{100},
1025 (2002);
D. Gazzillo and A. Giacometti, {Mol. Phys.} \textbf{100}, 3307
(2002);
M. A. Miller and D. Frenkel, { Phys. Rev. Lett.} {\bfseries 90},
135702 (2003);
D. Gazzillo and A. Giacometti, {J. Chem. Phys.} \textbf{120}, 4742
(2004);
R. Fantoni, D. Gazzillo, and A. Giacometti, Phys. Rev. E
\textbf{72}, 011503 (2005);
A. Jamnik, Chem. Phys. Lett. \textbf{423}, 23 (2006).

\bibitem{SG87}
A. J. Post and E. D. Glandt, { J. Chem. Phys.} \textbf{84}, 4585
(1986); N. A. Seaton and E. D. Glandt, J. Chem. Phys. \textbf{84},
4595 (1986); \textbf{86}, 4668 (1986); \textbf{87}, 1785 (1987).

\bibitem{MF04a}
M. A. Miller and D. Frenkel, { J. Phys.: Condens. Matter}
{\bfseries 16}, S4901 (2004).

\bibitem{MF04b}
M. A. Miller and D. Frenkel, { J. Chem. Phys.} {\bfseries 121},
535 (2004).

\bibitem{MYS06}
Al. Malijevsk\'y, S. B. Yuste, and A. Santos,  J. Chem. Phys.
\textbf{125}, 074507 (2006).

\bibitem{SYH98}
A. Santos, S. B. Yuste, and M. L\'opez de Haro,  J. Chem. Phys.
\textbf{109}, 6814 (1998).

\bibitem{YS93b}
S.~B. Yuste and A. Santos, { J. Stat. Phys.} {\bfseries 72}, 703
(1993).

\bibitem{YS93c}
S. B. Yuste and A. Santos, {  Phys. Rev. E} {\bfseries 48}, 4599
(1993).

\bibitem{YS94}
S. B. Yuste and A. Santos,  J. Chem. Phys. \textbf{101}, 2355
(1994).

\bibitem{AS01}
L. Acedo and A. Santos,  J. Chem. Phys. \textbf{115}, 2805 (2001).

\bibitem{A00}
L. Acedo, J. Stat. Phys. \textbf{99}, 707 (2000).

\bibitem{LSAS03}
J. Largo, J. R. Solana, L. Acedo, and A. Santos,  Mol. Phys.
\textbf{101}, 2981 (2003).

\bibitem{LSYS05}
J. Largo, J. R. Solana, S. B. Yuste, and A. Santos, J. Chem. Phys.
\textbf{122}, 084510 (2005).

\bibitem{FI81}
\bibinfo{author}{\bibfnamefont{C.}~\bibnamefont{Freasier}} \bibnamefont{and}
  \bibinfo{author}{\bibfnamefont{D.~J.} \bibnamefont{Isbister}},
  \bibinfo{journal}{Mol. Phys.} \textbf{\bibinfo{volume}{42}},
  \bibinfo{pages}{927} (\bibinfo{year}{1981}).

\bibitem{L84}
\bibinfo{author}{\bibfnamefont{E.}~\bibnamefont{Leutheusser}},
  \bibinfo{journal}{Physica A} \textbf{\bibinfo{volume}{127}},
  \bibinfo{pages}{667} (\bibinfo{year}{1984}).

\bibitem{RHS04}
M. Robles, M. L\'opez de Haro, and A. Santos,  J. Chem. Phys.
\textbf{120}, 9113 (2004).



\bibitem{CRR76}
D. G. Chae, F. H. Ree, and T. Ree, J. Chem. Phys. \textbf{50}, 1581
(1976).

\bibitem{YS93a}
S. B. Yuste and A. Santos,  J. Chem. Phys. \textbf{99}, 2020 (1993).

\bibitem{MC69}
G. A. Mansoori and F. B. Canfield, {J. Chem. Phys.}
 {\bf 51}, 4958 (1969).

\bibitem{MPC69}
G. A. Mansoori, J. A. Provine, and F. B. Canfield, {J. Chem. Phys.}
{\bf 51}, 5295 (1969).

\bibitem{RS70}
J. Rasaiah and G. Stell, {Mol. Phys.} {\bf 18}, 249 (1970).

\bibitem{BH67}
J. A. Barker and D. Henderson, {J. Chem. Phys.}
{\bf 47}, 2856 (1967).

\bibitem{note}
The macroscopic compressibility approach is only one of the
possibilities of approximation to the second order Barker--Henderson
perturbation theory term. Another successful approach is the
local-compressibility approximation (see Ref.\ \protect\cite{M76},
p.\ 308). This expresses the free energy  in terms of $\phi _{1}( r)
$ and HS quantities.

\bibitem{WCA71}
J. D. Weeks, D. Chandler, and H. C. Andersen, {J. Chem.
Phys.}  {\bf 53}, 149 (1971).

\bibitem{VW72}
A simple algorithm to compute a rather accurate approximation for
the HS diameter $\diam$ in the WCA theory has been given in L.
Verlet and J. J. Weis, {Phys. Rev. A}  {\bf 5}, 939 (1972).

\bibitem{HG75}
D. Henderson and E. W. Grundke, { J. Chem. Phys.} {\bf  63}, 601
(1975).

\bibitem{BS85}
J. A. Ballance and R. J. Speedy, { Mol. Phys.}
 {\bf 54}, 1035 (1985).

\bibitem{ZS88}
Y. Zhou and G. Stell, { J. Stat. Phys.}  {\bf 52}, 1389 (1988).

\bibitem{HV69}
J.-P. Hansen and L. Verlet, { Phys. Rev.} {\bf 184}, 151 (1969).

\bibitem{S94}
R. J. Speedy, J. Chem. Phys. \textbf{100}, 6684 (1994).

\bibitem{RH01}
M. Robles and M. L\'opez de Haro,   {Phys. Chem. Chem. Phys.}
\textbf{3}, 5528 (2001).

\bibitem{HR04}
M. L\'opez de Haro and M. Robles, J. Phys.: Condens. Matt.
\textbf{16}, S2089 (2004).

\bibitem{HR06}
M. L\'opez de Haro and M. Robles,  Physica A \textbf{372}, 307
(2006).

\bibitem{L01b}
C. N. Likos, Phys. Rep. \textbf{348},  267 (2001).

\bibitem{J99}
E. A. Jagla, J. Chem. Phys. \textbf{111}, 8980 (1999).






















\end{thebibliography}
%



\printindex
\end{document}